\documentclass{usbthesis}

\usepackage{graphicx,amsmath,amssymb,array,calc,rotating,epsfig,psfrag}
\usepackage[nosort]{cite}
\usepackage{color}
\usepackage{latex_package/appendix}

\title{Applications of instantons to hadronic processes}

\author{Valeriu Ioan Zetocha}

\month{May} \year{2004 \\ $\mbox{}$
\\$\mbox{}$  
}

\director{Thomas Schaefer}{Associate Professor, Nuclear physics theory
, SBU}
\chairman{Edward Shuryak}{Professor, Nuclear physics theory,
SBU }
\fstmember{Konstantin Likharev}{Distinguished Professor, Department of Physics,
SBU }
\outmember{Raju Venugopalan}{Physicist, Department of Physics, Brookhaven National Laboratory}

\begin{document}

\maketitle
\makeapproval

\begin{abstract}

Instantons constitute an important part of QCD as they provide a way 
to reach behind the perturbative region. In the introductory chapters we 
present, in the framework of a simple standard integral, the ideas that 
constitute the backbone of instanton computation.
 We explain why instantons are crucial for capturing non-perturbative 
aspects of any theory and get a feel for zero mode difficulties and moduli
 space. Within the same setting we explore the configuration space 
further by showing how constrained instantons and instanton valleys 
come into play.

We then turn our attention to QCD instantons and briefly show the steps 
to compute the effective lagrangian. We also show how single instanton 
approximation arises and how one can use it to evaluate correlation functions.
 By this we set the stage for the main parts of the thesis: computation 
of $\eta_c$ decay and evaluation of nucleon vector and axial vector couplings.

 Having understood the effective lagrangian we use it as a main tool for
 studying instanton contributions to hadronic decays of the
scalar glueball, the pseudoscalar charmonium state $\eta_c$,
and the scalar charmonium state $\chi_c$. Hadronic decays of
the $\eta_c$ are of particular interest. The three main decay
channels are $K\bar{K}\pi$, $\eta\pi\pi$ and $\eta'\pi\pi$,
each with an unusually large branching ratio $\sim 5\%$.
On the quark level, all three decays correspond to an instanton
type vertex $(\bar{c}c)(\bar{s}s)(\bar{d}d)(\bar{u}u)$. We show
that the total decay rate into three pseudoscalar mesons can
be reproduced using an instanton size distribution consistent
with phenomenology and lattice results. Instantons correctly
reproduce the ratio $B(\pi\pi\eta)/B(\pi\pi\eta')$ but
over-predict the ratio $B(K\bar{K}\pi)/B(\pi\pi\eta(\eta'))$.
We consider the role of scalar resonances and suggest that the
decay mechanism can be studied by measuring the angular
distribution of decay products.

 In the next part, motivated by measurements of the flavor singlet axial 
coupling constant of the nucleon in polarized deep inelastic 
scattering we study the contribution of instantons to OZI 
violation in the axial-vector channel. We consider, in particular, 
the $f_1-a_1$ meson splitting, the flavor singlet and triplet
axial coupling of a constituent quark, and the axial coupling
constant of the nucleon. We show that instantons provide a short 
distance contribution to OZI violating correlation functions
which is repulsive in the $f_1$ meson channel and adds 
to the flavor singlet three-point function of a constituent 
quark. We also show that the sign of this contribution
is determined by general arguments similar to the Weingarten
inequalities. We compute long distance contributions using 
numerical simulations of the instanton liquid. We find that 
the iso-vector axial coupling constant of a constituent quark 
is $(g_A^3)_Q=0.9$ and that of a nucleon is $g_A^3=1.28$, in good
agreement with experiment. The flavor singlet coupling of 
quark is close to one, while that of a nucleon is suppressed
$g_A^0=0.8$. This number is still significantly larger than 
the experimental value $g_A^0=(0.28-0.41)$.

Throughout the instanton computation in QCD one employs integration
 over the $SU(2)$ group as the $SU(2)$ parameters are part of the 
moduli space. The techniques of group integration are especially useful for
computing the effective lagrangian. 
We therefore present an algorithm for computation of integrals 
over compact groups 
that is both simple and easy to implement. The main idea was mentioned 
before by Michael Creutz 
but, to our knowledge, never carried out 
completely. We exemplify it on 
integrals over $SU(N)$ of type
$\int du (u {u}^\dagger)^n$, with $n=1,2,3$
as well as integrals of adjoint representation matrices  $\int du 
(R^{ab})^n, n=1,\ldots 4$, $R^{ab}=\frac{1}{2}Tr(\lambda^b u \lambda^a 
u^\dagger)$.

\end{abstract}

\tableofcontents

\begin{acknowledgements}

First of all, I am deeply indebted to my adviser Thomas Schaefer for
all his professional help and moral support during my Ph.D. research period.
I enjoyed learning under Thomas' guidance. While steering me in the right 
direction Thomas gave me a lot of freedom to broaden my knowledge and
get a better perspective on and beyond physics. The discussions with him
have always been fruitful, usually solving my problems on the spot with his
characteristic style of to-the-point remarks spiced with fine humor.
I would also like to thank Thomas for providing the numerical computations
of correlation functions in instanton liquid model. 

I have benefited greatly from discussions with professors at Stony Brook, 
like Sasha Abanov, Gerry Brown,  Madappa Prakash,
 Edward Shuryak and Ismail Zahed.
I would especially like to thank Edward Shuryak for eye-opening
discussions on instantons. I always felt that talking to Edward was like getting
a chance to take a bird eye look on physics. Everything started to relate and
suddenly I could see the wood despite the trees.

During the years at Stony Brook I learned a lot from the excellent lectures
of George Sterman, Edward Shuryak, Peter van Nieuwenhuizen and Ismail Zahed.
I also benefited a lot from talking to my colleagues like Pietro Faccioli,
 Radu Ionas, Tibor Kucs, Peter Langfelder, Achim
Schwenk and Diyar Talbayev.

Further I wish to thank my professors and teachers from Slovakia and Romania
for opening the doors to physics for me. I am especially grateful to
my high school teachers Adrian Rau-Lehoczki and Stefan Beraczko as well as
to my undergraduate adviser at Comenius University, Bratislava,
 prof. Peter Presnajder.

There is no one I am more indebted to than my family. My brother Klaudy
and my parents Ondrej and Albinka have always been a supportive pillar for 
my studies. I am sincerely grateful for all their help and sacrifices 
that made this Ph.D. thesis possible.

\end{acknowledgements}

\pagenumbering{arabic}


\renewcommand{\a}{\alpha}
\renewcommand{\d}{\delta}
\newcommand{\x}{\times}
\newcommand{\beq}{\begin{equation}}
\newcommand{\eeq}{\end{equation}}
\newcommand{\lb}{\lambda}
\newcommand{\g}{\gamma}
\newcommand{\e}{\epsilon}
\newcommand{\ts}{\theta^*}
\newcommand{\te}{\theta}
\newcommand{\st}{\sigma_T}
\newcommand{\s}{\sigma}
\renewcommand{\t}{\tau}
\newcommand{\kp}{\kappa}


\newcommand{\bea}{\begin{eqnarray}}
\newcommand{\eea}{\end{eqnarray}}
\newcommand{\be}{\begin{equation}}
\newcommand{\ee}{\end{equation}}
\newcommand{\Dslash}{D\!\!\!\!/\,}
\newcommand{\qslash}{q\!\!\!/}
\newcommand{\slashed}[1]{#1\!\!\!/}
\newcommand{\smn}{\sigma_{\mu\nu}}
\newcommand{\sng}{\sigma_{\nu\gamma}}
\newcommand{\sgm}{\sigma_{\gamma\mu}}
\renewcommand{\r}{\rho}
\newcommand{\gpm}{\gamma_{\pm}}
\newcommand{\gmp}{\gamma_{\mp}}
\newcommand{\gf}{\gamma_{5}}
\newcommand{\noto}{\to\hspace{-0.47cm}/\hspace{0.3cm}}
\newcommand{\ivec}[1]{\stackrel{\leftarrow}{#1}}
\newcommand{\ebe}{\eta\!\!\!\!\!-}

\newcommand {\pslash}{p\!\!\!/}
\newcommand {\muslash}{\mu\!\!\!/}

\renewcommand{\aleph}{\mathcal{N}}
\newcommand{\bi}{\begin{itemize}}
\newcommand{\ei}{\end{itemize}}
\newcommand{\vp}{\vec{\Phi}}
\newcommand{\gri}{\frac{\partial S}{\partial\Phi_i}}
\newcommand{\grj}{\frac{\partial S}{\partial\Phi_i}}
\newcommand{\D}{{\mathcal{D}}^2}
\newcommand{\pc}{\phi_{cl}}
\newcommand{\pt}{\phi_{\tau}}
\newcommand{\pts}{\phi_{{\tau}^*}}
\newcommand{\DA}{\mathbf{\int\!\!\mathcal{D}A}}
\newcommand{\Dm}{\ensuremath{\mathbf{D}_\mu}}
\newcommand{\Db}{\mathbf{\mathcal{D}b}}
\newcommand{\Dc}{\mathbf{\mathcal{D}c}}
\newcommand{\ebmna}{\bar\eta_{\mu\nu}^a}
\newcommand{\ebmnb}{\bar\eta_{\mu\nu}^a}
\newcommand{\emna}{\eta_{\mu\nu}^a}
\newcommand{\emnb}{\eta_{\mu\nu}^b}



\chapter{Introduction}
\label{intro_chapter}
The physics of fundamental interactions is dominated today by the Standard Model(SM), the combined theory of strong, weak and electromagnetic force. 
Since its development in 1970's it has proved extremely powerful in
 explaining the experimental results. 
The model has been so successful that, for thirty years, physicists have been
 desperately looking for a discrepancy between the model and experiment 
that would point them towards new physics "beyond the standard model".

Quantum chromodynamics(QCD), as a part of the Standard Model 
describes the gluon-mediated strong interactions between quarks. The origins of QCD date back to early 1960's. The myriad of observed particles and their mass spectrum played then the same role as Mendeleev's periodic table of elements a century ago: it pointed to the existence of underlying constituents, particles that represent a more elementary form of matter.

In 1964 Gell Mann and Zweig introduced spin-$\frac{1}{2}$ particles:
 up, down and strange quarks\footnote{charm, bottom and top were added later}
with fractional charge of $\frac{2}{3}$ for $u$ quark and $-\frac{1}{3}$ for 
$d$ and $s$ quarks.
 Assigning mesons to $\bar{q}q$ states and baryons to $qqq$ states 
using $SU(3)_{flavor}$ symmetry led to a good match of known 
particles and valuable predictions of new ones. (Valuable indeed, as Gell Mann was rewarded with Nobel Prize in 1969 for his "Eightfold Way").

However, the straight-forward quark model had to overcome the difficulty of 
reconciling the Pauli principle and 
the seemingly quark-exchange symmetric function of 
baryons made of 3 same-flavor quarks with the same spin, like 
$\Delta^{++}\sim u^{\uparrow}u^{\uparrow}u^{\uparrow}$.
The solution came with an extra quantum number: color, that made it possible to antisymmetrize the baryonic state.
Quarks would then come in three colors: red, green and blue and the hadrons
would be composed of white combinations of quarks. Color-anticolor pairs of
quark - antiquark would form mesons while baryons would be of type 
$\epsilon^{ijk}q_i q_j q_k$.

At this stage, the colored quarks were little more than mathematical
objects that explained the spectrum of observed hadrons. 
The unsuccessful search for free spin-$\frac{1}{2}$ particles with 
fractional charge presented a big obstacle for the quark model to become more
 than a nice mathematical description of hadronic spectra. In late 1960's, 
SLAC-MIT deep inelastic scattering experiments discovered point-like 
constituents, "partons", later identified with quarks.
Eventually the discovery of 
$J/\Psi$ in 1974 in both hadroproduction and $e^+e^-$ annihilation finalized
the conclusion that the quarks are real particles but are confined 
to colorless combinations.

The main conclusion of DIS experiments was the asymptotic freedom
of proton constituents, which basically means that the interaction of partons
is small for high energy transfer($\gtrsim 1\;{\rm GeV}$). This was an extra 
feature that any theory of strong interactions would need to possess.

QCD, the gauge theory of quarks based on $SU(3)$ group then emerged as the only viable candidate that would incorporate $N_c=3$ quarks, have a different
representation for antiquarks and would feature asymptotic freedom 
and confinement.

On its way to maturity, QCD underwent a long series of experimental tests, 
like $\pi^0 \to \gamma\gamma$ decay, deep inelastic scattering, $e^+e^- \to hadrons$ and so on. Most of the early successes were predominantly in the 
perturbative area of large momentum transfer and hence small coupling.
In this sector, the methods of computation had already been known from QED
and the Feynman diagrams techniques were readily available.

A completely different view was provided by lattice gauge theory methods, which
use discretization of large but finite volume 
of spacetime and evaluate the path integral using Monte Carlo techniques.
With increasing computational power the lattice QCD is becoming an
accessible laboratory for non-perturbative physics. The main drawback of
 lattice QCD is the absence of analytic results from which one could get a 
better insight into the physical picture.

In the non-perturbative regime different analytic approaches have been developed
 that describe the physics at different energy scales. For small 
momenta well below
1 GeV the energetically accessible degrees of freedom are pions and other 
low-lying mesons. Therefore effective theories like the ones based
on chiral lagrangians were natural tools successfully applied to physics
of low energy pions. On the other side of particle spectrum, one can use the 
fact, that low energy interactions are less likely to create new heavy
 quark-antiquark pairs due to energy gaps. Therefore
 heavy quarkonia are accessible through non-relativistic quantum mechanical 
models similar to positronium treatment.

Instantons bridge the gap between the effective field theories and perturbative
QCD. With their origins in the heart of QCD theory, they constitute one of the 
best understood non-perturbative tools.
Instantons saturate the $U(1)_A$ anomaly and provide dynamical chiral 
symmetry braking. However, they do not solve the problem of confinement,
which even after 30 years of work is still an open question.

The main tool for instanton computations is the instanton liquid model (ILM)
developed in 1980's. Is is based on the assumption that the vacuum is 
dominated by instantons.
The parameters of the model, the mean density of 
instantons $n \cong 1\; {\rm fm^{-4}}$ and the average size of instantons
$\rho \cong \frac{1}{3}\; {\rm fm}$ were fitted
 from the phenomenological values 
of quark and gluon condensates.

Among the successes of ILM are calculations of hadronic correlation functions,
hadronic masses and coupling constants. A better understanding of chiral
phase transition has also been achieved by studying instantons in 
finite temperature QCD.

In addition to that, we would like to identify direct instanton contributions
to hadronic processes. One possibility is to go to very high energy, for example in DIS. In this case, instantons are very rare, but they lead to special 
processes with multi-gluon and quark emission, analogous to the baryon number 
violating instanton process in the electroweak sector.

In this work we focus on another possibility. The instanton induced interaction 
has very peculiar spin and flavor correlations that distinguish instantons 
from perturbative forces. We study two systems in which unusual flavor and spin effects have been observed: the decay of $\eta_c$ and the so-called "proton
spin crisis".

First part of this work analysis the decay of $\eta_c$.
The dominant decay channels have a specific structure of final 
products in which all pairs of constituent quarks are present. The resulting 
vertex is of type $(\bar{c}c)(\bar{s}s)(\bar{d}d)(\bar{u}u)$ which points
to a possible instanton-induced mechanism behind the process.

The experimental evidence that quarks only carry about $30\%$ of the proton
spin is contrary to the naive quark model predictions and it implies
a large amount of OZI violation in the flavor singlet axial vector channel.
The second part of this work studies the instanton contribution to this process.



\section{High school instantons}

\label{toy_model_section}

\subsection{Toy model instantons}
Instanton is a finite-action solution of the euclidean equation of motion. Since it's discovery in  QCD in 1975 by Belavin et al \cite{Belavin:fg} 
it has been enjoying 
considerable attention. In this section we will try to convey a feel for 
instanton physics without dipping into the details of any particular theory.

Many of the features of instanton computation can be easily explained
on trivial examples that do not require more than high school integration.
We will use the standard integral as an analog for the path integral. This way 
the ideas of the computation will be unveiled in full light without the 
technical difficulties blocking the view.  

The single most important object in quantum field theory is the partition 
function, which can be represented as the path integral\footnote{For simplicity we take directly the euclidean spacetime}
$$
Z[J] = \int \mathcal{D} \Phi e^{-S[\Phi,J]}
$$
where $\Phi(x)$ is the field and $S[\Phi,J]$ is the action that also depends on the source $J$.
Any correlation function can be computed once we know the partition function. 
Schematically:
$$
<\Phi_1\Phi_2> = \delta_{J_1}\delta_{J_2} Z[J]
$$
where $J_1,J_2$ are the sources corresponding to $\Phi_1,\Phi_2$.

\begin{figure}
\begin{center}
\includegraphics[width=6cm,angle=-90]{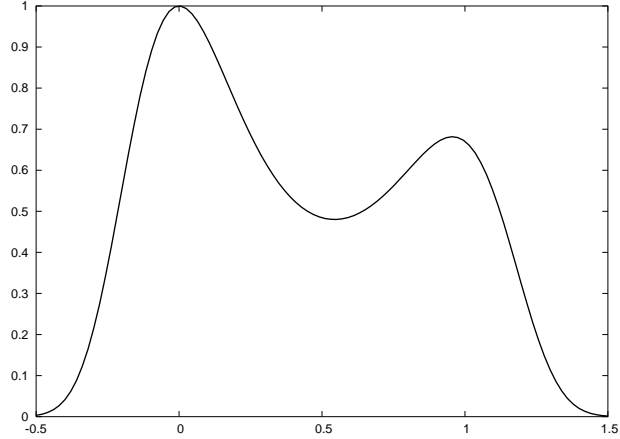}
\end{center}
\begin{center}
\caption{\label{Dwell_e_S_fig}
Exp(-S) for a tilted double well potential.
}
\end{center}
\end{figure}

It is usually not a trivial exercise to compute $Z[J]$. In fact, most of the 
time we are forced to rely on some kind of approximation. The standard 
technique is to expand the action around the trivial minimum ($\Phi\equiv 0$), 
keep the quadratic terms and treat the rest as a perturbation:
\be
e^{-S[\Phi]} = e^{-S[0] - \frac{1}{2}\Phi S''[0] \Phi + O(\Phi^3)}
= e^{ - \frac{1}{2}\Phi S''[0] \Phi }(1 + O(\Phi^3) + \cdots )
\ee
However, the perturbation approach usually does not, and can not, reach all
the 'dark corners' of the theory. The reason is the possible existence of 
other, non-trivial minima of the action, located in a different sector of the
configuration space.
\begin{figure}
\begin{center}
\includegraphics[width=6cm,angle=-90]{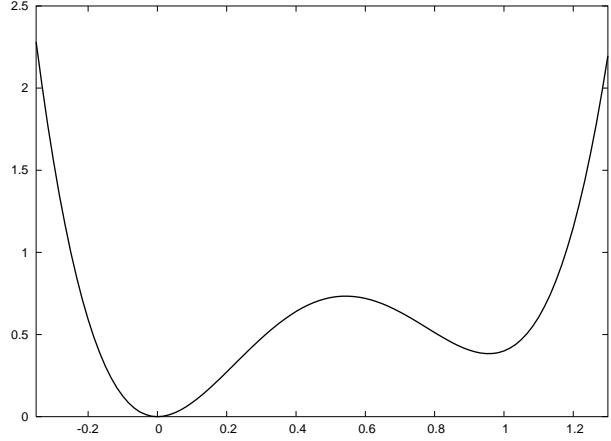}
\end{center}
\begin{center}
\caption{\label{double_well_fig}
Action of a tilted double well potential.
}
\end{center}
\end{figure}
To get a better feel about the approximations to the path integral, let us 
consider the most trivial toy-model for the path integral: QFT for a
 field in one point in spacetime.
The configuration space is now considerably shrunk to one single variable,
and the path integral is just a standard integral. As an example of theory
with 'instanton', let us take a tilted double well potential shown in Fig.
\ref{double_well_fig}:
\be
S(\Phi) = \Phi^2 (b^2(\Phi - a)^2 + m).
\ee
The partition function is:
\be
Z = \int_{-\infty}^{+\infty} d\Phi \: e^{-S(\Phi)}
\ee
Fig. \ref{Dwell_e_S_fig} presents the graph of $e^{-S}$. The goal is to 
compute the area under the curve.
The approach corresponding to standard perturbation theory is to keep the quadratic terms in exponent and expand the rest:
\be
e^{-S(\Phi)} = e^{-(m+a^2b^2)\Phi^2}[1+O(\Phi^3)+\cdots]
\ee

\begin{figure}
\begin{center}
\includegraphics[width=6cm,angle=-90]{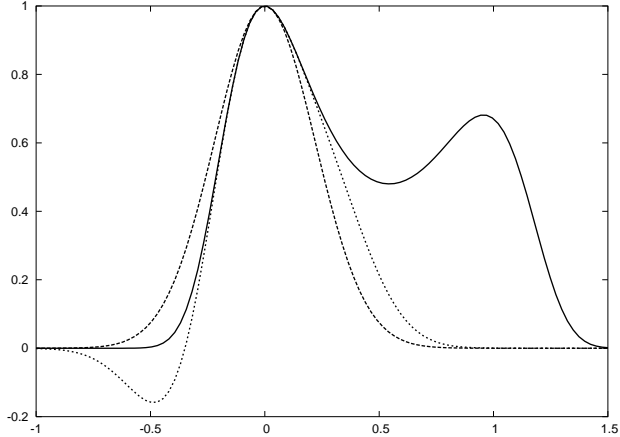}
\end{center}
\begin{center}
\caption{\label{Dwell_2expan_fig}
Zeroth (dashed line) and first (dotted line) order expansion 
around the trivial minima of the action
 are far away from giving a reasonable approximation to the $exp(-S)$. Bad news for perturbation theory: with higher order expansions, it gets worse before it
gets better.
}
\end{center}
\end{figure}

\begin{figure}
\begin{center}
\includegraphics[width=6cm,angle=-90]{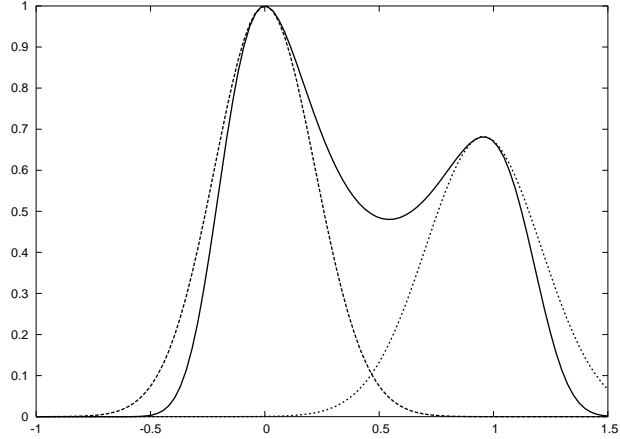}
\end{center}
\begin{center}
\caption{\label{Dwell_both_min_fig}
The combined contribution of zeroth order expansions around both minima gives
a clearly better approximation to the full integral. 
}
\end{center}
\end{figure}

The graphs of zeroth and first order expansions are shown in 
Fig. \ref{Dwell_2expan_fig}. It is clear that in order to capture the full integral 
one needs a large order expansion.\footnote{As we will see later, in QCD
the graph under the smaller 'bump' is not accessible at all this way, as the
bump happens in a disconnected part of configuration space.
} 
We can get a much better approximation of partition function by including the
'bump' from the very beginning: just compute the additional area by repeating
 the perturbation approach for the smaller bump. Mathematically, its peak is 
at the minimum of the action, which is nothing else than the instanton.
 This separate term is the 'instanton' contribution:
\be
Z_I = \int d\Phi\: e^{-S(\Phi_I) 
- \frac{1}{2}
\left.
\frac{d^2 S}{d\Phi^2}
\right|_{\Phi=\Phi_I}
(\Phi-\Phi_I)^2}(1 + \cdots )
\ee
The total area under the curve would be the sum of the results of perturbation
approach under both total and local minima of the action:
$$ Z = Z_0 + Z_I .$$
The zeroth order expansions under both minima are shown in 
Fig. \ref{Dwell_both_min_fig}. It is again obvious that this is a much better 
approximation of the integral than what we started with. A choice of 
$a=1$, $b=10$ and $m=0.4$ gives $Z_0=0.55$, $Z_I=0.43$ and $Z_0+Z_I=0.98$. The 
numerical integration gives $Z = 0.99$.

In real models trivial fields and instantons live in 
separate, disconnected parts of configuration space. Therefore one not only
 needs to add instantons to get a better approximation with less computation,
but has to account for instanton part in order to achieve the right result.

\subsection{Zero modes and moduli space}

There is a long shot from the trivial toy model to real world physics and many
 difficulties arise on the way. One of them, the zero modes and the moduli
 space is an omni-present feature that has to be dealt with in every instanton 
computation. The idea is easily explained on a 'next-to-trivial' model of
two spacetime points, i.e. a two-dimensional integral. 
\begin{figure}
\begin{center}
\includegraphics[width=7cm,angle=0]{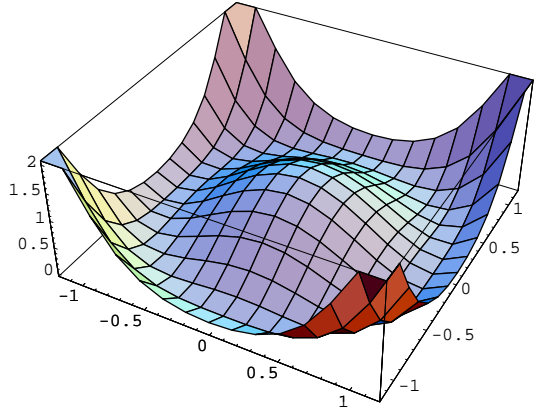}
\end{center}
\begin{center}
\caption{\label{sombrero_fig}
Sombrero action.
}
\end{center}
\end{figure}

\begin{figure}
\begin{center}
\includegraphics[width=7cm,angle=0]{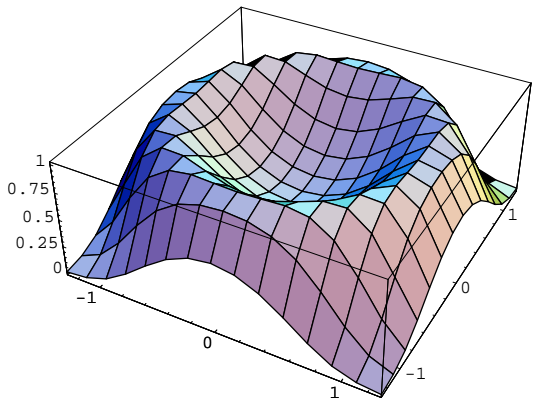}
\end{center}
\begin{center}
\caption{\label{sombrero_eS_fig}
Exp(-S) for the sombrero action.
}
\end{center}
\end{figure}

Let us consider the following 'sombrero' action, depicted in Fig. 
\ref{sombrero_fig}, with the graph of $exp(-S)$ in Fig. \ref{sombrero_eS_fig}:
$$
S = (x^2 + y^2 -1)^2
$$
As a result of the presence of rotational
symmetry, we have a continuum of minima of action. 
The set of all 'instantons' is a circle with radius 1. 
The parameters that describe it are called collective coordinates.
In our case we can use an angle $\phi \in (0,2\pi)$:
$$
(x,y)_I(\phi) = (\cos{\phi},\sin{\phi}) .
$$
The parameter space of all instantons is called the moduli space and in our case
has the topology of a circle.

Suppose we did not have the err function available and decided to compute
the integral numerically in a similar way we did it in the previous case of 
one-dimensional integral. We would need to expand around the minimum of
the action. But which one should we choose?
Let's say we just pick an arbitrary minimum. The zeroth order expansion will now
have a form of a two-dimensional distorted Gaussian bell, with different curvatures in different directions. If we press ahead and 
try to compute the volume under such a bell, we quickly run into difficulties:
the result is infinite. Obviously this is not a problem of the integral, but 
of the method we used: the radius in one direction of the approximating 
two-dimensional Gaussian bell is infinity(the surface is flat in one direction 
corresponding to the symmetry direction).
Therefore none of the minima alone is suited for an expansion. 

This kind of difficulty appears every time we deal with a symmetry of the
action that is broken by instanton. For each symmetry, there is an associated 
direction in the configuration space along which the action does not change.
The vector that points in this direction is the zero mode. In our case it is
the vector corresponding to a rotation, i.e. $\partial_\phi$, with $\phi$
being the azimuthal angle\footnote{In general one obtains the zero mode by
differentiating the instanton solution with respect to the collective
coordinate}:
$$\vec{z}_\phi=
\frac{d}{d\phi}\left(x,y\right)_I(\phi) = (-\sin{\phi},\cos{\phi})
$$
The solution of the problem is obvious: one needs to turn the symmetry into an advantage, not a liability. The change of variables to polar coordinates is in
order. To use the lingo of instanton physics, we will integrate over 
the direction of the zero mode non-perturbatively, leaving the perturbation
method for the modes perpendicular to this. In other words, in one direction
nothing changes, so we should just compute the one dimensional integral in
the perpendicular direction and then multiply the result by the volume of 
the space in the zero mode direction.
\begin{figure}
\begin{center}
\includegraphics[width=8cm,angle=0]{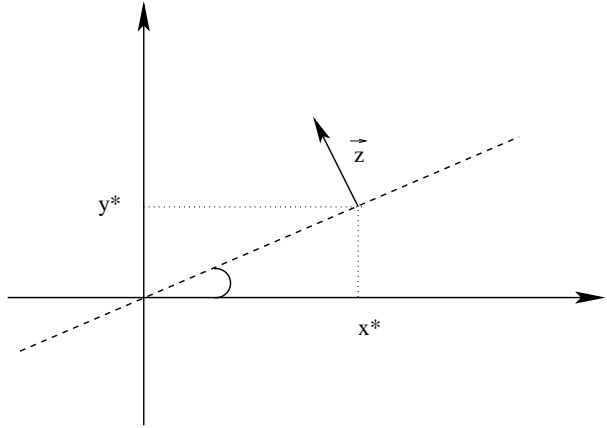}
\end{center}
\begin{center}
\caption{\label{ray_fig}
At any point $(x^\star,y^\star)$ there is a zero mode direction and 
a 'perturbative' direction. We integrate perturbatively along the ray while the 
integration over the azimuthal angle just gives the volume $2\pi$ of the moduli
 space. 
}
\end{center}
\end{figure} 

Let us take as a starting point of our integration an arbitrary point
$(x^\star,y^\star)\ne (0,0)$. The one-dimensional space perpendicular 
to the zero mode in this point is the line 
$(x,y)\cdot\vec{z}_{\phi}\equiv -x\sin{\phi} + y\cos{\phi} = 0$
with $\tan{\phi}=\frac{y^\star}{x^\star}$ - see Fig \ref{ray_fig}.
For every point on the line, we only want to integrate radially, and leave the
perpendicular direction for later. This can be achieved by introducing
the following delta function in the integral:
$$
\delta((x,y)\cdot \vec{z}_{\phi})
=\delta(-x\sin{\phi} + y\cos{\phi})
$$
which forces the integration points lay on the line.
To obtain the contribution over the whole plane we only need to integrate
over the angle, with a weight that makes the whole insertion a unity
\footnote{This is nothing else than the famous Faddeev-Popov unity insertion}:
$$
\Delta(x,y)\int_0^{2\pi}\delta(-x\sin{\phi} + y\cos{\phi})d\phi = 1
$$
It is easy to compute $\Delta(x,y) =\frac{1}{2}\sqrt{x^2+y^2}$. The 
whole integral then is:
$$
\int\! dxdy\;e^{-S(x,y)}\frac{1}{2}\sqrt{x^2+y^2}
\int_0^{2\pi}\delta(-x\sin{\phi} + y\cos{\phi})d\phi
$$
After a change of variables
\bea
x'& =& x\cos{\phi} - y\sin{\phi}\nonumber\\
y'& =& -x\sin{\phi} + y\cos{\phi}\nonumber
\eea
and evaluation of integral over $y'$ one obtains
\be\label{2D_result}
\int_0^{2\pi}\!\d\phi \x \frac{1}{2}\int_{-\infty}^{+\infty}
\!dx'\;e^{-S(x',0)}\sqrt{x'^2}
\ee
The result is nothing else but the volume of moduli space multiplied
by the integral in radial direction(integral over non-zero modes).
Now we could expand the radial integral around the 2 bumps it contains
in the very same way we did it in the example from the beginning of this 
section.

The main step in dealing with zero modes is 
their separation from 'perturbative' direction by requiring 
the scalar product to be zero:
\be\label{scalar_prod}
(x,y)\cdot \vec{z}_\phi = 0
\ee 
Another interesting interpretation is the following. The question is around 
which of the continuum of instantons one should expand. A very intuitive
approach would be to expand different parts of space around different
instanton, mainly, the $closest$ one. Let us take again the starting point
$(x^\star,y^\star)$. The closest instanton is given by the minimization of
the distance
$$
min_{\phi} ||(x^\star,y^\star) - (\cos{\phi},\sin{\phi}||^2
$$
A differentiation w.r.t. $\phi$ gives the same condition \ref{scalar_prod}. 
One would then separate the configuration space into blocks,
 each one of them being dominated
by the closest instanton. In our case a block would be a ray with angle $\phi$
and the points on the ray would be under the 'jurisdiction' of the instanton
at $(\cos{\phi},\sin{\phi})$. Adding the contributions of all rays would lead
to the same result \ref{2D_result}.

\subsection{Valley instantons and all that}

We should be on a pretty good footing now that we know how to deal with
a continuum of minima. For any integral we would find all minima and expand 
around them, paying special attention to treatment of zero modes. 
This kind of 'turning the crank' could actually lead to a poor result
in some cases, when there are a lot of important points without 
satisfying the condition of minimum.

\begin{figure}
\begin{center}
\includegraphics[width=7cm,angle=0]{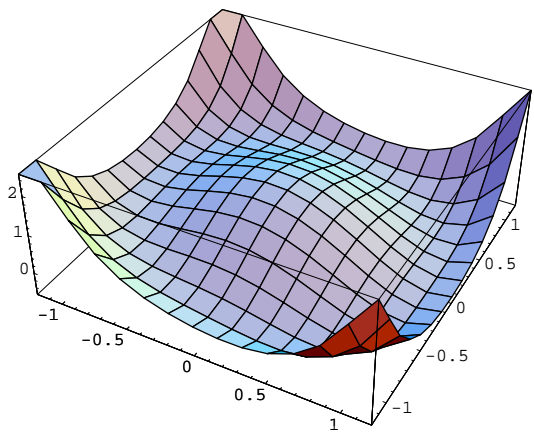}
\end{center}
\begin{center}
\caption{\label{tilted_sombrero_S_fig}
Tilted sombrero action.
}
\end{center}
\end{figure}

\begin{figure}
\begin{center}
\includegraphics[width=7cm,angle=0]{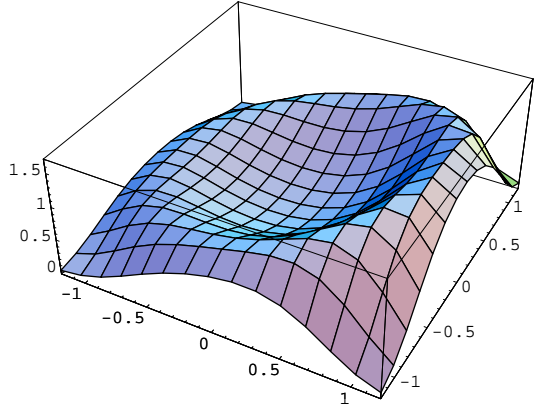}
\end{center}
\begin{center}
\caption{\label{tilted_sombrero_e_S_fig}
exp(-S) for tilted sombrero action.
}
\end{center}
\end{figure}

The presence of an 'almost zero mode', a direction with very low gradient, 
would make the Gaussian bell around the minima a poor approximation 
of the integral. To get a 
better understanding, let us take again the sombrero action and slightly
tilt it:
$$
S = (x^2+y^2 - 1)^2 -\epsilon x
$$
The graph of the tilted sombrero action with $\epsilon = \frac{1}{2}$ is
shown in Fig. \ref{tilted_sombrero_S_fig}. The surface now exhibits a single 
global minimum - the instanton - close to the point $(1,0)$ and a saddle
point close to $(-1,0)$. The exponent of tilted action is shown in Fig.
 \ref{tilted_sombrero_e_S_fig}.

If $\epsilon$ is small, the volume under the surface of $e^{-S}$ of the 
tilted sombrero will not differ significantly from the original one. In fact,
one could obtain an $\epsilon$-expansion of the integral. However, our method
of expanding around minima would fail, as the single Gaussian bell centered
at the global minimum of $S$ would not describe well the whole surface.

Since $\epsilon$ is small, we should, in principle do something similar to
$\epsilon = 0$ case discussed in the previous section. For that reason, we 
first need to identify the important points - 'valley instantons' -
that are dominant, in some way, for ${\rm exp}(-S)$ 
in their restricted vicinity. Let us postpone for a moment the strict 
definition of the valley points.
Intuitively it is clear, that in our case, the special points would form a loop $\epsilon$-close
to the former circle of instantons. Once we have the valley trajectory, we 
would integrate perturbatively in the sector perpendicular to valley in the 
same way we treated the non-zero modes before. Now the final integral along the
valley trajectory would not simply give the volume of the would-be moduli
space, as the result of integration over non-valley directions 
depends on the parameter of the given point on trajectory.

We still have not defined precisely the valley trajectory. There are actually
two approaches with slightly different results: the streamline 
\cite{Yung:1987zp,Balitsky:qn} 
and the so-called proper valley \cite{Aoyama:1995ca} method. 

The streamline method requires a starting point located on the trajectory but
 different 
from the global minimum. In our case it could be the saddle point. Having the
starting point, the instanton valley is constructed dynamically by following
the highest slope downwards. It is exactly the path a stream of water 
would follow - hence the name.

The proper valley method features more stability and does not require
any starting point. The valley is now made of points with the lowest gradient
along the contour of equal action. One should therefore draw the contour lines
('izoactas') and for each of them identify the point with lowest gradient.
Joining these points would give the trajectory.

Once we have the trajectory, we can start turning the crank again: at each 
point of the valley integrate perturbatively over modes perpendicular to the 
trajectory and then integrate over the trajectory of the valley.

Another approach to the same problem uses so-called 'constrained instantons'
 \cite{Affleck:1980mp,Nielsen}.
The idea is very simple: slice the configuration space in surfaces given by a 
family of functions. In our case one could take the slices generated
by the family of vertical planes $x=\alpha$, with 
$\alpha \in (-\infty,+\infty)$. For example, the slice generated by
$x=0$ is just the standard double well potential in $y$ direction.
Each slice would feature constrained minima which one could use to
compute the integral over the slice. At the end, one would only need to
sum over the slices, i.e. integrate over $\alpha$.

The ambiguity of choosing the slicing function makes this approach less
appealing. One has to have some physical intuition to use it with success. In
our case, the almost-symmetry of the graph would point to slicing by rays
$-x\sin{\phi} + y\cos{\phi} = 0$. 

There is a lot one can learn about the methods of path integration from a 
standard 2 dimensional high-school integral. We will now take the earned 
intuition and apply it to instantons in QCD. 

\section{Instantons in QCD}
In this section we will focus on providing the main ideas for the derivation
 of the effective Lagrangian as well as on setting the stage for using the
 single instanton approximation for computing correlation functions.
 A thorough review on instantons in QCD can be found in \cite{Schafer:1996wv}.

As mentioned before, instantons are finite action 
solutions of equation of motion in Euclidean spacetime.
 For a Yang-Mills theory, 
the partition function reads: 
$$Z=\DA e^{-S_{YM}}$$  
where
$$S_{YM}=\frac{1}{4g^2}\int d^4x 
F^a_{\mu\nu}F^a_{\mu\nu}$$
and $F_{\mu\nu}=\partial_\mu A_\nu - \partial_\nu A_\mu + [A_\mu,A_\nu]$ is the field strength. 
Let us for simplicity consider $SU(2)$ YM theory.
The requirement of a finite action leads to fields that tend to pure gauge at
infinity: $A_{\mu} \to u^{-1}\partial_\mu u $, with $u\in SU(2)$. The 
infinity is topologically a 
3-sphere. The instanton at infinity is therefore a map from the spatial 
$S^3$ to $S^3$ of $SU(2)$ parameters. Every such a mapping is characterized 
by a winding number, an integer that shows how many times one sphere is wrapped around the other by the map.

\begin{figure}
\begin{center}
\includegraphics[width=10cm,angle=0]{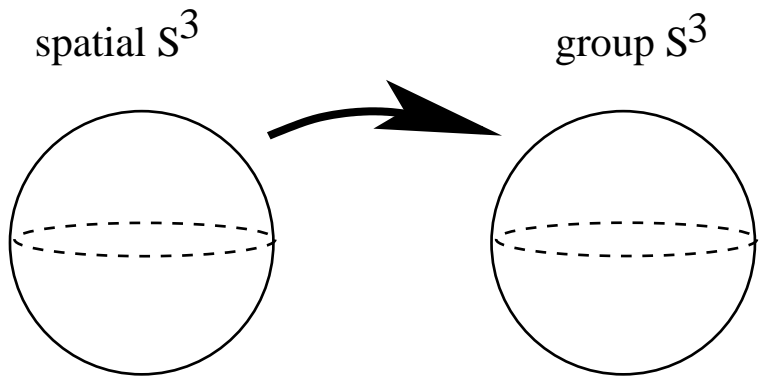}
\end{center}
\begin{center}
\caption{\label{winding_fig}
The instanton at infinity is a mapping from spatial 3-sphere to 3-sphere 
of $SU(2)$ parameters. 
Each such mapping is characterized by integer winding number.
}
\end{center}
\end{figure} 

\begin{figure}
\begin{center}
\includegraphics[width=8cm,angle=0]{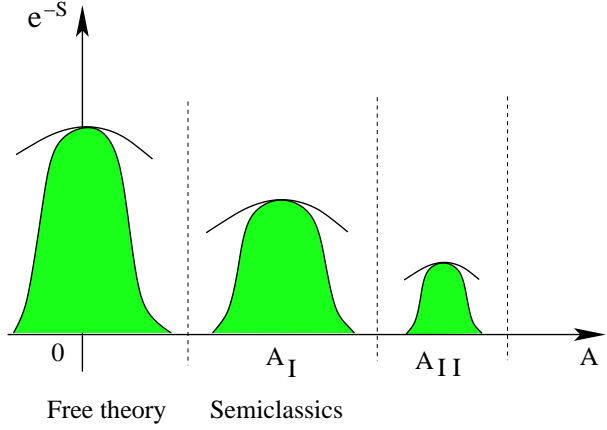}
\end{center}
\begin{center}
\caption{\label{inst_gauss_fig}
Schematic representation of quadratic approximation in different sectors of the
configuration space.
}
\end{center}
\end{figure}

\begin{figure}
\begin{center}
\includegraphics[width=6cm,angle=0]{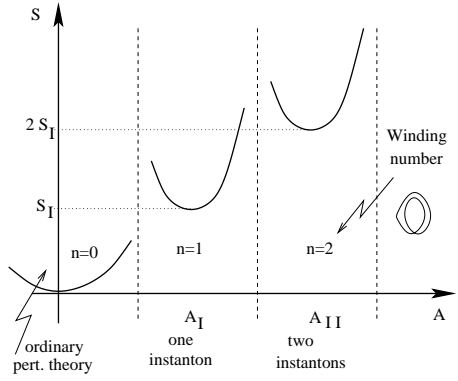}
\end{center}
\begin{center}
\caption{\label{inst_winding_fig}
YM action in different winding number sectors 
}
\end{center}
\end{figure}

The whole configuration space of finite-action gauge fields is then 
separated into distinct sectors
characterized by different values of winding number. 
Figs. {\ref{inst_gauss_fig}}
 and {\ref{inst_winding_fig}} 
show schematically the action $S$ and $e^{-S}$ in the sectors of $n=0$ (zero winding number), $n=1$ (one instanton) and $n=2$ (two instantons).
The sectors are completely separated by infinite action walls, in the sense that 
any trajectory from one sector to another will have points where 
$S_{YM}=\infty$. It is exactly because of this separation that one has to 
account for instanton contributions, as one can not retrieve them from $n=0$ 
sector.

Before plunging into any instanton computation, one has to deal first 
with gauge invariance. As any other symmetry, it induces zero modes and flat
directions, derailing the perturbative approach.
The way around this difficulty was explained on the toy model of 
section {\ref{toy_model_section}}: integrate over zero modes direction 
non-perturbatively, then compute quantum fluctuations perpendicular to zero 
modes. In the case of gauge invariance, this means fixing the space
for quantum fields by requiring
$$\Dm^{I}A_\mu^{qu} = 0$$
where $\Dm^{I}$ is the covariant derivative in a background field.
After exponentiating the Fadeev-Popov determinant, the action becomes a functional of ghost fields $b$ and $c$:
$$Z=\DA\Db\Dc\: e^{-\mathcal{S}[A,b,c]}$$
$$\mathcal{S}=\int d^4x\left\{ \frac{1}{4g^2}F^a_{\mu\nu}F^a_{\mu\nu} +
\frac{1}{2}(\Dm^IA_\mu^a)^2 + \Dm^Ib^a\Dm^Ic^a\right\}$$

\subsection{Effective Lagrangian  for gauge fields} 

The instanton solution of the equation of motion reads:
\bea
A_\mu^{(cl)}& =& 
2\rho^2\frac{\ebmna(x-x_0)_\nu}
{(x-x_0)^2((x-x_0)^2+\rho^2)}u\frac{\tau^a}{2}\bar{u} \nonumber\\
b^a &=& c^a =0
\eea

where $A_\mu^{(cl)}$ is the instanton configuration with the center at 
$x_0^\mu$ and orientation given by the $SU(2)$ matrix $u$. Here $\tau^a$ 
is the Pauli matrix and $\bar{\eta}^a_{\mu\nu}$ is the t'Hooft tensor 
with properties given e.g. in \cite{Schafer:1996wv}.

Let us now explore the semiclassical approach, i.e. perturbation
theory around the minima of the action. Expanding the action in the
functional Taylor series and keeping the terms up to second order
we obtain, for the gauge sector:
$$e^{-\mathcal{S}}=e^{-\mathcal{S}[A^{(cl)}] -\frac{1}{2} A 
M_A A 
+ ...}$$

Even after fixing the gauge, we are still left with a rigid gauge symmetry that
brings $4N - 5$ zero modes, where $N$ is the number of colors. Besides that, there are 5 flat directions 
corresponding to scaling and translational invariance. The remaining $SO(4)$
rotational symmetry does not bring any new zero modes, as any such rotation 
can be undone by a gauge rotation. In other words, the $SO(4)$ rotational
zero mode points to the space unavailable to quantum fluctuations,
as they were made perpendicular to gauge zero modes by fixing the gauge. 
Therefore altogether there are $4N$zero modes. 

The non-perturbative integration over the zero mode directions 
and subsequent computation of quantum fluctuations limited to Gaussian
approximation gives

$$\frac{Z}{Z_0}=\int \prod_{i=1}^{4N}\frac{d\gamma^i}{\sqrt{2\pi}}
\left[ 
2^4\rho^{2(4N-5)}\mathcal{S}_{cl}^{4N}
\right]^{\frac{1}{2}}
e^{-\mathcal{S}_{cl}} \times$$
$$\times
\left(
\frac{\det' M_A}{\det M_A^0}
\right) ^{-1/2}_{reg} 
\left(
\frac{\det' M_{gh}}{\det M_{gh}^0}
\right)_{reg} ,
$$ 
where prime on determinants means that zero eigenvalues are not included. 
The computation of determinants was performed in \cite{'tHooft:fv} 
and the full result is:
$$
\frac{Z}{Z_0}=C\int d^4x_0\frac{d\rho}{\rho^5}\int d\Omega 
\frac{1}{g^{4N}} e^{-\frac{8\pi^2}{g^2(\rho)}} \equiv \int d\mu_I
$$
For completeness, the constant is
$$
C=\frac{2^{4N+2}\pi^{4N-2}}{(N-1)!(N-2)!}\: e^{-\alpha(1) - 
2(N-2)\alpha(\frac{1}{2})}
$$
with
$\alpha(1)=0.443$ and
$\alpha(\frac{1}{2})=0.145$


Let us now push the computation one step further and obtain the 
instanton-induced effective Lagrangian. We are interested in computing 
the correlations of the type: 
$$
<A_\mu^a(x)A_\nu^b(y)>_I=\frac{1}{Z}\DA\Db\Dc\: 
e^{-\mathcal{S}[A,b,c]}A_\mu^a(x)A_\nu^b(y)$$
Expanding near instanton 
$A_\mu=A_\mu^{(cl)}+A_\mu^{(qu)}$ we obtain: 
$$<A_\mu^a(x)A_\nu^b(y)>_I=\int\!\!d\mu_I A_\mu^{a,(cl)}(x)A_\nu^{b,(cl)}(y) +
 <A_\mu^{a,(qu)}(x)A_\nu^{b,(qu)}(y)>
$$
Instead of computing the correlations this way, we are interested in an
effective potential $\mathbf{V}_I$ such that for large distance,
$|x-x_0|^2>>\rho^2$, one retrieves the instanton fields: 
$$
<A_\mu^a(x)A_\nu^b(y)>_{
I
} =
 <A_\mu^a(x)A_\nu^b(y) 
\mathbf{V}_I
>_{
pert
}
$$
One can check, that Callan Dashen and Gross (CDG) potential \cite{Callan:1977gz}:
$$
\mathbf{V}_I \equiv V_{CDG} = \int\!\!d\mu_I 
\exp{ 
\left[-\frac{\pi^2\rho^2}{g^2}\emnb\:
Tr[\tau^b\bar{u}\tau^a u]F^a_{\mu\nu}
\right]
}
$$
provides exactly this. To see that this is indeed the case, let us compute 
the gauge field propagator in the instanton background for large distance.
 The second order expansion in $\eta F$ gives, schematically:
$$
 <A_\mu(x)A_\nu(y)
 \mathbf{V}_I> =
\int\!\!d\mu_I <A_\mu(x)A_\nu(y) \eta F \eta F>
$$
Disregarding the vacuum bubbles and next order corrections in the coupling 
constant, one is left with terms like
$$
<A_\mu^a(x)F^b_{\alpha\beta}(x_0)>
=
\frac{2g^2}{(2\pi)^2}\delta^{ab}
\frac{[\delta_{\mu\beta}(x-x_0)_\alpha - 
\delta_{\mu\alpha}(x-x_0)_\beta]}{(x-x_0)^4}
$$

This way every $A_\mu(x)$ field will contribute a factor of 
$$
2\rho^2\frac{\emna(x-x_0)_\nu}{(x-x_0)^4} u\frac{\tau^a}{2}\bar{u}
$$
which is just the large distance limit of instanton field.

The anti-instanton fields lead to similar expressions with the substitutions
$u\leftrightarrow \bar{u}$ , $\eta \leftrightarrow \bar\eta$.

\subsection{Fermions and t'Hooft vertex}
We will introduce now the fermionic degrees of freedom and study how
this affects the computation in the instanton background.

t'Hooft's discovery of left-handed zero mode of the Dirac operator in the presence of instanton came a little bit as a surprise with huge implications.
First of all the zero mode renders the tunneling amplitude zero in case of
massless quarks, as the tunneling is proportional to the determinant of
Dirac operator. However, the instanton contribution to some Green's functions 
is non-zero and clearly distinguishable from the perturbative contribution.
In these correlations the small mass parameter $m$ from the Dirac determinant
cancels against $\frac{1}{m}$ from the zero mode propagator, as we will 
show below.

The specific helicity of zero mode also pointed to chiral symmetry braking, and
in fact solved the $\eta'$ puzzle. 

Let us now dip into some details of t'Hooft's computation. 
The fermionic part of QCD action reads:
$$
\mathcal{S}[\psi,J]=\!\!\int dx\: \bar\psi(x)(\slash\!\!\!\!D + 
m)\psi(x) 
-
\int\!\!\!\! \int dx\:dy \:\: \bar\psi_s(x)J_{st}(x,y)\psi_t(y)$$
where $J_{st}$ is the source used to generate bilinear fermionic fields.

The massless Dirac operator in the instanton background has a left-handed zero
 mode:
$$\psi^0(x)=\frac{\rho}{\pi}\frac{1}{(x^2+\rho^2)^{3/2}}
\frac{\slash\!\!\!x}{\sqrt{x^2}} 
\frac{1+\gamma_5}{2}\Phi
$$
with $\Phi^{\alpha m}=\frac{\epsilon^{\alpha m}}{\sqrt{2}}$.
The semiclassical result will now have an extra determinant:
$$
\frac{Z}{Z_0}=\!\int\!\!d\mu_I\rho^{N_f} 
m^{N_f}
 e^{[-\frac{2}{3}N_f 
\ln(\mu_0)\rho + 2N_f\alpha(1/2)]}\equiv \int\!\!d\mu_{I,f}
$$
This shows that the tunneling is suppressed in the presence of light fermions
by a factor of $m^{N_f}$.

To compute the Green's functions we differentiate with respect to the source:
$$
<\bar\psi^s_{\alpha}(x)\psi^t_\beta(y)>=
\left. 
\frac{\delta}{\delta 
J^{st}_{\alpha \beta} (x,y)} \frac{Z[J]}{Z[0]}
\right|_{J=0}.
$$
$Z[J]$ depends on $J$ through $\det M_\psi$:
$$
\det M_\psi = \det (-\slash\!\!\!\!D \delta_{st} + J_{st}).
$$

Let us take for simplicity the propagator for one flavor.
Then the determinant of Dirac operator with the source is 
given in terms of the eigenvalues $\lambda_n^J$:
$$\det M_\psi=\prod_n \lambda_n^J,$$
where $\lambda_n^J$ satisfy
$(-\slash\!\!\!\!D +J)\psi = \lambda_n^J\psi$.
Expanding $\lambda_n^J$ in powers of J we get:
$$\lambda_n^J=\lambda_n^{J=0} + \alpha_n\cdot J + \ldots$$
where $\alpha_n\cdot J =\int\!\!\int dx\:dy\:\: \alpha_n(x,y) \: J(x,y)$. 
The propagator then involves:
$$\delta_{J_{\alpha \beta}}
\left. \prod_n(\lambda_n^{J=0} + \alpha_n\cdot 
J)\right|_{J=0}=
\left. \delta_{J_{\alpha\beta}}\left[
 (\alpha_0\cdot J)
(\lambda_1 + 
\alpha_1\cdot 
J)\ldots\right]\right|_{J=0}\;.
$$
The final evaluation at $J=0$ renders all but one term zero. After simplification with $Z[0]$ we obtain 
$$<\bar{\psi}_\alpha\psi_\beta>=\frac{1}{m}\alpha_{0{\beta\alpha}}\;,$$
which is nothing else but the first correction to zero energy due to source, i.e. standard 
perturbation theory
$$(\alpha_0\cdot J)=<\psi^0|J|\psi^0>\;.$$
Then the zero mode part of the propagator reads:
$$<\bar\psi_{\alpha}(x)\psi_\beta(y)>=
\frac{1}{m}{\psi^0}^*_\alpha(x)\psi^0_\beta(y)\;.
$$
Note the $\frac{1}{m}$ factor that will cancel with $m$ from the Dirac 
determinant to give finite correlation functions.

\begin{figure}
\begin{center}
\includegraphics[width=6cm,angle=0]{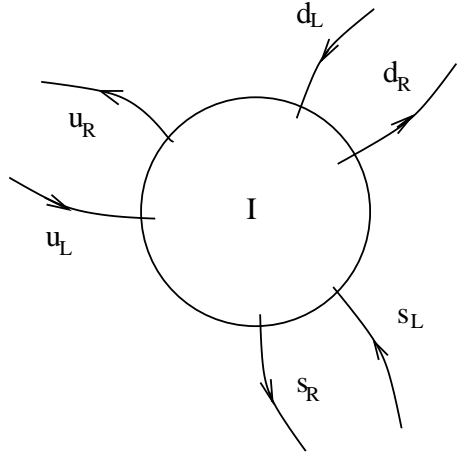}
\end{center}
\begin{center}
\caption{\label{inst_vertex_fig}
Instanton vertex contains all the pairs of light quarks. 
Particles are left-handed while anti-particles are right handed. 
}
\end{center}
\end{figure}

In the case of many massless quarks,
the operator $(-\slash\!\!\!\!D \delta_{st} + J_{st})$ has $N_f$ times 
degenerate ground state.
The surviving term in a Green's function is then the product of perturbed 
eigenvalues, which can be written as the determinant of 
'perturbation matrix':
\be
\prod_{i=1}^{N_f}(\alpha_0^i\cdot J)=det_{st}<\psi^0|J_{st}|\psi^0>
=det_{st}\int\!\!dxdy\; {\psi^0}^*(x) J_{st}(x,y)\psi^0(y)\; .\nonumber
\ee
This clearly points to the following properties of the Green's functions in 
the zero mode approximation of the instanton computation:

\begin{itemize}
\item at least 
$2N_f$ fermions
participate (otherwise the contribution is of the higher order 
than zero modes).

\item It has a determinantal structure in flavor, 
all flavors must be there in pairs

\item Quarks propagate in the left-handed mode, 
while anti-quarks in the right-handed mode.

\item One can not have 2 fermions propagating in the same zero mode( this is
the instantonic version of Pauli principle)
\end{itemize}
The graphical representation of the instanton vertex is shown in the 
Fig. \ref{inst_vertex_fig}.
Let us now construct the fermionic effective lagrangian. The idea is the same as for the gauge field effective vertex:
we need to find 
$L_{t'H}$ such that:
$$
<\bar{\psi}_\alpha(x)\psi_\beta(y)L_{t'H}>_{pert}=
\langle \bar{\psi}\psi \rangle_{inst} =
\frac{1}{m}{\psi}^*_\alpha(x)\psi_\beta(y)
$$
\be\label{zmprop}
=\frac{\rho^2}{m\pi^2}
\frac
{
\left
((\slash\!\!\!x-\slash\!\!\!x_0)\frac{1+\gamma_5}{2}\Phi
\right)^+_{\alpha}
}
{(x-x_0)^4}
\frac{
\left(
(\slash\!\!\!y-\slash\!\!\!x_0)\frac{1+\gamma_5}{2}\Phi
\right)_\beta
}
{(y-x_0)^4}
\ee
t'Hooft proposed a Lagrangian of the type:
$$S_{t'H}=\int\!\!d\mu_{I,f}K(\bar\psi\omega)(\bar\omega\psi)$$
with K a constant and $\omega$ a spinor to be determined.
First order expansion of the exponent of action gives for fermion propagator:
$$<\bar{\psi}_\alpha(x)\psi_\beta(y)(\bar\psi\omega)(\bar\omega\psi)>=
\left(
\frac{\slash\!\!\!y - 
\slash\!\!\!x_0}{2\pi^2(z-x_0)^4}
\right)
_{\beta\gamma}
\omega_\gamma\bar\omega_\delta 
\left(
\frac{\slash\!\!\!x - 
\slash\!\!\!x_0}{2\pi^2 (x-x_0)^4}
\right)
_{\delta\alpha}
$$
which leads to the r.h.s. of \ref{zmprop} if we choose
$\omega=\frac{1+\gamma_5}{2}\Phi$ , $K=4\rho^2\pi^2/m$

To make it gauge symmetric we average over all possible orientations of
 $\omega$ to get:
$$
S_{t'H}=\int d \mu_{I,f}\left( 
4\pi^2\rho^3 
\bar\psi^a(x_0)\left(\frac{1+\gamma_5}{2}\right)\psi^a(x_0)
\right)
$$

Combining the many-flavors fermionic and gluonic effective lagrangians, the 
large-distance effects of instantons can be represented by the following 
lagrangian \cite{Callan:1977gz,'tHooft:up,Shifman:nz,Shifman:uw}:
\bea
\label{tot_lagr}
{\cal L}_I &=&\int  \prod_q \left[
        m_q\r - 2\pi^2\r^3 \bar{q}_R
        \left(U\mathbf{1}_2U^{\dagger} +
        \frac{i}{2}t^aR^{aa'}\bar{\eta}^{a'}_{\mu\nu}\s^{\mu\nu}
        \right)q_L  \right] \nonumber \\
 & &\times \exp\left(
        -\frac{2\pi^2}{g}\r^2 \bar{\eta}^{b'}_{\g\delta} R^{b'b}G^{b,\g\d}
 \right)dz\frac{d_0(\r)}{\r^5}d\r\;dU  .
\eea
where $t^a=\frac{1}{2}\lambda^a$ with ${\rm tr}[\lambda^a\lambda^b]
=2\delta^{ab}$ are $SU(3)$ generators, ${1}_2 = {\rm diag}
(1,1,0)$, $\eta^a_{\mu\nu}$ is the 't Hooft symbol and $\smn=
\frac{1}{2}[\g_\mu,\g_\nu]$. The instanton is characterized by
$4N_c$ collective coordinates, the instanton position $z$, the
instanton size $\r$, and the color orientation $U\in SU(N_c)$.
We also define the rotation matrix $R^{ab}$ by $R^{aa'}\lambda^{a'}
=U\lambda^aU^{\dagger}$. For an anti-instanton we have to replace
$L\leftrightarrow R$ and $\bar{\eta}\leftrightarrow \eta$. The
semi-classical instanton density $d(\rho)$ is given by
\be
\label{drho}
  d(\rho) = \frac{d_0(\rho)}{\rho^5}=
   \frac{0.466\exp(-1.679N_c)1.34^{N_f}}{(N_c-1)!(N_c-2)!}\,
  \left(\frac{8\pi^2}{g^2}\right)^{2N_c}
 \rho^{-5}\exp\left[-\frac{8\pi^2}{g(\rho)^2}\right],
\ee
where $g(\rho)$ is the running coupling constant. For small
$\rho$ we have $d(\rho)\sim \rho^{b-5}$ where $b=(11N_c)/3
-(2N_f)/3$ is the first coefficient of the beta function.

The need for this lagrangian is not obvious from computations of correlation
functions, as one can do well without it(better actually), since direct use of 
instanton solution and fermionic zero mode is correct at any distance. However,
the effective lagrangian becomes a great tool if one tries to 
compute the instanton contribution to matrix elements of type
$<\eta_c|K\bar{K}\pi>$.
Our inability to write pion fields in terms of more fundamental quark
 creation/annihilation operators renders 
the straight-forward approach of computing correlation functions inapplicable. 
As we shall see in chapter \ref{etac_chapter}, 
there are still ways to employ the 
effective lagrangian and compute the matrix element as
$$<\eta_c|K\bar{K}\pi>_{inst} 
= <\eta_c|\int\!\! d\mu_{I,f}V_{t'H}V^{CDG}|K\bar{K}\pi>.
$$
Another great advantage of the effective lagrangian is that it displays
transparently the physical properties of the interaction. This way one 
easily gains physical intuition by thinking in
terms of Feynman diagrams.

\subsection{Dilute gas approximation}
\label{dil_gas_appr}
The instanton is a theoretically tractable non-perturbative object.
However, the QCD vacuum proved to be much more complicated. A good description
has been achieved by constructing the vacuum from instantons and 
anti-instantons. Within this framework,
dilute gas approximation leads to a manageable computation, where 
multiple instanton effects can be computed from single instanton. 
In this section, following \cite{Andrei:xg}, \cite{Nason:1993ak} we will
 present the main ideas of dilute gas and single instanton approximation.

Configuration space of finite-action gauge fields consists of disjunct 
subspaces characterized by different winding numbers. The 
vacuum expectation value of an operator $\Pi$ is given as the  sum over 
these sectors of different homotopy number n: 
\be \label{path_int}
\langle vac | \Pi | vac \rangle =
\frac{
\sum_{n=-\infty}^{+\infty} \int \left[
D A_\mu D\psi D\bar{\psi}
\right]_n e^{-S(A,\psi,\bar{\psi})}
\Pi (A,\psi,\bar{\psi})
}{
\sum_{n=-\infty}^{+\infty} \int \left[
D A_\mu D\psi D\bar{\psi}
\right]_n e^{-S(A,\psi,\bar{\psi})}
}
\ee
Semiclassical approximation amounts to getting the 
minimum of action in each sector (solution of the 
equations of motion) and performing the Gaussian approximation around this solution. In each sector, a multi-instanton configuration is the true minimum of 
the action. However, it has been argued that the superposition of well 
separated $n_+$ instantons and $n_-$ anti-instantons, such that 
$n_+-n_-=n$ has a much higher entropy then the true minima for the sector
with winding number $n$, therefore dominating the path integral.
. The 
expansion around these approximate solutions provide a very good 
description of the true vacuum of QCD. Moreover, the Gaussian 
approximation proves to be expressible through the measure of a single 
instanton, in case one neglects the interaction between the instantons.
Neglecting other minima, the path integral  
 (\ref{path_int})  can therefore be 
written as a dilute gas:
\be \label{dil_gas}
\langle vac | \Pi | vac \rangle =
\frac{
\sum_{n_{+},n_-}\frac{1}{n_+!n_-!} \int (d\mu_{+})^{n_+}(d\mu_{-})^{n_-}
\langle \Pi \rangle_{A_{n_+,n_-}}
}{
\sum_{n_{+},n_-}\frac{1}{n_+!n_-!} \int (d\mu_{+})^{n_+}(d\mu_{-})^{n_-} 
}
\ee
where $d\mu_{\pm}$ is instanton measure of moduli space.
Phenomenological estimates give the density of instantons to be of the 
order of $n \approx 1\;{\rm fm}^{-4}$ while the mean 
instanton size has been found
to be $\bar{\rho} \approx 0.3 \; {\rm fm}$.
The dimensionless parameter $\bar{\rho}^4n \cong 0.008$ shows that the 
instanton liquid is dilute and therefore one can take it as a parameter
for expansion of the path integral.
The same parameters show that for distances  
$\ll 1{\rm fm}$ it is therefore 
reasonable to expect 
the one instanton to provide the dominant contribution.
One should not, however, that small distance requirement and diluteness of 
the liquid are two independent aspects. The diluteness renders the expansion
in density of instantons meaningful, while the small distance justifies the use
of a single instanton approximation.

Neglecting the second and higher orders of 
the density, one arrives at:
\be\label{sia}
\langle vac | \Pi | vac \rangle = 
\langle  \Pi  \rangle_0 + \sum_{+-}\int d\mu_{\pm} 
[\langle \Pi \rangle_{\pm} - 
\langle \Pi \rangle_{0} ] + O(\mu_{\pm}^2)\equiv 
\langle  \Pi  \rangle_0 + \delta \Pi
\ee
The above expression represents the single instanton approximation (SIA) and 
we will be using it throughout the present work
\footnote{
Let us briefly comment on SIA versus other first order dilute gas approximation.
SIA does not account for any correlation between instantons. This can be
achieved, simply speaking, by using an effective mass $m^\star$ instead of
the current quark mass $m$, with the instanton interactions being funneled
into the value of $m^\star$. For more details see \cite{Faccioli:2001ug}
}.

\subsection{SIA versus t'Hooft effective lagrangian}
\label{sia_thoft_section}
Before tackling correlators in the SIA approach, let us comment on 
some generic features of the calculations and highlight the link to the 
t'Hooft effective lagrangian.

As the name points out, the effective lagrangian induced by instantons 
is only valid for large distances compared to the width of instanton: 
$\frac{\rho^2}{x^2}\ll 1$. With the typical size of the order of 
$.3{\rm fm}$, the above condition is satisfied with $\simeq 10\%$
accuracy  already for distances of $\simeq 1{\rm fm}$. One is therefore 
endowed with an additional way of computing large distance correlators. 
More importantly, the effective lagrangian provides an intuition as to 
what diagrams are dominant and which ones disappear completely. The 
purpose of this small section is to show how well-known characteristics 
of t'Hooft lagrangian are reflected in SIA approach.

Let us then recall the main features of effective vertex. First of all, 
it is based on zero modes only, so there are always additional terms 
besides the ones given by t'Hooft Lagrangian. However, once the zero 
modes give a non-zero contribution, all the non-zero mode terms  are 
suppressed by the quark mass $m$ and can be neglected in the $m\to0$ 
limit.

 One can read all the fundamental features of the zero mode propagation 
from the form of the 
 lagrangian: $\prod_{u,d,s} \bar{q}_R \left( 1 + 
\frac{i}{4}\sigma^{\mu\nu}\eta^a_{\mu\nu}\sigma^a\right)q_L $. {\it All} 
massless quarks have to participate, they come in $\bar{q}_R q_L$ 
{\it pairs} and their {\it helicity flips}. The determinantal structure 
also restrains the species of quarks from participating with more than 
one pair ({\it Pauli principle}).

 None of the above rules have to be imposed by hand in SIA - they are already incorporated by means of chirality of the zero mode or rules of Wick contractions. It is instructive to see how it works on some simple examples.

Helicity of the quarks flips through the instanton vertex. A vector insertion
at point $y$ into a quark propagator from $x$ to $z$ gives rise to 
\be
S^{ZM}(x,y)\gamma^\mu S^{ZM}(y,z) = \frac{1}{m}\Psi_0(x)\Psi_0^+(y)
\gamma^\mu \frac{1}{m}\Psi_0(y)\Psi_0^+(z)
\ee
The chirality of the zero mode $\Psi_0(y)=\gpm \Psi_0(y)$
 and anticommutation relation $\{\g^\mu,\g^5\} = 0$ renders the 
diagram zero since
$$\Psi_0^+(y)\g^\mu\Psi_0(y) = 
\Psi_0^+(y)\gpm\g^\mu\gpm\Psi_0(y) = 
\Psi_0^+(y)\g^\mu\gmp\gpm\Psi_0(y) = 0
$$

All massless quarks participate - is just a statement that, whenever 
possible, the zero mode propagation is favored to non-zero mode due to 
$\frac{1}{m}$ factor.

The most interesting feature of t'Hooft Lagrangian is the Pauli 
Principle: no 2 identical quarks can propagate in the same state. This 
stems of course from anticommutation of fermion operators. But that is 
true independent on the species. What makes it work is that whenever 
there are 2 identical quarks propagating in the zero 
mode, there is an 
additional diagram that gives exactly the same contribution with 
opposite sign. To illustrate this, consider the correlation of scalar 
operator $\bar{u}u$:
\be
\langle \bar{u}u(x) \bar{u}u(y) \rangle = -Tr[ S(x,y) S(y,x) ] + Tr 
[S(x,x)] Tr[S(y,y)]
\ee
For both $u$-quarks propagating in the zero mode we get, due to trace 
cyclicity:
\bea
\langle \bar{u}u(x) \bar{u}u(y) \rangle &=&
-\frac{1}{m^2}
\bigl\{
Tr[\Psi_0(x)\Psi_0^+(y) \Psi_0(y)\Psi_0^+(x)]
\bigr.
\nonumber\\
& &
\hspace{0.5cm}
\bigl.
- Tr[\Psi_0(x)\Psi_0^+(x)] Tr[\Psi_0(y)\Psi_0^+(y)]
\bigr\} = 0 \nonumber
\eea
enforcing thus the Pauli principle.



\chapter{$\eta_c$ decay}
\label{etac_chapter}
\section{Introduction}

  The charmonium system has played an important role in shaping
our knowledge of perturbative and non-perturbative QCD. The 
discovery of the $J/\psi$ as a narrow resonance in $e^+e^-$ 
annihilation confirmed the existence of a new quantum number,
charm. The analysis of charmonium decays in $e^+e^-$ pairs, 
photons and hadrons established the hypothesis that the 
$J/\psi$ and $\eta_c$ are, to a good approximation, 
non-relativistic $^3S_1$ and $^1S_0$ bound states of heavy 
charm and anti-charm quarks. However, non-perturbative 
dynamics does play an important role in the charmonium
system \cite{Novikov:dq,Shifman:nx}. For example, an analysis 
of the $\psi$ spectrum lead to the first determination of the 
gluon condensate.
  
  The total width of charmonium is dominated by short 
distance physics and can be studied in perturbative 
QCD \cite{Appelquist:zd}. The only non-perturbative input 
in these calculations is the wave function at the origin. 
A systematic framework for these calculations is provided 
by the non-relativistic QCD (NRQCD) factorization method
\cite{Bodwin:1994jh}. NRQCD facilitates higher order calculations 
and relates the decays of states with different quantum numbers. 
QCD factorization can also be applied to transitions of the 
type $\psi'\to\psi+X$ \cite{Gottfried:1977gp,Voloshin:hc}. 

  The study of exclusive decays of charmonium into light 
hadrons is much more complicated and very little work in this 
direction has been done. Perturbative QCD implies some helicity 
selection rules, for example $\eta_c \noto \rho\rho,p\bar{p}$
and $J/\psi \noto \rho\pi,\rho a_1$ 
\cite{Brodsky:1981kj,Chernyak:1983ej}, but these rules 
are strongly violated \cite{Anselmino:yg}. The $J/\psi$ 
decays mostly into an odd number of Goldstone bosons. 
The average multiplicity is $\sim (5-7)$, which is consistent 
with the average multiplicity in $e^+e^-$ annihilation away
from the $J/\psi$ peak. Many decay channels have been 
observed, but none of them stand out. Consequently, we 
would expect the $\eta_c$ to decay mostly into an even 
number of pions with similar multiplicity. However, the 
measured decay rates are not in accordance with this 
expectation. The three main decay channels of the $\eta_c$ 
are $K\bar{K}\pi$, $\eta\pi\pi$ and $\eta'\pi\pi$, each 
with an unusually large branching ratio of $\sim 5$\%. 
Bjorken observed that these three decays correspond to 
a quark vertex of the form $(\bar{c}c)(\bar{s}s)(\bar{d}d)
(\bar{u}u)$ and suggested that $\eta_c$ decays are a 
``smoking gun'' for instanton effects in heavy quark 
decays \cite{Bjorken:2000ni}. 
  
 We shall try to follow up on this idea by 
performing a more quantitative estimate of the instanton 
contribution to $\eta_c$ and $\chi_c$ decays. 
In section \ref{sec_eff} we 
review the instanton induced effective lagrangian. 
In the following sections we apply the effective lagrangian 
to the decays of the scalar glueball, eta charm, and chi 
charm. We should note that this investigation should be 
seen as part of a larger effort to identify ``direct'' 
instanton contributions in hadronic reactions, such as 
deep inelastic scattering, the $\Delta I=1/2$ rule, or 
$\eta$ production in $pp$ scattering
\cite{Balitsky:1993jd,Moch:1996bs,Kochelev:2001pp,Kochelev:1999tc}.

\section{Effective Lagrangians}
\label{sec_eff}

  Instanton effects in hadronic physics have been studied
extensively \cite{Diakonov:1995ea,Schafer:1996wv}. Instantons 
play an important role in understanding the $U(1)_A$ anomaly 
and the mass of the $\eta'$. In addition to that, there is also 
evidence that instantons provide the mechanism for chiral 
symmetry breaking and play an important role in determining
the structure of light hadrons. All of these phenomena are 
intimately related to the presence of chiral zero modes in
the spectrum of the Dirac operator in the background field 
of an instanton. The situation in heavy quark systems is 
quite different. Fermionic zero modes are not important 
and the instanton contribution to the heavy quark potential
is small \cite{Callan:1978ye}. 

 This does not imply that instanton effects are not relevant. 
The non-perturbative gluon condensate plays an important 
role in the charmonium system \cite{Novikov:dq,Shifman:nx},
and instantons contribute to the gluon condensate. In general, 
the charmonium system provides a laboratory for studying 
non-perturbative glue in QCD. The decay of a charmonium 
state below the $D\bar{D}$ threshold involves an intermediate 
gluonic state. Since the charmonium system is small, $r_{c\bar{c}}
\sim (v m_c)^{-1} < \Lambda_{QCD}^{-1}$, the gluonic system
is also expected to be small. For this reason charmonium decays 
have long been used for glueball searches. 

  Since charmonium decays produce a small gluonic system we
expect that the $c\bar{c}$ system mainly couples to instantons
of size $\rho\sim r_{c\bar{c}}\sim (vm_c)^{-1}$. In this limit 
the instanton effects can be summarized in terms of an effective 
lagrangian \ref{tot_lagr} discussed in chapter \ref{intro_chapter}: 
\bea
\label{main_lagr}
{\cal L}_I &=&\int  \prod_q \left[
        m_q\r - 2\pi^2\r^3 \bar{q}_R
	\left(U\mbox{\boldmath{1}}_2U^{\dagger} + 
	\frac{i}{2}t^aR^{aa'}\bar{\eta}^{a'}_{\mu\nu}\s^{\mu\nu}
	\right)q_L  \right] \nonumber \\
 & &\times \exp\left(
	-\frac{2\pi^2}{g}\r^2 \bar{\eta}^{b'}_{\g\delta} R^{b'b}G^{b,\g\d}
 \right)dz\frac{d_0(\r)}{\r^5}d\r\;dU  .
\eea

  Expanding the effective lagrangian in powers of the external
gluon field gives the leading instanton contribution to different 
physical matrix elements. If the instanton size is very small,
$\rho\ll m_c^{-1}$, we can treat the charm quark mass as light 
and there is an effective vertex of the form $(\bar{u}u)(\bar{dd})
(\bar{s}s)(\bar{c}c)$ which contributes to charmonium decays. 
Since the density of instantons grows as a large power of $\rho$
the contribution from this regime is very small. In the realistic
case $\rho\sim (vm_c)^{-1}$ we treat the charm quark as heavy 
and the charm contribution to the fermion determinant is 
absorbed in the instanton density $d(\rho)$. The dominant
contribution to charmonium decays then arises from expanding
the gluonic part of the effective lagrangian to second order 
in the field strength tensor. This provides effective vertices
of the form $(G\tilde G)(\bar{u}\gamma_5 u)(\bar{d}\gamma_5 d)
(\bar{s}\gamma_5 s)$, $(G^2)(\bar{u}\gamma_5 u)(\bar{d}\gamma_5
d)(\bar{s}s)$, etc. 

  We observe that the $N_f=3$ fermionic lagrangian combined
with the gluonic term expanded to second order in the field
strength involves an integral over the color orientation of the 
instanton which is of the form $\int dU(U_{ij}U_{kl}^\dagger)^5$.
This integral gives $(5!)^2$ terms. A more manageable result
is obtained by using the vacuum dominance approximation. We
assume that the coupling of the initial charmonium or glueball
state to the instanton proceeds via a matrix element of the 
form $\langle 0^{++}|G^2|0\rangle$ or $\langle 0^{-+}|G\tilde{G}
|0\rangle$. In this case we can use
\be
\langle 0^{++}|G^a_{\mu\nu}G^b_{\alpha\beta}|0\rangle=
 \frac{1}{12(N_c^2-1)}\delta^{ab} 
  (\delta_{\mu\alpha}\delta_{\nu\beta}-\delta_{\mu\beta}
  \delta_{\nu\alpha})
 \langle 0^{++}|G^{a'}_{\rho\sigma}G^{a'}_{\rho\sigma}|0\rangle
\ee 
in order to simplify the color average. The vacuum dominance 
approximation implies that the color average of the fermionic 
and gluonic parts of the interaction can be performed independently. 
In the limit of massless quarks the instanton ($I$) and anti-instanton
($A$) lagrangian responsible for the decay of scalar and pseudoscalar 
charmonium decays is given by
\be
\label{main}
{\cal L}_{I+A} = \int\!\!dz\frac{d_0(\r)}{\r^5}d\r
\frac{\pi^3\r^4}{(N_c^2-1)\alpha_s} 
\left\{
 \left(G^2-G\tilde{G}\right)\times L_{f,I} +
 \left(G^2+G\tilde{G}\right)\times L_{f,A}
\right\}. 
\ee
Here, ${\cal L}_{f,IA}$ is the color averaged $N_f=3$ fermionic 
effective lagrangian \cite{Shifman:uw,Diakonov:1995ea,Schafer:1996wv}.

\section{Scalar glueball decays}
\label{sec_glue}

  Since the coupling of the charmonium state to the instanton
proceeds via an intermediate gluonic system with the quantum 
numbers of scalar and pseudoscalar glueballs it is natural
to first consider direct instanton contributions
to glueball decays. This problem is of course important in its 
own right. Experimental glueball searches have to rely on identifying
glueballs from their decay products. The successful identification
of a glueball requires theoretical calculations of glueball mixing
and decay properties. In the following we compute the direct
instanton contribution to the decay of the scalar $0^{++}$ 
glueball state into $\pi\pi$, $K\bar{K}$, $\eta\eta$ and 
$\eta\eta'$.

 Since the initial state is parity even only the $G^2$ term in
equ.~(\ref{main}) contributes. The relevant effective interaction
is given by
\bea
\label{glue_lagr}
  {\cal L}_{I+A}
   &=&\int\!\! dz\int\!\! 
	d_0(\rho)\frac{d\rho}{\rho^5}\frac{1}{N_c^2-1} 
	\left( 
	  \frac{\pi^3\rho^4}{\alpha_s}\right) G^2 
	  (-\frac{1}{4}) \left( \frac{4}{3}\pi^2\rho^3 
	\right)^3 \times  \nonumber \\
 &  &\biggl\{
	[
	(\bar{u}u)(\bar{d}d)(\bar{s}s) +
	(\bar{u}\gamma^5u)(\bar{d}\gamma^5d)(\bar{s}s) \nonumber\\
 & &
	+(\bar{u}\gamma^5u)(\bar{d}d)(\bar{s}\gamma^5s) +
	(\bar{u}u)(\bar{d}\gamma^5d)(\bar{s}\gamma^5s) 
	]     \nonumber \\
 &  &+\frac{3}{8}
      {\biggl[}
	(\bar{u}t^au)(\bar{d}t^ad)(\bar{s}s) +
	(\bar{u}t^a \gamma^5u)(\bar{d}t^a \gamma^5d)(\bar{s}s)\nonumber\\
 & &
	+(\bar{u}t^a \gamma^5u)(\bar{d}t^a d)(\bar{s}\gamma^5s)  
       +(\bar{u}t^a u)(\bar{d}t^a\gamma^5d)(\bar{s}\gamma^5s) \nonumber\\
 & &  \hspace{0.1cm} 
       -\frac{3}{4}\bigl[
	(\bar{u}t^a \sigma_{\mu\nu} u)(\bar{d}t^a \smn d)(\bar{s}s) +
	(\bar{u}t^a \smn \gamma^5u)(\bar{d}t^a \smn\gamma^5d)(\bar{s}s)
           \nonumber \\
 &  &	\hspace{0.5cm}
      +(\bar{u}t^a \smn\gamma^5u)(\bar{d}t^a\smn d)(\bar{s}\gamma^5s) +
	(\bar{u}t^a\smn u)(\bar{d}t^a\smn\gamma^5d)(\bar{s}\gamma^5s)\bigr]  
            \nonumber \\
 &  & \hspace{0.1cm}
       -\frac{9}{20}d^{abc} \bigl[ 
	(\bar{u}t^a \sigma_{\mu\nu} u)(\bar{d}t^b \smn d)(\bar{s}t^cs) +
	(\bar{u}t^a \smn \gamma^5u)(\bar{d}t^b\smn\gamma^5d)(\bar{s}t^cs) 
            \nonumber \\
 &  & \hspace{0.5cm}
       +(\bar{u}t^a \smn\gamma^5u)(\bar{d}t^b\smn d)(\bar{s}t^c\gamma^5s) +
	(\bar{u}t^a\smn u)(\bar{d}t^b \smn\gamma^5d)(\bar{s}t^c\gamma^5s)  
       \bigr]\nonumber \\
 &  &  \hspace{0.5cm} + (2\; {\it cyclic\; permutations}\; 
           u\leftrightarrow d \leftrightarrow s)
      {\biggr]}\nonumber \\
 &  & \hspace{0.1cm}
-\frac{9}{40}d^{abc}
       \biggl[  
	(\bar{u}t^au)(\bar{d}t^bd)(\bar{s}t^cs) +
	(\bar{u}t^a\gamma^5u)(\bar{d}t^b\gamma^5d)(\bar{s}t^cs) \nonumber\\
 &  &	\hspace{0.5cm}
	+(\bar{u}t^a\gamma^5u)(\bar{d}t^bd)(\bar{s}\gamma^5t^cs)   
       +(\bar{u}t^au)(\bar{d}t^b\gamma^5d)(\bar{s}t^c\gamma^5s) 
         \biggr] \nonumber  \\
 &  & \hspace{0.1cm} 
 - \frac{9}{32}if^{abc} \biggl[   
	(\bar{u}t^a \sigma_{\mu\nu} u)(\bar{d}t^b \sng d)
	(\bar{s}t^c\sgm s)\nonumber  \\
 &  & \hspace{1.5cm}  
       +(\bar{u}t^a \smn\gamma^5u)(\bar{d}t^b\sng\gamma^5d) 
	(\bar{s}t^c\sgm s) \nonumber \\
 &  & \hspace{1.5cm}
     + (\bar{u}t^a\smn\gamma^5u)(\bar{d}t^b\sng d) 
	(\bar{s}t^c\sgm\gamma^5s)
	\nonumber  \\
 &  & \hspace{1.5cm}
       +(\bar{u}t^a\smn u)
	(\bar{d}t^b\sng\gamma^5d)(\bar{s}t^c\sgm\gamma^5s)  
      \biggr]
   \biggr\}
\eea
Let us start with the process $0^{++} \rightarrow \pi\pi$. 
In practice we have Fierz rearranged equ.~(\ref{glue_lagr}) 
into structures that involve the strange quark condensate 
$\bar{s}s$ as well as operators with the quantum numbers of 
two pions. In order to compute the coupling of these operators 
to the pions in the final state we have used PCAC relations
\bea
\langle 0|\bar{d}\g^5u|\pi^+\rangle
  &=& \frac{i\sqrt{2}m_\pi^2f_\pi}{m_u+m_d}\equiv K_\pi ,  \\
\langle0|\bar{s}\g^5u|K^+\rangle
  &=&\frac{i\sqrt{2}m_{K}^2f_K}{m_u+m_s}\equiv K_{K}  .
\eea
The values of the decay constants are $f_\pi=93$ MeV, $f_K=113$ 
MeV \cite{PDG}. We also use $Q_u\equiv \langle\bar{u}u\rangle = 
-(248\,{\rm MeV})^3$ and $Q_d=Q_u$ as well as $Q_s=0.66Q_u$ 
\cite{Narison:2002hk}. The coupling of the $\eta'$ meson is 
not governed by chiral symmetry. A recent analysis of $\eta-\eta'$ 
mixing and the chiral anomaly gives \cite{Feldmann:1999uf} 
\bea
 \langle 0|\bar{u}\gamma_5 u |\eta\rangle  &=& 
    \langle 0|\bar{d}\gamma_5 d|\eta\rangle = 
      - i(358\; {\rm  MeV} )^2\equiv K_\eta^q , \\
 \langle 0|\bar{u}\gamma_5 u |\eta'\rangle  &=& 
     \langle 0|\bar{d}\gamma_5 d|\eta'\rangle = 
      - i(320\; {\rm  MeV} )^2\equiv K_{\eta'}^q ,\\
 \langle 0|\bar{s}\gamma_5 s| \eta\rangle &=& i(435\;{\rm MeV})^2 
       \equiv K_{\eta}^s ,\\ 
 \langle 0|\bar{s}\gamma_5 s| \eta'\rangle &=& 
        -i(481\;{\rm MeV})^2 \equiv K_{\eta'}^s . 
\eea
Finally, we need the coupling of the glueball state to the 
gluonic current. This quantity has been estimated using QCD 
spectral sum rules \cite{Novikov:va,Narison:1996fm}
and the instanton model \cite{Schafer:1994fd}. We use
\be
 \langle0^{++}|g^2G^2|0\rangle
\equiv \lb_0 = 15\,{\rm GeV}^3.
\ee
We can now compute the matrix element for $0^{++}\rightarrow
\pi^+\pi^-$. The interaction vertex is 
\be
\label{0++pipi}
{\cal L}_{I+A}^{\pi^+\pi^-}
  = \int\!\! dz
   \int\!\!\frac{d\r}{\r^5}d_0(\r)
  \frac{1}{N_c^2-1}\left(\frac{\pi^3\r^4}{\alpha_s^2}\right)
\left(
\frac{4}{3}\pi^2\r^3
\right)^3 
 \x \frac{1}{4}
   (\a_sG^2)(\bar{s}s)(\bar{u}\g^5d)(\bar{d}\g^5u). 
\ee
The integral over the position of the instanton leads to a momentum 
conserving delta function, while the vacuum dominance approximation
allows us to write the amplitude in terms of the coupling constants
introduced above. We find
\be
\langle0^{++}(q)|\pi^+(p^+)\pi^-(p^-)\rangle
    = (2\pi)^4\delta^4(q-p^+-p^-)
      \frac{A}{16\pi}\lb_0 Q_s K_\pi^2,
\ee
where 
\be
\label{A_int}
 A=\int\frac{d\r}{\r^5}d_0(\r)\frac{1}{N_c^2-1}
 \left(\frac{\pi^3\r^4}{\a_s^2}\right) 
 \left(\frac{4}{3}\pi^2\r^3\right)^3.
\ee
The instanton density $d_0(\rho)$ is known accurately only
in the limit of small $\r$. For large $\rho$ higher loop 
corrections and non-perturbative effects are important.
The only source of information in this regime is lattice QCD 
\cite{Michael:1995br,Smith:1998wt,deForcrand:1997sq,DeGrand:1997gu}.
A very rough caricature of the lattice results is provided
by the parameterization
\be
\frac{d_0(\r)}{\r^5}=\frac{1}{2}n_0\d(\r-\r_c),
\ee
with $n_0 \simeq 1\,{\rm fm}^{-4}$ and $\r_c \simeq 0.33\,
{\rm fm}$. This parameterization gives a value of 
$A=(379\,{\rm MeV})^{-9}$.
Another way to compute $A$ is to regularize the integral over
the instanton size by replacing $d(\rho)$ with $d(\rho)\exp(-
\alpha\rho^2)$. The parameter $\alpha$ can be adjusted in order 
to reproduce the size distribution measured on the lattice. 
We notice, however, that whereas the instanton density scales 
as $\rho^{b-5}\sim \rho^4$, the decay amplitude scales as 
$\rho^{b+8}\sim \rho^{17}$. This implies that the results 
are very sensitive to the density of large instantons. We 
note that when we study the decay of a small-size bound 
state the integral over $\rho$ should be regularized by 
the overlap with the bound state wave function. We will 
come back to this problem in section \ref{sec_eta} below. 

 We begin by studying ratios of decay rates. These ratios 
are not sensitive to the instanton size distribution. The 
decay rate $0^{++}\to\pi^+\pi^-$ is given by
\be
\Gamma_{0^{++}\rightarrow \pi^+\pi^-}
 =\frac{1}{16\pi}\frac{\sqrt{m_{0^{++}}^2-4m_{\pi}^2}}{m_{0^{++}}^2}
 \left[ \frac{A}{16\pi}\lb_0Q_sK_{\pi}^2 \right]^2 .
\ee
The decay amplitude for the process $0^{++}\rightarrow\pi_0\pi_0$ 
is equal to the $0^{++}\rightarrow\pi^+\pi^-$ amplitude as required 
by isospin symmetry. Taking into account the indistinguishability 
of the two $\pi_0$ we get the total $\pi\pi$ width
\be
\Gamma_{0^{++}\rightarrow \pi\pi} =
  \frac{3}{32\pi}
   \frac{\sqrt{m_{0^{++}}^2-4m_{\pi}^2}}{m_{0^{++}}^2}
   \left[\frac{A}{16\pi}\lb_0Q_sK_{\pi}^2\right]^2 .
\ee
In a similar fashion we get the decay widths for the $K\bar{K}$, 
$\eta\eta$, $\eta\eta'$ and $\eta'\eta'$ channels
\bea
\Gamma_{0^{++}\rightarrow K\bar{K}}
  &=& 2\frac{1}{16\pi}
   \frac{\sqrt{m_{0^{++}}^2-4m_{K}^2}}{m_{0^{++}}^2}
    \left[ \frac{A}{16\pi}\lb_0Q_uK_K^2 \right]^2, \nonumber \\
\Gamma_{0^{++}\rightarrow \eta\eta}
  &=&  \frac{1}{32\pi}
  \frac{\sqrt{m_{0^{++}}^2-4m_{\eta}^2}}{m_{0^{++}}^2}
  \left[ \frac{A}{16\pi}\lb_0
    K_{\eta}^q\;2(Q_sK_{\eta}^q+(Q_u+Q_d)K_{\eta}^s)
  \right]^2,   \nonumber\\
\Gamma_{0^{++}\rightarrow \eta\eta'}
 &=&\frac{1}{16\pi}
  \frac{\sqrt{[m_{0^{++}}^2-(m_{\eta}+m_{\eta'})^2]
  [m_{0^{++}}^2-(m_\eta-m_{\eta'})^2]}}{m_{0^{++}}^3} \nonumber \\
  & & \x \left[
   \frac{A}{16\pi}\lb_0(2Q_sK_{\eta}^qK_{\eta'}^q+
   (Q_u+Q_d)(K_{\eta}^qK_{\eta'}^s+K_{\eta}^sK_{\eta'}^q)
   \right]^2  \nonumber\\
\Gamma_{0^{++}\rightarrow \eta'\eta'}
 &=&  \frac{1}{32\pi}
   \frac{\sqrt{m_{0^{++}}^2-4m_{\eta'}^2}}{m_{0^{++}}^2}
   \left[ \frac{A}{16\pi}\lb_0 K_{\eta'}^q2(Q_sK_{\eta'}^q
       +(Q_u+Q_d)K_{\eta'}^s) \right]^2 .\nonumber
\eea
Here, $\bar{K}K$ refers to the sum of the $K^+K^-$ and 
$\bar{K}_0K_0$ final states. We note that in the chiral 
limit the instanton vertices responsible for $\pi\pi$ 
and $\bar{K}K$ decays are identical up to quark interchange. 
As a consequence, the ratio of the decay rates $\Gamma_{0^{++}
\rightarrow \pi\pi}/ \Gamma_{0^{++}\rightarrow K\bar{K}}$ 
is given by the phase space factor multiplied by the ratio 
of the coupling constants
\be
\label{0++ratio}
\frac{\Gamma_{0^{++}\rightarrow \pi\pi}} 
     {\Gamma_{0^{++}\rightarrow K\bar{K}}}
   = \frac{3}{4}\x \frac{Q_s^2 K_\pi^4}{Q_u^2 K_K^4}\x
      \sqrt{\frac{m_{0^{++}}^2-4m_{\pi}^2}{m_{0^{++}}^2-4m_K^2}}=
  (0.193 \pm 0.115)
      \sqrt{\frac{m_{0^{++}}^2-4m_{\pi}^2}{m_{0^{++}}^2-4m_K^2}}.
\ee
\begin{figure}
\begin{center}
\includegraphics[width=6cm,angle=-90]{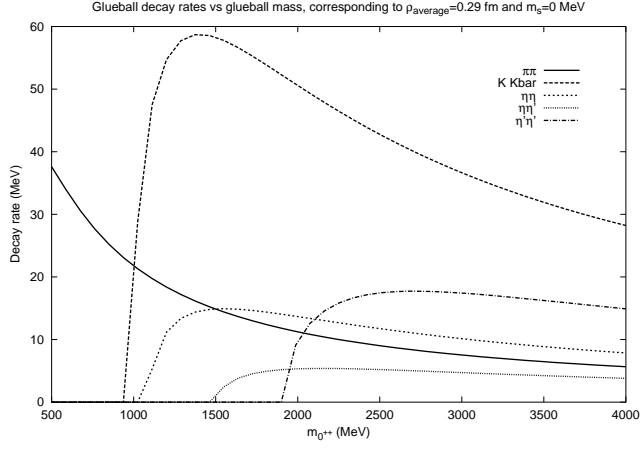}
\end{center}
\begin{center} 
\caption{\label{fig_0++}
Scalar glueball decay rates plotted a function of the mass
of the scalar glueball. The rates shown in this figure were
computed from the instanton vertex in the chiral limit. The
average instanton size was taken to be $\bar{\rho}=0.29$ fm.}
\end{center} 
\end{figure}
The main uncertainty in this estimate comes from the value of 
$m_s$, which is not very accurately known. We have used 
$m_s=(140\pm 20)\,{\rm MeV}$. The ratio of $\pi\pi$ to $\eta\eta$ 
decay rates is not affected by this uncertainty,
\be
\label{0++ratio_pi_eta}
\frac{\Gamma_{0^{++}\rightarrow \pi\pi}} 
     {\Gamma_{0^{++}\rightarrow \eta\eta}}
  =  0.69
   \sqrt{\frac{m_{0^{++}}^2-4m_{\pi}^2}{m_{0^{++}}^2-4m_\eta^2}}.
\ee
In Fig.\ref{fig_0++} we show the decay rates as functions of 
the glueball mass. We have used $\Lambda_{QCD}=300\,{\rm MeV}$ 
and adjusted the parameter $\alpha$ to give the average instanton 
size $\bar{\rho}=0.29$ fm. We observe that for glueball masses
$m_{0^{++}}>1$ GeV the $K\bar{K}$ phase space suppression 
quickly disappears and the total decay rate is dominated
by the $K\bar{K}$ final state. We also note that for 
$m_{0^{++}}>1.5$ GeV the $\eta\eta$ rate dominates over
the $\pi\pi$ rate. 

\begin{figure}
\begin{center}
\includegraphics[width=6cm,angle=-90]{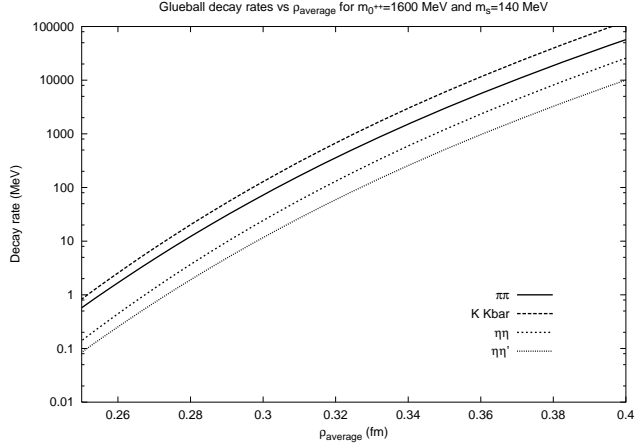}
\end{center}
\begin{center}
\caption{\label{fig_glue_rho_ms}
Dependence of glueball decay rates on the average instanton size.
The results shown in this figure correspond to the instanton
vertex with $m_s\neq 0$ terms included. The strange quark mass
was taken to be $m_s=140$ MeV.}
\end{center}
\end{figure}

 In deriving the effective instanton vertex equ.~(\ref{0++pipi}) 
we have taken all quarks to be massless. While this is a good 
approximation for the up and down quarks, this it is not necessarily 
the case for the strange quark. The $m_s\ne 0$ contribution to the 
effective interaction for $0^{++}$ decay is given by
\bea
 {\cal L}_{m_s} 
  &=& \int\, \frac{d\rho}{\rho^5}d_0(\rho) \frac{1}{N_c^2-1}
     \frac{\pi^3\rho^4}{\alpha_s^2}\left(\frac{4}{3}\pi^2\rho^3 \right)^2
      m_s\rho (\alpha_sG^2)\x \nonumber \\
  & & \frac{1}{2} 
   \left\{ 
     (\bar{u}u)(\bar{d}d) + (\bar{u}\gamma^5u)(\bar{d}\gamma^5d) +
   \frac{3}{8} \left[
     (\bar{u}t^au)(\bar{d}t^ad) + 
     (\bar{u}\gamma^5t^au)(\bar{d}\gamma^5t^ad)\right.  
    \right.    \nonumber \\ 
  & & \mbox{}\hspace{0.2cm}-
    \left.\left.
     \frac{3}{4}(\bar{u}\sigma_{\mu\nu}t^au)(\bar{d}\sigma_{\mu\nu}t^ad) 
    - \frac{3}{4}(\bar{u}\sigma_{\mu\nu}\gamma^5t^au)  
       (\bar{d}\sigma_{\mu\nu}\gamma^5t^a d) 
      \right]  \right\}. \nonumber 
\eea
There is no $m_s\neq0$ contribution to the $K\bar{K}$ channel.
The $m_s\neq 0$ correction to the other decay channels is
\bea
\label{glue_pipi_ms} 
\Gamma_{0^{++}\rightarrow \pi\pi} 
  &=& \frac{3}{32\pi}
   \frac{\sqrt{m_{0^{++}}^2-4m_{\pi}^2}}{m_{0^{++}}^2}
   \left[\frac{1}{16\pi}\lb_0K_{\pi}^2(AQ_s-2Bm_s)\right]^2,\!\!\!\!
	\nonumber
\eea
\bea
\Gamma_{0^{++}\rightarrow \eta\eta}
  &=&\frac{1}{32\pi}
  \frac{\sqrt{m_{0^{++}}^2-4m_{\eta}^2}}{m_{0^{++}}^2}\times\nonumber\\
 &  & 
  \left[\frac{1}{16\pi}\lb_0\;2
  [(AQ_s-2Bm_s)(K_{\eta}^q)^2+A(Q_u+Q_d)K_{\eta}^sK_{\eta}^q)]
  \right]^2 , \nonumber \\
\Gamma_{0^{++}\rightarrow \eta\eta'}
 &=&\frac{1}{16\pi}\frac{\sqrt{[m_{0^{++}}^2-(m_{\eta}+m_{\eta'})^2]
        [m_{0^{++}}^2-(m_\eta-m_{\eta'})^2]}}{m_{0^{++}}^3}\x \nonumber \\
&&  \hspace{-0.8cm} \left[
     \frac{1}{16\pi}\lb_0[2\;(AQ_s-2Bm_s)K_{\eta}^qK_{\eta'}^q
     +A(Q_u+Q_d)(K_{\eta}^qK_{\eta'}^s+K_{\eta'}^qK_{\eta}^s)]
     \right]^2 ,  \nonumber \\
\label{glue_etaPetaP_ms}
\Gamma_{0^{++}\rightarrow \eta'\eta'}
 &=& \frac{1}{32\pi}
    \left[
    \frac{1}{16\pi}\lb_02[(AQ_s-2Bm_s)(K_{\eta'}^q)^2
    +A(Q_u+Q_d)K_{\eta'}^sK_{\eta'}^q]
    \right]^2, \nonumber
\eea
where 
\be
\label{B_int}
B=\int\frac{d\r}{\r^5}d_0(\r)\frac{1}{N_c^2-1}
  \left(\frac{\pi^3\r^4}{\a_s^2}\right) 
  \left(\frac{4}{3}\pi^2\r^3\right)^2 \r \; .
\ee
\begin{figure}
\begin{center}
\includegraphics[width=6cm,angle=-90]{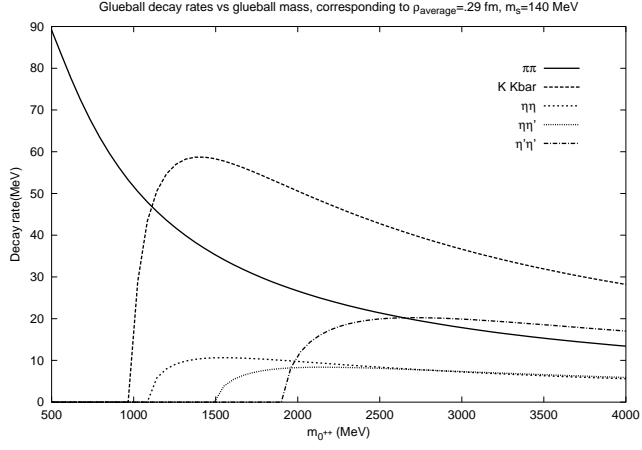}
\end{center}
\begin{center}
\caption{\label{fig_0++_ms}
Same as Fig.~\ref{fig_0++} but with $m_s\neq 0$ corrections
in the instanton vertex taken into account. The results
shown in this figure correspond to $m_s=140$ MeV.}
\end{center}
\end{figure}
The decay rates with the $m_s\neq 0$ correction to the instanton
vertex taken into account are plotted in Fig.~\ref{fig_0++_ms}.
We observe that effects due to the finite strange quark mass are 
not negligible. We find that the $\pi\pi$ , $\eta\eta'$, and $\eta'\eta'$
channels are enhanced 
whereas the $\eta\eta$ channel
is reduced. For a typical glueball mass $m_{0^{++}}=(1.5-1.7)$
GeV the ratio $r=B(\pi\pi)/B(K\bar{K})$ changes from $r\simeq 
0.25$ in the case $m_s=0$ to $r\simeq 0.55$ for $m_s\neq 0$. In 
Fig.~\ref{fig_glue_rho_ms} we show the dependence of the decay 
rates on the average instanton size $\bar{\rho}$. We observe 
that using the phenomenological value $\bar{\rho}=0.3$ fm gives 
a total width $\Gamma_{0^{++}}\simeq 100$ MeV. We note, however, 
that the decay rates are very sensitive to the value of $\bar{\rho}$. 
As a consequence, we cannot reliably predict the total decay rate. 
On the other hand, the ratio of the decay widths for different 
final states does not depend on $\bar{\rho}$ and provides a 
sensitive test for the importance of direct instanton effects.  

  In Tab.~\ref{table_glue} we show the masses and decay widths 
of scalar-isoscalar mesons in the (1-2) GeV mass range. These
states are presumably mixtures of mesons and glueballs. This 
means that our results cannot be directly compared to experiment
without taking into account mixing effects. It will be interesting
to study this problem in the context of the instanton model, but 
such a study is beyond the scope of this work. It is nevertheless
intriguing that the $f_0(1710)$ decays mostly into $K\bar{K}$. 
Indeed, a number of authors have suggested that the $f_0(1710)$ 
has a large glueball admixture 
\cite{Sexton:1995kd,Lee:1999kv,Minkowski:1998mf,Close:1996yc}.
\begin{table}
\begin{tabular}{|c|c|c|c|}\hline\hline
  resonance &   full width   $\Gamma({\rm MeV})$   
  & Mass (MeV) & decay channels \\ 
\hline\hline
$f_0(1370)$ & 200-500 & 1200-1500 & 
  \begin{tabular}{c}
	$\r\r$ dominant \\
	$\pi\pi,K\bar{K},\eta\eta$ seen \\ 
  \end{tabular}
\\ \hline
$f_0(1500)$ & $109\pm 7$ & $1507\pm5$ & 
  \begin{tabular}{lcl}
     $\frac{\Gamma_{K\bar{K}} }{\Gamma_{\pi\pi}}$&=& $0.19 \pm 0.07 $ \\
     $\frac{\Gamma_{\eta\eta'}}{\Gamma_{\pi\pi}}$&=& $0.095 \pm 0.026 $\\ 
     $\frac{\Gamma_{\eta\eta}}{\Gamma_{\pi\pi}}$ &=& $0.18 \pm 0.03  $\\
  \end{tabular}
  \\ \hline
$f_0(1710)$ & $125 \pm 10$ & $ 1713\pm 6$ & 
  \begin{tabular}{lcl}
     $\frac{\Gamma{\pi\pi}}{\Gamma_{K\bar{K}} }$  &=& $0.39 \pm 0.14 $ \\
     $\frac{\Gamma_{\eta\eta}}{\Gamma_{K\bar{K}}}$&=& $0.48 \pm 0.15 $\\ 
  \end{tabular}
\\ \hline
\end{tabular}
\caption{
\label{table_glue}
Masses, decay widths, and decay channels for 
scalar-isoscalar mesons with masses in the $(1-2)$ GeV
range. The data were taken from \cite{PDG}.}
\end{table}

\section{Eta charm decays}
\label{sec_eta}

  The $\eta_c$ is a pseudoscalar $J^{PC}=0^{-+}$ charmonium
bound state with a mass $m_{\eta_c}=(2979\pm 1.8)$ MeV. The total 
decay width of the $\eta_c$ is $\Gamma_{\eta_c}=(16\pm 3)$ MeV.
In perturbation theory the total width is given by
\be
 \Gamma(\eta\to 2g) = \frac{8\pi\alpha^2_s|\psi(0)|^2}{3m_c^2} 
   \left( 1 + 4.4\frac{\alpha_s}{\pi}\right).
\ee
Here, $\psi(0)$ is the $^1S_0$ ground state wave function at the
origin. Using $m_c=1.25$ GeV and $\alpha_s(m_c)=0.25$ we get 
$|\psi(0)|\simeq 0.19\,{\rm GeV}^{3/2}$, which is consistent with 
the expectation from phenomenological potential models. Exclusive
decays cannot be reliably computed in perturbative QCD. As
discussed in the introduction Bjorken pointed out that $\eta_c$
decays into three pseudoscalar Goldstone bosons suggest that 
instanton effects are important \cite{Bjorken:2000ni}. The 
relevant decay channels and branching ratios are $B(K\bar{K}\pi)
=(5.5\pm 1.7)\%$, $B(\eta\pi\pi)=(4.9\pm1.8)\%$ and $B(\eta'\pi
\pi)=(4.1\pm 1.7\%)$. These three branching ratios are anomalously 
large for a single exclusive channel, especially given the 
small multiplicity. The total decay rate into these three 
channels is $(14.5\pm 5.2)\%$ which is still a small fraction 
of the total width. This implies that the assumption that 
the three-Goldstone bosons channels are instanton dominated 
is consistent with our expectation that the total width is 
given by perturbation theory. For comparison, the next most 
important decay channels are $B(2(\pi^+\pi^-))=(1.2\pm 0.4)\%$ 
and $B(\r\r)=(2.6\pm 0.9)\%$. These channels do not receive 
direct instanton contributions. 

\begin{figure}
\begin{center}
\includegraphics[width=8cm]{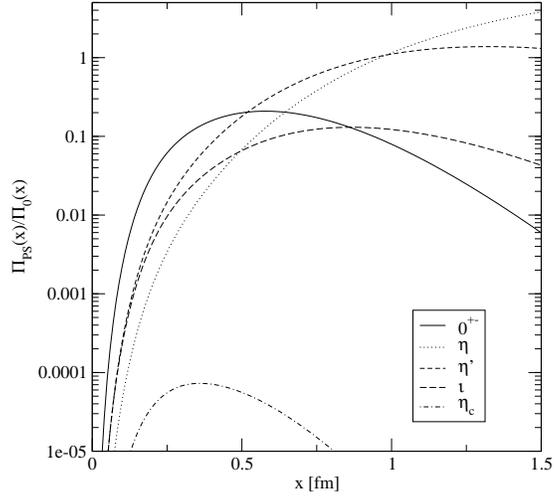}
\includegraphics[width=8cm]{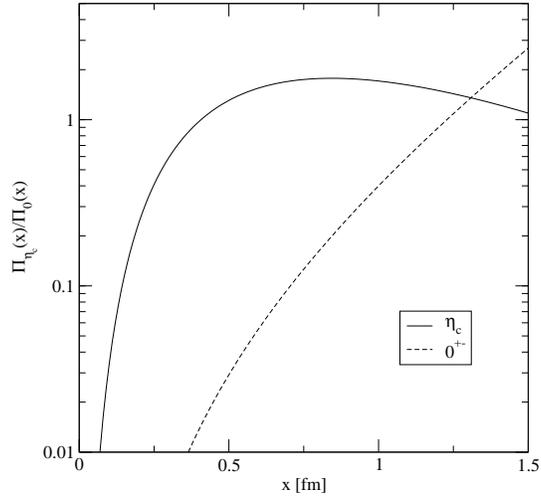}
\end{center}
\begin{center}
\caption{\label{fig_gg}
Resonance contributions to the pseudoscalar glueball
correlation function $\langle g^2G\tilde{G}(0)g^2G\tilde{G}
(x)\rangle$ and the charmonium correlator $\langle \bar{c}
\gamma_5 c(0)\bar{c}\gamma_5 c(x)\rangle$. Both correlation
functions are normalized to free field behavior. In the
case of the gluonic correlation function we show the
glueball contribution compared to the $\eta$, $\eta'$,
$\eta(1440)$ and $\eta_c$ contribution. For the charmonium
correlation function we show the $\eta_c$ and glueball contribution.}
\end{center}
\end{figure}

  The calculation proceeds along the same lines as the glueball
decay calculation. Since the $\eta_c$ is a pseudoscalar only the 
$G\tilde{G}$ term in equ.~(\ref{main}) contributes. The relevant 
interaction is 
\bea
\label{eta_c_lagr}
 &{\cal L}_{I+A}&
   = \int\!\! dz\int\!\! 
	d_0(\rho)\frac{d\rho}{\rho^5}\frac{1}{N_c^2-1} 
	\left( 
	  \frac{\pi^3\rho^4}{\alpha_S}\right) G\tilde{G} 
	  (\frac{1}{4}) \left( \frac{4}{3}\pi^2\rho^3 
	\right)^3 \times  \nonumber \\
 &  &\biggl\{
	[
	(\bar{u}\gamma^5u)(\bar{d}d)(\bar{s}s) +
	(\bar{u}u)(\bar{d}\gamma^5d)(\bar{s}s) \nonumber\\
 &  &
	+(\bar{u}u)(\bar{d}d)(\bar{s}\gamma^5s) +
	(\bar{u}\gamma^5u)(\bar{d}\gamma^5d)(\bar{s}\gamma^5s) 
	]     \nonumber \\
 &  &+\frac{3}{8}
      {\biggl[}
	(\bar{u}t^a\gamma^5u)(\bar{d}t^ad)(\bar{s}s) +
	(\bar{u}t^a u)(\bar{d}t^a \gamma^5d)(\bar{s}s) +
	(\bar{u}t^a u)(\bar{d}t^a d)(\bar{s}\gamma^5s)   \nonumber \\
 &  &  \hspace{0.5cm}
       +(\bar{u}t^a\gamma^5 u)(\bar{d}t^a\gamma^5d)(\bar{s}\gamma^5s) 
        \nonumber \\
 &  &  - \frac{3}{4}\bigl[
	(\bar{u}t^a \sigma_{\mu\nu}\gamma^5 u)(\bar{d}t^a \smn d)(\bar{s}s) +
	(\bar{u}t^a \smn u)(\bar{d}t^a \smn\gamma^5d)(\bar{s}s) 
          \nonumber \\ 
 &  &  \hspace{0.5cm}+
       (\bar{u}t^a \smn u)(\bar{d}t^a\smn d)(\bar{s}\gamma^5s)
	+ (\bar{u}t^a\smn\gamma^5u)
	(\bar{d}t^a\smn\gamma^5d)(\bar{s}\gamma^5s)\bigr]  \nonumber 
\eea
\bea
 &  &	-\frac{9}{20}d^{abc}
       \bigl[ 
	(\bar{u}t^a \sigma_{\mu\nu}\gamma^5 u)
        (\bar{d}t^b \smn d)(\bar{s}t^cs) +
	(\bar{u}t^a \smn u)(\bar{d}t^b\smn\gamma^5d)(\bar{s}t^cs) \nonumber \\
 &  &  \hspace{0.5cm}
       +(\bar{u}t^a \smn u)(\bar{d}t^b\smn d)(\bar{s}t^c\gamma^5s) +
	(\bar{u}t^a\smn\gamma^5 u)(\bar{d}t^b \smn\gamma^5d)
        (\bar{s}t^c\gamma^5s)  
       \bigr]\nonumber \\
 &   & \hspace{0.5cm}+ (2\,{\it  cyclic\; permutations}\; 
            u\leftrightarrow d \leftrightarrow s)
      {\biggr]}\nonumber \\
 &  &-\frac{9}{40}d^{abc}
       \biggl[  
	(\bar{u}t^a\gamma^5u)(\bar{d}t^bd)(\bar{s}t^cs) +
	(\bar{u}t^a u)(\bar{d}t^b\gamma^5d)(\bar{s}t^cs) \nonumber\\
 &  &\hspace{0.5cm} 
	+(\bar{u}t^a u)(\bar{d}t^bd)(\bar{s}\gamma^5t^cs)  
    +(\bar{u}t^a\gamma^5u)(\bar{d}t^b\gamma^5d)(\bar{s}t^c\gamma^5s) 
        \biggr] 
 - \frac{9}{32}if^{abc}\times\nonumber \\
 & & \x       \biggl[   
	(\bar{u}t^a \sigma_{\mu\nu}\gamma^5 u)(\bar{d}t^b \sng d)
	(\bar{s}t^c\sgm s) 
       +(\bar{u}t^a \smn u)(\bar{d}t^b\sng\gamma^5d) 
	(\bar{s}t^c\sgm s) \nonumber \\
 &  &  \hspace{0.4cm}
       +(\bar{u}t^a\smn u)(\bar{d}t^b\sng d) 
	(\bar{s}t^c\sgm\gamma^5s)\nonumber\\
 & &   \hspace{0.4cm}
       +(\bar{u}t^a\smn\gamma^5 u)
	(\bar{d}t^b\sng\gamma^5d)(\bar{s}t^c\sgm\gamma^5s)  
      \biggr]
   \biggr\}  
\eea
The strategy is the same as in the glueball case. We Fierz-rearrange 
the lagrangian (\ref{eta_c_lagr}) and apply the vacuum dominance 
and PCAC approximations. The coupling of the $\eta_c$ bound state
to the instanton involves the matrix element
\be
\label{l_etac_def}
\lambda_{\eta_c}=\langle\eta_c|g^2G\tilde{G}|0\rangle .
\ee
We can get an estimate of this matrix element using a simple
two-state mixing scheme for the $\eta_c$ and pseudoscalar glueball. 
We write
\bea
\label{mix1}
 |\eta_c\rangle\;\; &=& \;\;\cos(\theta)|\bar{c}c\rangle
                   +\sin(\theta)|gg\rangle, \\
\label{mix2}
 |0^{-+}\rangle &=& -\sin(\theta)|\bar{c}c\rangle
                   +\cos(\theta)|gg\rangle .
\eea
The matrix element $f_{\eta_c}=\langle 0|2m_c\bar{c}\gamma_5 c|
\eta_c\rangle\simeq 2.8\,{\rm GeV}^3$ is related to the charmonium 
wave function at the origin. The coupling of the topological charge
density to the pseudoscalar glueball was estimated using QCD spectral
sum rules, $\lambda_{0^{-+}} = \langle 0|g^2G\tilde{G}|0^{-+}\rangle 
\simeq 22.5\,{\rm GeV}^3$ \cite{Narison:1996fm}. Using the two-state 
mixing scheme the two ``off-diagonal'' matrix elements $f_{0^{-+}}=
\langle 0| 2m_c\bar{c}\gamma_5 c |0^{-+}\rangle$ and $\lambda_{\eta_c}
=\langle 0|g^2G\tilde{G}|\eta_c\rangle$ are given in terms of one 
mixing angle $\theta$. We can estimate this mixing angle by computing 
the charm content of the pseudoscalar glueball using the heavy quark 
expansion. Using \cite{Franz:2000ee}
\be
\label{hqexp}
\bar{c}\gamma^5c = \frac{i}{8\pi m_c}\a_s G\tilde{G} 
    + O\left(\frac{1}{m_c^3}\right),
\ee
we get $f_{0^{-+}}\simeq 0.14\,{\rm GeV}^3$ and a mixing angle 
$\theta \simeq 3^0$. This mixing angle corresponds to 
\be
\label{l_etac}
\lambda_{\eta_c}\simeq 1.12\,{\rm GeV}^3.
\ee 
The uncertainty in this estimate is hard to assess. Below we 
will discuss a perturbative estimate of the instanton coupling 
to $\eta_c$. In order to check the phenomenological consistency 
of the estimate equ.~(\ref{l_etac}) we have computed the 
$\eta_c$ contribution to the $\langle g^2G\tilde{G}(0)g^2G\tilde{G}
(x)\rangle$ correlation function. The results are shown in 
Fig.~\ref{fig_gg}. The contribution of the pseudoscalar
glueball is determined by the coupling constant $\lambda_{0^{-+}}$
introduced above. The couplings of the $\eta$, $\eta'$ and 
$\eta(1440)$ resonances can be extracted from the decays
$J/\psi\to \gamma\eta$ \cite{Novikov:uy}. We observe that 
the $\eta_c$ contribution is strongly suppressed, as one 
would expect. We also show the $\eta_c$ and $0^{-+}$ glueball
contributions to the $\langle \bar{c}\gamma_5 c(0)\bar{c}
\gamma_5 c(x)\rangle$ correlation function. We observe that 
even with the small mixing matrix elements obtained from 
equs.~(\ref{mix1}-\ref{hqexp}) the glueball contribution
starts to dominate the $\eta_c$ correlator for $x>1$ fm. 

  We now proceed to the calculation of the exclusive decay 
rates. There are four final states that contribute to the 
$K\bar{K}\pi$ channel, $\eta_c\to K^+K^-\pi^0$, $K^0\bar{K}^0
\pi^0$, $K^+\bar{K}^0\pi^-$ and $K^-K^0\pi^+$. Using isospin
symmetry it is sufficient to calculate only one of the 
amplitudes. Fierz rearranging equ.~(\ref{eta_c_lagr}) we 
get the interaction responsible for the $\eta_c\to K^+K^-\pi^0$
\bea
{\cal L}_{I+A}^{K^+K^-\pi^0} &=& \int dz \int\frac{d\r}{\r^5}
   d_0(\r)\frac{1}{N_c^2-1}
   \left(\frac{\pi^3\r^4}{\alpha_s^2}\right)
   \left(\frac{4}{3}\pi^2\r^3\right)^3\nonumber\\
  & & \x \frac{1}{4}
   (\alpha_s G\tilde{G})
   (\bar{s}\gamma^5u)(\bar{u}\g^5s)(\bar{d}\g^5d).
\eea
The decay rate is given by
\be
\label{KKpi}
 \Gamma_{K^+K^-\pi^0} = 
   \int({\rm phase\;\; space})\x |M|^2 =
\left[ \frac{1}{16\pi\sqrt{2}}A\lambda_{\eta_c}K_\pi K_K^2 \right]^2
 \x \left( 0.111\,{\rm MeV} \right),
\ee
with $A$ given in equ.~(\ref{A_int}). Isospin symmetry implies that 
the other $K\bar{K}\pi$ decay rates are given by
\be
 \Gamma_{K^+K^-\pi^0}
 =\Gamma_{K^0\bar{K}^0\pi^0}
 = \left(\frac{1}{\sqrt{2}}\right)^2 \Gamma_{K^0K^-\pi^+}
 = \left(\frac{1}{\sqrt{2}}\right)^2 \Gamma_{K^+\bar{K}^0\pi^-} .
\ee
The total $K\bar{K}\pi$ decay rate is
\be
 \Gamma_{K\bar{K}\pi}=6\x 
 \left[ \frac{1}{16\pi\sqrt{2}}A\lambda_{\eta_c} K_\pi K_K^2 \right]^2
 \x \left(0.111\,{\rm MeV}\right).
\ee  
In a similar fashion we obtain
\bea
 \Gamma_{\eta\pi\pi}
  &=& \frac{3}{2}\x
   \left[\frac{1}{16\pi}
     A\lambda_{\eta_c} K_\eta^{s}K_\pi^2 \right]^2
     \x \left(0.135\,{\rm MeV}\right),  \\
\Gamma_{\eta'\pi\pi}
  &=& \frac{3}{2}\x
   \left[\frac{1}{16\pi}
   A\lambda_{\eta_c} K_{\eta'}^{s}K_\pi^2 \right]^2
   \x \left( 0.0893\,{\rm MeV}\right),\\
\Gamma_{K\bar{K}\eta}
  &=& 2 \x
   \left[\frac{1}{16\pi}
   A\lambda_{\eta_c} K_{\eta}^{q}K_K^2 \right]^2
   \x \left( 0.0788\,{\rm MeV}\right),
\eea
\bea
\Gamma_{K\bar{K}\eta'}
  &=& 2 \x
   \left[\frac{1}{16\pi}
   A\lambda_{\eta_c} K_{\eta'}^{q}K_K^2 \right]^2
   \x \left( 0.0423\,{\rm MeV}\right),\\
\label{etaetaeta}
\Gamma_{\eta\eta\eta}
  &=& \frac{1}{6} \x
   \left[\frac{3!}{16\pi}
   A\lambda_{\eta_c} (K_{\eta}^{q})^2K_{\eta}^s \right]^2
   \x \left( 0.0698\,{\rm MeV}\right).
\eea
Here, the first factor is the product of the isospin 
and final state symmetrization factors. The second factor 
is the amplitude and the third factor is the phase-space 
integral.
\begin{figure}
\begin{center}
\includegraphics[width=6cm,angle=-90]{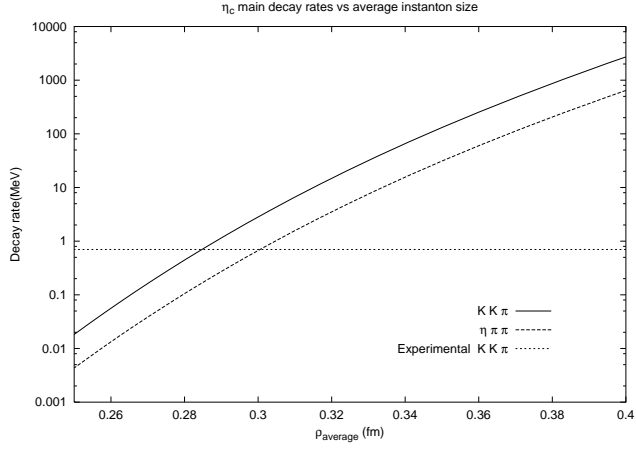}
\end{center}
\begin{center}
\caption{\label{fig_etac_rho}
Decay widths $\eta_c\to KK\pi$ and $\eta_c\to \eta\pi\pi$
as a function of the average instanton size $\rho$. The
short dashed line shows the experimental $KK\pi$ width.}
\end{center}
\end{figure}

  In Fig.~\ref{fig_etac_rho} we show the dependence of the 
decay rates on the average instanton size. We observe that 
the experimental $K\bar{K}\pi$ rate is reproduced for $\bar{\rho}
=0.29$ fm. This number is consistent with the phenomenological 
instanton size. However, given the strong dependence on the 
average instanton size it is clear that we cannot reliably
predict the decay rate. On the other hand, the following 
ratios are independent of the average instanton size
\bea
 \frac{\Gamma_{K\bar{K}\pi}}{\Gamma_{\eta\pi\pi}} &=&
    4\x \left[
  \frac{K_K^2}
       {\sqrt{2}K_\eta^{s}K_\pi}\right]^2
   \x\left(\frac{0.111}{0.135}\right)=4.23 \pm 1.27 ,   
       \\
 \frac{\Gamma_{\eta\pi\pi}}{\Gamma_{\eta'\pi\pi}} &=&
  \left(\frac{K_\eta^{s}}{K_{\eta'}^{s}} \right)^2
  \x \left(\frac{0.135}{0.0893} \right)=1.01 ,
       \\
 \frac{\Gamma_{K\bar{K}\eta}}{\Gamma_{K\bar{K}\pi}} &=&
  \frac{1}{3}\x \left[
  \frac{\sqrt{2}K_\eta^{q}}
       {K_\pi}\right]^2
   \x\left(\frac{0.0788}{0.111}\right)=0.141 \pm 0.042 , 
       \\
 \frac{\Gamma_{K\bar{K}\eta}}{\Gamma_{K\bar{K}\eta'}} &=&
  \left(\frac{K_\eta^{q}}{K_{\eta'}^{q}} \right)^2
  \x \left(\frac{0.0788}{0.0423} \right)=2.91 ,
       \\ 
 \frac{\Gamma_{\eta\eta\eta}}{\Gamma_{K\bar{K}\pi}} &=&
  \frac{1}{36}\x \left[
  \frac{3!\sqrt{2}(K_\eta^{q})^2K_{\eta}^s}
       {K_\pi K_K^2}\right]^2
   \x\left(\frac{0.0698}{0.111}\right)=0.011 \pm 0.003,
\eea
where we have only quoted the error due to the uncertainty 
in $m_s$. These numbers should be compared to the experimental 
results 
\bea
 \left.\frac{\Gamma_{K\bar{K}\pi}}{\Gamma_{\eta\pi\pi}}
 \right|_{exp} &=&  1.1\pm 0.5\\
\left.\frac{\Gamma_{\eta\pi\pi}}{\Gamma_{\eta'\pi\pi}} 
 \right|_{exp} &=&  1.2\pm 0.6.
\eea
We note that the ratio $B(\eta\pi\pi)/B(\eta'\pi\pi)$
is compatible with our results while the ratio $B(K\bar{K}
\pi)/B(\eta\pi\pi)$ is not. This implies that either there 
are contributions other than instantons, or that the PCAC 
estimate of the ratio of coupling constants is not reliable,
or that the experimental result is not reliable. The branching 
ratios for $\eta\pi\pi$ and $\eta'\pi\pi$ come from MARK
II/III experiments  
\cite{Partridge:1980vk,Baltrusaitis:1985mr}. We 
observe that our results for $B(K\bar{K}\eta)/B(K\bar{K}\pi)$ 
and $B(K\bar{K}\eta')/B(K\bar{K}\pi)$ are consistent with 
the experimental bounds. 

 Another possibility is that there is a significant 
contribution from a scalar resonance that decays into 
$\pi\pi$. Indeed, instantons couple strongly to the 
$\sigma(600)$ resonance, and this state is not resolved
in the experiments. We have therefore studied the direct
instanton contribution to the decay $\eta_c \to \sigma\eta$. 
After Fierz rearrangement we get the effective vertex
\bea
 {\cal L}_{\sigma\eta} &=& \int dA\;
 (\a_s G\tilde{G})\; \frac{1}{4}
 \left[
  (\bar{u}\gamma^5 u)(\bar{d}d)(\bar{s}{s}) +
  (\bar{u} u)(\bar{d}\gamma^5d)(\bar{s}{s}) +
  (\bar{u} u)(\bar{d}d)(\bar{s}\gamma^5{s}) 
 \right] \nonumber \\
&  & \mbox{}-\int dB\;
 m_s (\a_s G\tilde{G})\;\frac{1}{2}
 \left[
  (\bar{u}\gamma^5 u)(\bar{d}d) +
  (\bar{u} u)(\bar{d}\gamma^5d) 
 \right ] ,
\eea
where the integrals $A$ and $B$ are defined in
equ.~(\ref{A_int},\ref{B_int}). The only new matrix element we 
need is $f_\sigma=\langle\sigma|\bar{u}u + \bar{d}d|0\rangle 
\simeq (500\,{\rm MeV})^2$ \cite{Shuryak:1992ke}. We get
\bea
 \Gamma_{\eta_c\to\sigma\eta} &=& \frac{1}{16\pi m_{\eta_c}^3}
  \sqrt{
   [m_{\eta_c}^2 - (m_\sigma + m_\eta)^2]
   [m_{\eta_c}^2 - (m_\sigma - m_\eta)^2]
   } \nonumber \\
  & &\x
  \left[
    \frac{1}{16\pi}f_\sigma \lambda_{\eta_c} 
    [(AQ_s - 2Bm_s)K_\eta^q + AK_\eta^sQ_d]
  \right]^2 .
\eea
Compared to the direct decay $\eta_c\to\eta\pi\pi$ the 
$\eta_c\to\eta\sigma$ channel is suppressed by a factor 
$\sim (2\pi^2/m_{\eta_c}^2)\cdot (Q_qf_\sigma/K_\pi^2)^2
\sim 1/100$. Here, the first factor is due to the difference 
between two and three-body phase space and the second factor
is the ratio of matrix elements. We conclude that the direct 
production of a $\sigma$ resonance from the instanton does
not give a significant contribution to $\eta_c\to\eta(\eta')
\pi\pi$. This leaves the possibility that the $\pi\pi$ 
channel is enhanced by final state interactions.
\begin{figure}
\begin{center}
\includegraphics[width=10cm,angle=0]{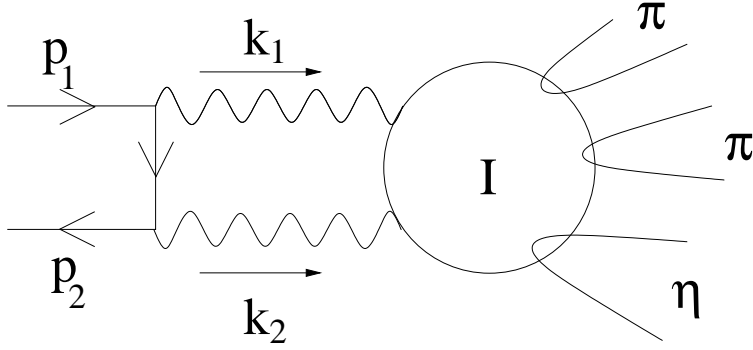}
\end{center}
\begin{center}
\caption{\label{Feynman_diag_fig}
The Feynman diagram corresponding to the perturbative treatment of
charmonium decay.}
\end{center}
\end{figure}

  Finally, we present a perturbative estimate of the coupling
of the $\eta_c$ to the instanton. We follow the method used
by Anselmino and Forte in order to estimate the instanton
contribution to $\eta_c\to p\bar{p}$ \cite{Anselmino:1993bd}.
The idea is that the charmonium state annihilates into two 
gluons which are absorbed by the instanton. The Feynman diagram 
for the process is shown in Fig.\ref{Feynman_diag_fig}. The 
amplitude is given by
\bea
\label{A_ccI}
 A_{c\bar{c}\to I}  &=& 
 g^2 \int \frac{d^4k_1}{(2\pi)^4} \int \frac{d^4k_2}{(2\pi)^4}
 \, (2\pi)^4 \delta^4(p_1+p_2-k_1-k_2)\x \nonumber \\
 & &\hspace{-0.3cm} 
   \bar{v}(p_2)\left[ \gamma_\mu \frac{\lambda^a}{2}
   \frac{1}{ {p\!\!\!/}_1-{k\!\!\!/}_1-m_c}
   \gamma_\nu\frac{\lambda^b}{2}\right] u(p_1)\;
   A_\mu^{a,cl}(k_2)A_\nu^{b,cl}(k_1),
\eea
where $u(p)$ and $\bar{v}(p)$ are free particle charm 
quark spinors and $A_\mu^{a,cl}(k)$ is the Fourier transform
of the instanton gauge potential 
\be
\label{A(k)}
 A_\mu^{a,cl}(k) = -i \frac{4\pi^2}{g} 
  \frac{\bar{\eta}^a_{\mu\nu}k^\nu}{k^4}\Phi(k),
  \hspace{1cm} 
  \Phi(k) = 4\left( 1 -\frac{1}{2}K_2(k\rho)(k\rho)^2
    \right).
\ee
The amplitude for the charmonium state to couple to an 
instanton is obtained by folding equ.~(\ref{A_ccI}) with 
the $\eta_c$ wave function $\psi(p)$. In the non-relativistic
limit the amplitude only depends on the wave function at 
the origin. 

  The perturbative estimate of the transition rate is 
easily incorporated into the results obtained above by
replacing the product $A\lambda_{\eta_c}$ in 
equs.~(\ref{KKpi}-\ref{etaetaeta}) according to
\be 
\label{sub}
 A\lambda_{\eta_c} \to 
    \int\frac{d\r}{\r^5}d_0(\r)
    \left(\frac{4}{3}\pi^2\r^3\right)^3
     \left(4\pi\right) \frac{8m_c^{3/2}}{\sqrt{6}}
      |\psi(0)| I_{\eta_c}(\rho)
      \x \frac{g^2(m_c^{-1})}{g^2(\rho)},
\ee 
with 
\be 
\label{I_etac}
 I_{\eta_c}(\rho) = \int d^4k \;
   \frac{\vec{k}^2\Phi(k)\Phi(k-2p_c)}
        {k^4(k-2p_c)^4((k-p_c)^2+m_c^2)}.
\ee
Here, $p_c=(m_c,0)\simeq (M_{\eta_c}/2,0)$ is the momentum 
of the charm quark in the charmonium rest frame. We note that
because of the non-perturbative nature of the instanton 
field higher order corrections to equ.~(\ref{sub}) are 
only suppressed by $g^2(m_c^{-1})/g^2(\rho)$. 

\begin{figure}
\begin{center}
\includegraphics[width=6cm,angle=-90]{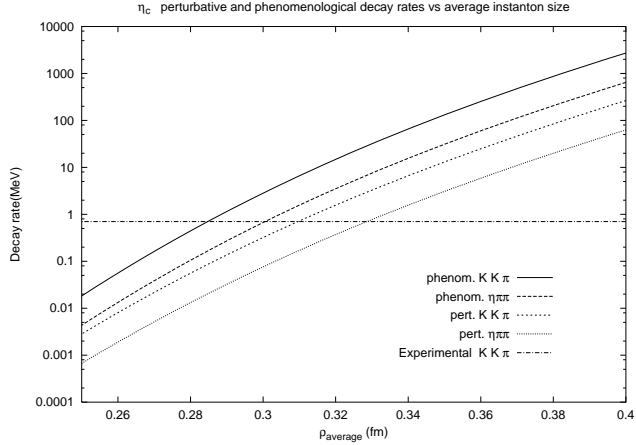}
\end{center}
\begin{center}
\caption{\label{fig_etac_pert}
Decay rates $\Gamma(\eta_c\to K\bar{K}\pi)$ and
$\Gamma(\eta_c\to \eta\pi\pi)$ as a function of the
average instanton size $\bar{\rho}$. We show both the
results from a phenomenological and a perturbative
estimate of the $\bar{c}c$ coupling to the instanton. }
\end{center}
\end{figure}

The integral $I_{\eta_c}$ cannot be calculated analytically. 
We use the parameterization
\be 
   I_{\eta_c}(\rho) \simeq 
   \frac{\pi^2\;A_0\;\rho^4\log(1+1/(m_c\rho))}
        {1 + B_0\;(m_c\rho)^4\log(1+1/(m_c\rho))},
\ee
which incorporates the correct asymptotic behavior. We
find that $A_0=0.213$ and $B_0=0.124$ provides a good
representation of the integral. In Fig.~\ref{fig_etac_pert} 
we show the results for the $\eta_c$ decay rates as a 
function of the average instanton size. We observe that 
the results are similar to the results obtained from the 
phenomenological estimate equ.~(\ref{l_etac}). The 
effective coupling $(A\lambda_{\eta_c})$ differs from
the estimate equ.~(\ref{l_etac}) by about a factor of 3.
The experimental $K\bar{K}\pi$ rate is reproduced for 
$\bar{\rho}=0.31$ fm. 

\section{Chi charm decays}
\label{sec_chi}
 
  Another interesting consistency check on our results is 
provided by the study of instanton induced decays of the 
$\chi_c$ into pairs of Goldstone bosons. The $\chi_c$ is 
a scalar charmonium bound state with mass $m_{\chi_c}=
3415$ MeV and width $\Gamma_{\chi_c}= 14.9$ MeV. In a 
potential model the $\chi_c$ corresponds to the $^3P_0$
state. In perturbation theory the total decay rate is
dominated by $\bar{c}c\to 2g$. The main exclusive decay 
channels are $\chi_c\to 2(\pi^+\pi^-)$ and $\chi_c\to
\pi^+\pi^-K^+K^-$ with branching ratios $(2.4\pm 0.6)\%$
and $(1.8\pm 0.6)\%$, respectively. It would be very 
interesting to know whether these final states are 
dominated by scalar resonances. We will concentrate 
on final states containing two pseudoscalar mesons. 
There are two channels with significant branching ratios,
$\chi_c\to\pi^+\pi^-$ and $\chi_c\to K^+K^-$ with 
branching ratios $(5.0\pm 0.7)\cdot 10^{-3}$ and 
$(5.9\pm 0.9)\cdot 10^{-3}$.

\begin{figure}
\begin{center}
\includegraphics[width=6cm,angle=-90]{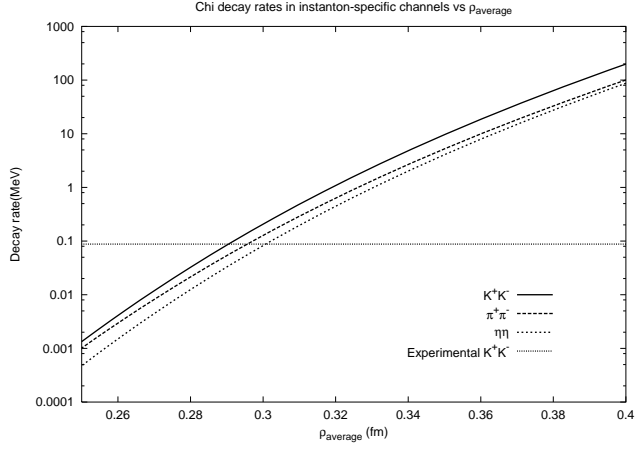}
\end{center}
\begin{center}
\caption{\label{fig_chi_dec}
Decay widths $\chi_c \to K^+K^-,\pi^+\pi^-$ and $\eta\eta$
as a function of the average instanton size $\rho$. The
short dashed line shows the experimental $K^+K^-$ width.}
\end{center}
\end{figure}

\begin{figure}
\begin{center}
\includegraphics[width=6cm,angle=-90]{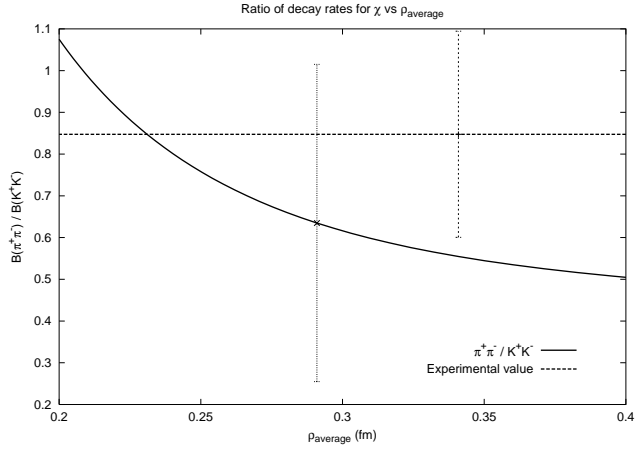}
\end{center}
\begin{center}
\caption{\label{fig_chi_ratio}
Ratio $B(\chi_c\to\pi^+\pi^-)/B(\chi_c \to K^+K^-)$ of
decay rates as a function of the average instanton size.
The dashed line shows the experimental value 0.84. We
also show the experimental uncertainty, as well as the
uncertainty in the instanton prediction due to the
the value of the strange quark mass. }
\end{center}
\end{figure}

 The calculation of these two decay rates proceeds along 
the same lines as the calculation of the $0^{++}$ glueball 
decays. The only new ingredient is the $\chi_c$ coupling to 
the gluon field strength $G^2$. We observe that the total
$\chi_c$ decay rate implies that $\langle 0|2m_c\bar{c}c|\chi_c
\rangle = 3.1\,{\rm GeV}^3\simeq \langle 0|2m_c\bar{c}i\gamma_5c
|\eta_c\rangle$. This suggests that a rough estimate of the
$\chi_c$ coupling to $G^2$ is given by
\be
\lambda_{\chi_c}\equiv \langle \chi_c | g^2 G^2 | 0 \rangle  
 \simeq \lambda_{\eta_c} = 1.12\; {\rm GeV}^3.
\ee
Using this result we can obtain the $\chi_c$ decay rates
by rescaling the scalar glueball decay rates 
equ.~(\ref{glue_pipi_ms}-\ref{glue_etaPetaP_ms}) 
according to 
\be
\Gamma_{\chi_c \rightarrow m1,m2} = 
  \Gamma_{0^{++} \rightarrow m1,m2}  \x 
  \left.   \left( \frac{\lambda_{\chi_c}}{\lambda_{0^{++}}} \right)^2
  \right|_{m_{0^{++}}\to m_{\chi_c}},
\ee
where $m1,m2$ labels the two-meson final state. In Fig.~\ref{fig_chi_dec}
we show the dependence of the $\chi_c$ decay rates on the average 
instanton size $\bar{\rho}$. We observe that the experimental 
$\pi^+\pi^-$ decay rate is reproduced for $\bar{\rho}=0.29$ fm.
In Fig. \ref{fig_chi_ratio} we plot the ratio of decay rates 
for $\pi^+\pi^-$ and $K^+K^-$. Again, the experimental value
is reproduced for $\bar{\rho}\sim 0.3$ fm.

\begin{figure}
\begin{center}
\includegraphics[width=6cm,angle=-90]{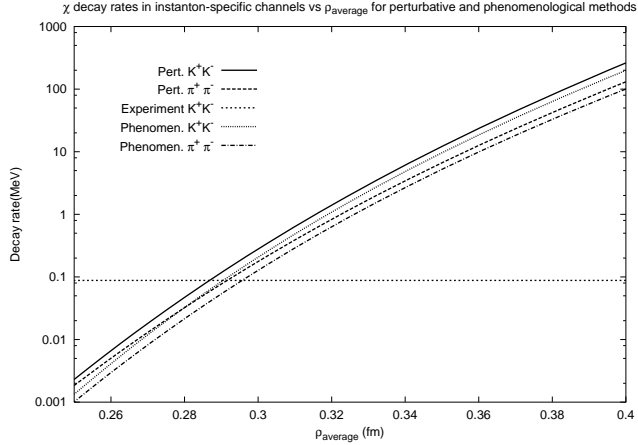}
\end{center}
\begin{center}
\caption{\label{fig_chi_c_pert}
Decay rates $\Gamma(\chi_c\to \pi^+\pi^-)$ and
$\Gamma(\chi_c\to K^+ K^-)$ as a function of the
average instanton size $\bar{\rho}$. We show both the
results from a phenomenological and a perturbative
estimate of the $\bar{c}c$ coupling to the instanton. }
\end{center}
\end{figure}

  Finally, we can also estimate the $c\bar{c}$ coupling 
to the instanton using the perturbative method introduced
in section \ref{sec_eta}. In the case of the $\chi_c$ we use
\bea
 \frac{1}{4\pi}\lambda_{\chi_c}A 
 &\rightarrow &
  \frac{1}{2\sqrt{3\pi}}
 \sqrt{M_\chi}R'(0)\int \frac{d_0(\rho)}{\rho^5}d\rho
 \left(\frac{4}{3}\pi^2
 \rho^3\right)^3 \frac{g^2(m_c)}{g^2(\rho)}\x I_\chi(\rho),
  \nonumber\\
 \frac{1}{4\pi}\lambda_{\chi_c}B 
  &\rightarrow & 
 \frac{1}{2\sqrt{3\pi}}
 \sqrt{M_\chi}R'(0)\int \frac{d_0(\rho)}{\rho^5}d\rho
 \left(\frac{4}{3}\pi^2
 \rho^3\right)^2 \rho\, \frac{g^2(m_c)}{g^2(\rho)}\x I_\chi(\rho)\:,
 \nonumber
\eea
where $R'(0)\simeq 0.39\,{\rm GeV}^{5/2}$ is the derivative of 
the $^3P_0$ wave function at the origin and $I_{\chi_c}$ is the 
loop integral
\be
  I_\chi(\rho)=\int\! d^4 k\: \frac{\Phi(k)\Phi(|2p_c-k|)}
{k^4(2p_c-k)^4}\:\:
 \frac{15(k-p_c)^2 +3m_c^2 + 4\vec{k}^2 }{(k-p_c)^2+m_c^2}.
\ee
In Fig.~\ref{fig_chi_c_pert} we compare the perturbative 
result with the phenomenological estimate. Again, the results
are comparable. The experimental $\pi^+\pi^-$ rate is reproduced
for $\bar{\rho}=0.29$ fm.

\section{Conclusions}
\label{sec_summary}
 
 In summary we have studied the instanton contribution to
the decay of a number of ``gluon rich'' states in the 
(1.5-3.5) GeV range, the scalar glueball, the $\eta_c$ and
the $\chi_c$. In the case of charmonium instanton induced
decays are probably a small part of the total decay rate, 
but the final states are very distinctive. In the case 
of the scalar glueball classical fields play an important 
role in determining the structure of the bound state and 
instantons may well dominate the total decay rate. 

 We have assumed that the gluonic system is small and that 
the instanton contribution to the decay can be described
in terms of an effective local interaction. The meson 
coupling to the local operator was determined using PCAC.
Using this method we find that the scalar glueball decay 
is dominated by the $K\bar{K}$ final state for glueball
masses $m_{0^{++}}>1$ GeV. In the physically interesting 
mass range $1.5\,{\rm GeV}<m_{0^{++}}<1.75\,{\rm GeV}$
the branching ratios satisfy $B(\eta\eta):B(\pi\pi):
B(\bar{K}K)=1:(3.3\pm 0.3):(5.5\pm 0.5)$.
 
 Our main focus in this work are $\eta_c$ decays into 
three pseudoscalar Goldstone bosons. We find that the 
experimental decay rate $\Gamma(\eta_c\to K\bar{K}\pi)$
can be reproduced for an average instanton size $\bar{\rho}
=0.31$, consistent with phenomenological determinations
and lattice results. This in itself is quite remarkable, 
since the phenomenological determination is based on
properties of the QCD vacuum.

The ratio of decay rates $B(\eta'\pi\pi):B(\eta\pi\pi):B(K\bar{K}\pi)
=1:1:(4.2\pm 1.3)$ is insensitive to the average 
instanton size. While the ratio $B(\eta'\pi\pi):B(\eta
\pi\pi)=1:1$ is consistent with experiment, the ratio
$B(\eta\pi\pi):B(K\bar{K})=1:(4.2\pm 1.3)$ is at best 
marginally consistent with the experimental value
$1.1\pm 0.5$. We have also studied $\chi_c$ decays 
into two pseudoscalars. We find that the absolute decay 
rates can be reproduced for $\bar{\rho}=0.29$ fm. 
Instantons are compatible with the measured ratio
$B(K^+K^-):B(\pi^+\pi^-)=1.2$

  There are many questions that remain to be answered. 
On the experimental side it would be useful if additional
data for the channels $\eta_c\to\eta'\pi\pi,\eta\pi\pi$
were collected. One important question is whether $(\pi\pi)$
resonances are important. It should also be possible to 
identify the smaller decay channels $\eta_c\to K\bar{K}\eta, 
K\bar{K}\eta'$. In addition to that, it is interesting to 
study the distribution of the final state mesons in all
three-meson channels. Instantons predict that the production 
mechanism is completely isotropic and that the final state 
mesons are distributed according to three-body phase space.

  In addition to that, there are a number of important 
theoretical issues that remain to be resolved. In the limit
in which the scalar glueball is light the decay $0^{++}\to \pi\pi
(\bar{K}K)$ can be studied using effective lagrangians based on 
broken scale invariance \cite{Shifman:1988zk,Jaminon:ac,Jin:2002up}.
Our calculation based on direct instanton effects is valid
in the opposite limit. Nevertheless, the instanton liquid 
model respects Ward identities based on broken scale invariance
\cite{Schafer:1996wv} and one should be able to recover the
low energy theorem. In the case $0^{++}\to \pi\pi(\bar{K}K)$ 
one should also be able to study the validity of the PCAC 
approximation in more detail. This could be done, for example,
using numerical simulations of the instanton liquid. Finally 
we need to address the question how to properly compute the 
overlap of the initial $\bar{c}c$ system with the instanton. 
This, of course, is a more general problem that also affects 
calculations of electroweak baryon number violation in high 
energy $p\bar{p}$ collisions \cite{Ringwald:1989ee,Espinosa:qn} 
and QCD multi-particle production in hadronic collisions 
\cite{Nowak:2000de}.

\chapter{Instantons and the spin of the nucleon}

\section{Introduction}
\label{sec_intro}

 In this chapter we will try to understand whether the so-called
"nucleon spin crises" can be related to instanton effects.
The current interest in the spin structure of the nucleon
dates from the 1987 discovery by the European Muon Collaboration
that only about 30\% of the spin of the proton is carried by 
the spin of the quarks \cite{Ashman:1987hv}. This result is
surprising from the point of view of the naive quark model, 
and it implies a large amount of OZI (Okubo-Zweig-Iizuka rule)
violation in the flavor singlet axial vector channel. The axial 
vector couplings of the nucleon are related to polarized quark 
densities by
\bea
 g_A^3 &=& \Delta u - \Delta d, \\
 g_A^8 &=& \Delta u + \Delta d - 2\Delta s,
\eea
\bea
 g_A^0 & \equiv & \Delta \Sigma = \Delta u + \Delta d + \Delta s.
\eea
The first linear combination is the well known axial vector coupling 
measured in neutron beta decay, $g_A^3=1.267\pm 0.004$.
The hyperon decay constant is less well determined. A conservative
estimate is $g_A^8= 0.57\pm 0.06$. Polarized deep inelastic
scattering is sensitive to another linear combination of the 
polarized quark densities and provides a measurement of the 
flavor singlet axial coupling constant $g_A^0$. Typical results 
are in the range $g_A^0=(0.28-0.41)$, see \cite{Filippone:2001ux} 
for a recent review. 

 Since $g_A^0$ is related to the nucleon matrix element of the 
flavor singlet axial vector current many authors have speculated 
that the small value of $g_A^0$ is in some way connected to the 
axial anomaly, see \cite{Dorokhov:ym,Bass:1993bs,Bass:2003vp} 
for reviews. The axial anomaly relation
\be
\label{anom}
\partial^\mu A_\mu^0 = \frac{N_fg^2}{16\pi^2}
 G_{\mu\nu}^a \tilde{G}_{\mu\nu}^a
 + \sum_f 2m_f\bar{q}_fi\gamma_5q_f
\ee
implies that matrix elements of the flavor singlet axial 
current $A_\mu^0$ are related to matrix elements of the 
topological charge density. The anomaly also implies that
there is a mechanism for transferring polarization from 
quarks to gluons. In perturbation theory the nature of the 
anomalous contribution to the polarized quark distribution 
depends on the renormalization scheme. The first moment of 
the polarized quark density in the modified minimal
subtraction $(\overline{MS})$ scheme is related to the 
first moment in the Adler-Bardeen $(AB)$ scheme by 
\cite{Altarelli:1988nr}
\be
\Delta \Sigma_{\overline{MS}} = \Delta \Sigma_{AB} 
 -N_f\frac{\alpha_s(Q^2)}{2\pi} \Delta G(Q^2),
\ee
where $\Delta G$ is the polarized gluon density. Several 
authors have suggested that $\Delta \Sigma_{AB}$ is more 
naturally associated with the ``constituent'' quark spin 
contribution to the nucleon spin, and that the smallness 
of $\Delta \Sigma_{\overline{MS}}$ is due to a cancellation 
between $\Delta\Sigma_{AB}$ and $\Delta G$. The disadvantage
of this scheme is that $\Delta\Sigma_{AB}$ is not associated
with a gauge invariant local operator \cite{Jaffe:1989jz}. 

 Non-perturbatively the anomaly implies that $g_A^0=\Delta
\Sigma$ can be extracted from nucleon matrix elements 
of 
the pseudoscalar 
density $m\bar{\psi}i\gamma_5\psi$ and
the topological charge density $g^2G^a_{\mu\nu}
\tilde{G}^a_{\mu\nu}/(32\pi^2)$ 
. The nucleon matrix
element of the topological charge density is not known, 
but the matrix element of the scalar density $g^2G^a_{\mu\nu}
G^a_{\mu\nu}$ is fixed by the trace anomaly. We have 
\cite{Shifman:zn}
\be 
\langle N(p)| \frac{g^2}{32\pi^2}G^a_{\mu\nu}G^a_{\mu\nu}
 | N(p')\rangle = C_S(q^2)m_N\bar{u}(p)u(p'),
\ee
with $C_S(0)=-1/b$ where $b=11-2N_f/3$ is the first coefficient 
of the QCD beta function. Here, $u(p)$ is a free nucleon spinor. 
Anselm suggested that in an instanton model of the QCD vacuum 
the gauge field is approximately self-dual, $G^2=\pm G\tilde{G}$, 
and the nucleon coupling constants of the scalar and pseudoscalar 
gluon density are expected to be equal, $C_S(0)\simeq C_P(0)$
\cite{Anselm:1992wz}, see also \cite{Kuhn:1990df}. Using $g_A^0=
N_f C_P(0)$ in the chiral limit we get $g_A^0 \simeq - N_f/b \simeq 
-0.2$, which is indeed quite small. 

 A different suggestion was made by Narison, Shore, and 
Veneziano \cite{Narison:hv}. Narison et al.~argued that 
the smallness of $\Delta \Sigma=g_A^0$ is not related
to the structure of the nucleon, but a consequence
of the $U(1)_A$ anomaly and the structure of the 
QCD vacuum. Using certain assumptions about the 
nucleon-axial-vector current three-point function 
they derive a relation between the singlet and octet 
matrix elements,
\be
\label{vs}
 g_A^0 = g_A^8 \frac{\sqrt{6}}{f_\pi} \sqrt{\chi'_{top}(0)}.
\ee
Here, $f_\pi=93$ MeV is the pion decay constant and 
$\chi'_{top}(0)$ is the slope of the topological charge 
correlator
\be 
\label{chi}
\chi_{top}(q^2) = \int d^4x\, e^{iqx} \,
 \langle Q_{top}(0) Q_{top}(x) \rangle ,
\ee
with $Q_{top}(x)=g^2G^a_{\mu\nu}\tilde{G}^a_{\mu\nu}/(32\pi^2)$.
In QCD with massless fermions topological charge is screened
and $\chi_{top}(0)=0$. The slope of the topological charge 
correlator is proportional to the screening length. In QCD 
we expect the inverse screening length to be related to the 
$\eta'$ mass. Since the $\eta'$ is heavy, the screening 
length is short and $\chi'_{top}(0)$ is small. Equation 
(\ref{vs}) relates the suppression of the flavor singlet 
axial charge to the large $\eta'$ mass in QCD. 

 Both of these suggestions are very interesting, but the status 
of the underlying assumptions is somewhat unclear. In this 
work we would like to address the role of the anomaly in 
the nucleon spin problem, and the more general question
of OZI violation in the flavor singlet axial-vector channel, 
by computing the axial charge of the nucleon and the axial-vector 
two-point function in the instanton model. There are several reasons 
why instantons are important in the spin problem. First of all, 
instantons provide an explicit, weak coupling, realization of 
the anomaly relation equ.~(\ref{anom}) and the phenomenon 
of topological charge screening \cite{Schafer:1996wv,Shuryak:1994rr}.
Second, instantons provide a successful phenomenology
of OZI violation in QCD \cite{Schafer:2000hn}. Instantons
explain, in particular, why violations of the OZI rule 
in scalar meson channels are so much bigger than OZI
violation in vector meson channels. And finally, the 
instanton liquid model gives a very successful description
of baryon correlation functions and the mass of the 
nucleon \cite{Schafer:1993ra,Diakonov:2002fq}.

 Our ideas are organized as follows. In Sect.~\ref{sec_anom}
we review the calculation of the anomalous contribution to the 
axial-vector current in the field of an instanton. In 
Sect.~\ref{sec_pia} and \ref{sec_gaq} we use this result 
in order to study OZI violation in the axial-vector correlation 
function and the axial coupling of a constituent quark. Our 
strategy is to compute the short distance behavior of the 
correlation functions in the single instanton approximation 
and to determine the large distance behavior using numerical 
simulations. In Sect.~\ref{sec_gan} we present numerical 
calculations of the axial couplings of the nucleon and in 
Sect.~\ref{sec_sum} we discuss our conclusions. Some 
results regarding the spectral representation of nucleon 
three-point functions are collected in appendix \ref{app_spect_rep}.

\section{Axial Charge Violation in the Field of an Instanton}
\label{sec_anom}

 We would like to start by showing explicitly how the axial
anomaly is realized in the field of an instanton. This 
discussion will be useful for the calculation of the OZI
violating part of the axial-vector correlation function 
and the axial charge of the nucleon. The flavor singlet 
axial-vector current in a gluon background is given by    
\be 
\label{axcur}
A_\mu(x) = {\rm Tr}\left[\gamma_5\gamma_\mu S(x,x) \right]
\ee
where $S(x,y)$ is the full quark propagator in the background 
field. The expression on the right hand side of equ.~(\ref{axcur})
is singular and needs to be defined more carefully. We will 
employ a gauge invariant point-splitting regularization
\be
\label{ax_reg}
 {\rm Tr}\left[\gf\gamma_\mu S(x,x)\right] 
 \equiv \lim_{\epsilon\to 0}
  {\rm Tr}\left[\gf\gamma_\mu 
    S(x+\epsilon,x-\epsilon)
    P\exp\left(-i\int_{x-\epsilon}^{x+\epsilon}A_\mu(x)dx\right)
    \right].
\ee
In the following we will consider an (anti) instanton in singular 
gauge. The gauge potential of an instanton of size $\rho$ and 
position $z=0$ is given by
\be 
 A_\mu^a =  \frac{2\rho^2}{x^2+\rho^2}
  \frac{x^\nu}{x^2}
  \,R^{ab}\bar\eta^b_{\mu\nu}.
\ee
Here, $\bar{\eta}^a_{\mu\nu}$ is the 't Hooft symbol and
$R^{ab}$ characterizes the color orientation of the instanton.
The fermion propagator in a general gauge potential can be 
written as 
\be
\label{prop_sum}
S(x,y)=\sum_{\lambda}\frac{\Psi_\lb(x)\Psi_\lb^{+}(y)}{\lb-m},
\ee
where $\Psi_\lb(x)$ is a normalized eigenvector of the Dirac 
operator with eigenvalue $\lambda$, $\Dslash \Psi_\lb(x)=\lb
\Psi_\lb(x)$. We will consider the limit of small quark masses. 
Expanding equ.~(\ref{prop_sum}) in powers of $m$ gives
\be
\label{prop_expan}
S_{\pm}(x,y)=-\frac{\Psi_0(x)\Psi_0^{+}(y)}{m} 
  +  \sum_{\lb\ne 0}\frac{\Psi_\lb(x)\Psi_\lb^{+}(y)}{\lb}
  + m\sum_{\lb\ne 0}\frac{\Psi_\lb(x)\Psi_\lb^{+}(y)}{\lb^2} 
  + O(m^2).
\ee
Here we have explicitly isolated the zero mode propagator. 
The zero mode $\Psi_0$ was found by 't Hooft and is given by
\be 
\Psi_0(x) = \frac{\rho}{\pi} \frac{1}{(x^2+\rho^2)^{3/2}}
  \frac{\gamma\cdot x}{\sqrt{x^2}} \gamma_\pm \phi.
\ee
Here, $\phi^{a\alpha}=\epsilon^{a\alpha}/\sqrt{2}$ is a constant 
spinor and $\gpm=(1\pm\gf)/2$ for an instanton/ anti-instanton. 
The second term in equ.~(\ref{prop_expan}) is the non-zero mode 
part of the propagator in the limit $m\to 0$ \cite{Brown:1977eb} 
\be
\label{vec_prop}
 S_{\pm}^{NZ}(x,y) 
   \equiv  \sum_{\lb\ne 0}\frac{\Psi_\lb(x)\Psi_\lb^{+}(y)}{\lb}
   =  \vec{\Dslash}_x\Delta_{\pm}(x,y)\gpm + 
              \Delta_\pm(x,y)\ivec{\Dslash}_y \gmp
\ee
where $D^\mu=\partial^\mu -iA_\pm^\mu$ and $\Delta_{\pm}(x,y)$ is 
the propagator of a scalar field in the fundamental representation.
Equ.~(\ref{vec_prop}) can be verified by checking that $S_{\pm}^{NZ}$
satisfies the equation of motion and is orthogonal to the zero mode.
The scalar propagator can be found explicitly
\be
\label{scal_prop}
\Delta_{\pm}(x,y)  =  \Delta_0(x,y)
   \frac{1}{\sqrt{1+\frac{\rho^2}{x^2}}}
   \left(1+\frac{\rho^2\sigma_{\mp}\cdot x
    \sigma_{\pm}\cdot y} {x^2 y^2}\right)
   \frac{1}{\sqrt{1+\frac{\rho^2}{y^2}}} ,
\ee
where $\Delta_0=1/(4\pi^2\Delta^2)$ with $\Delta = x-y$
is the free scalar propagator. The explicit form of the 
non-zero mode propagator can be obtained by substituting 
equ.~(\ref{scal_prop}) into equ.~(\ref{vec_prop}). We find 
\bea\label{NZ_prop}
 S^{NZ}_{\pm}(x,y) &=&
  \frac{1}{\sqrt{1+\frac{\rho^2}{x^2}}}
  \frac{1}{\sqrt{1+\frac{\rho^2}{y^2}}} 
  \left\{ S_0(x-y)\left(1+\frac{\rho^2\sigma_\mp\cdot x\sigma_{\pm}\cdot y} 
      {x^2y^2}\right) \right. \nonumber \\
 & & \hspace{0.3cm}  \mbox{}
  -  \frac{\Delta_0(x-y)}{x^2y^2}
     \left(\frac{\rho^2}{\rho^2+x^2}\sigma_\mp\cdot x 
         \sigma_\pm\cdot \gamma
         \sigma_\mp\cdot \Delta \sigma_\pm\cdot y \gpm 
    \right. \nonumber \\
 & & \hspace{2cm}  \mbox{}\left.\left.
  + \frac{\rho^2}{\rho^2+y^2}\sigma_\mp\cdot x \sigma_\pm\cdot 
      \Delta \sigma_\mp\cdot \gamma \sigma_\pm\cdot y \gmp
 \right) \right\}
\eea
Here, $S_0=-\slashed{\Delta}/(2\pi^2\Delta^4)$ denotes the free 
quark propagator. As expected, the full non-zero mode
propagator reduces to the free propagator at short distance. The 
linear mass term in equ. (\ref{prop_expan}) can be written in 
terms of the non-zero mode propagator 
\be
\label{int_SS}
 \sum_{\lb\ne 0}\frac{\Psi_\lb(x)\Psi_{\lb}^+(y)}{\lb^2}=
  \int d^4z S^{NZ}_\pm(x,z) S^{NZ}_\pm(z,y) 
  =-\Delta_\pm(x,y)\gpm -\Delta^M_\pm(x,y) \gmp ,
\ee
where $\Delta_\pm(x,y)$ is the scalar propagator and $\Delta^M_\pm
(x,y)=\langle x|(D^2+\sigma\cdot G/2)^{-1}|y\rangle$ is the 
propagator of a scalar particle with a chromomagnetic moment.
We will not need the explicit form of $\Delta^M_\pm(x,y)$
in what follows. We are now in the position to compute the 
regularized axial current given in equ.~(\ref{ax_reg}). We 
observe that neither the free propagator nor the zero mode 
part will contribute. Expanding the non-zero mode propagator 
and the path ordered exponential in powers of $\epsilon$ we find
\be
\label{ax_cur_inst}
{\rm Tr}\left[\gf\gamma^\mu S(x,x)\right]
 =\pm\frac{2\rho^2x^\mu}{\pi^2(x^2+\rho^2)^3},
\ee
which shows that instantons act as sources and sinks for the 
flavor singlet axial current. We can now compare this result 
to the anomaly relation equ.~(\ref{anom}). The divergence of 
equ.~(\ref{ax_cur_inst}) is given by 
\be 
\partial^\mu A_\mu(x) = \pm
  \frac{2\rho^2}{\pi^2}\frac{4\rho^2-2x^2}{(x^2+\rho^2)^4}.
\ee
The topological charge density in the field of an 
(anti) instanton is 
\be 
\label{qtop_i}
\frac{g^2}{16\pi^2}G^a_{\mu\nu} \tilde{G}^a_{\mu\nu}
 = \pm\frac{12\rho^4}{\pi^2(x^2+\rho^2)^4}.
\ee
We observe that the divergence of the axial current given in 
equ.~(\ref{ax_cur_inst}) does not agree with the topological 
charge density. The reason is that in the field of an instanton 
the second term in the anomaly relation, which is proportional 
to $m\bar{\psi}\gamma_5 \psi$, receives a zero mode contribution 
and is enhanced by a factor $1/m$. In the field of an (anti) 
instanton we find
\be 
\label{mq5q}
2m\bar{\psi}i\gamma_5 \psi = \mp
 \frac{4\rho^2}{\pi^2(x^2+\rho^2)^3}.
\ee
Taking into account both equ.~(\ref{qtop_i}) and (\ref{mq5q})
we find that  the anomaly relation (\ref{anom}) is indeed 
satisfied. 

\section{OZI Violation in Axial-Vector Two-Point Functions}
\label{sec_pia}

\begin{figure}
\begin{center}
\includegraphics[width=13cm,angle=0,clip=true]{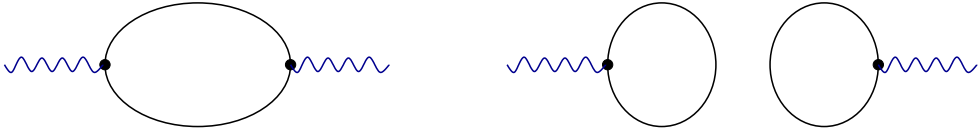}
\end{center}
\caption{\label{fig_vec_cor}
Quark line diagrams that contribute to the
vector and axial-vector two-point function in the
iso-vector and iso-singlet channel. The solid lines
denote quark propagators in a gluonic background field.
The two diagrams show the connected and disconnected
contribution.}
\end{figure}

 In this section we wish to study OZI violation in the axial-vector 
channel due to instantons. We consider the correlation functions 
\be 
\Pi_{\mu\nu}^{j}(x,y) =\langle j_\mu(x)j_\nu(y)\rangle ,
\ee
where $j_\mu$ is one of the currents
\be 
\begin{array}{rcllrcll}
V_\mu^a &=& \bar{\psi}\gamma_\mu\tau^a\psi & (\rho), \hspace{1.5cm}&
V_\mu^0 &=& \bar{\psi}\gamma_\mu\psi & (\omega), \\
A_\mu^a &=& \bar{\psi}\gamma_\mu\gamma_5\tau^a\psi & (a_1), 
\hspace{1.5cm}&
A_\mu^0 &=& \bar{\psi}\gamma_\mu\gamma_5\psi & (f_1), 
\end{array}
\ee
where in the brackets we have indicated the mesons with the 
corresponding quantum numbers. We will work in the chiral limit 
$m_u=m_d\to 0$. The iso-vector correlation functions only 
receive contributions from connected diagrams. The iso-vector
vector ($\rho$) correlation function is 
\be
\label{triplet}
(\Pi_V^3)_{\mu\nu}(x,y) = 2(P^{con}_V)_{\mu\nu}(x,y)
 = -2 \langle {\rm Tr}\left[ 
  \gamma_\mu S(x,y)\gamma_\nu S(y,x)\right]\rangle .
\ee
The iso-singlet correlator receives additional, disconnected, 
contributions, see Fig.~\ref{fig_vec_cor}. 
The iso-singlet 
vector ($\omega$) correlator is given by
\be
\label{singlet}
(\Pi_V^0)_{\mu\nu}(x,y)  = 2(P^{con}_V)_{\mu\nu}(x,y) 
  + 4(P^{dis}_V)_{\mu\nu}(x,y)
\ee
with 
\be
(P^{dis}_V)_{\mu\nu}(x,y) = \langle 
  {\rm Tr}\left[\gamma_\mu S(x,x)\right] 
  {\rm Tr}\left[\gamma_\nu S(y,y)\right] \rangle .
\ee
The axial-vector correlation functions are defined analogously. 
At very short distance the correlation functions are dominated 
by the free quark contribution $\Pi_A^3=\Pi_A^0=\Pi_V^3=\Pi_V^0
\sim 1/x^6$. Perturbative corrections to the connected correlators
are $O(\alpha_s(x)/\pi)$, but perturbative corrections to the 
disconnected correlators are very small, $O((\alpha_s(x)/\pi)^2)$. 
In this section we will compute the instanton contribution 
to the correlation functions. At short distance, it is 
sufficient to consider a single instanton. For the connected 
correlation functions, this calculation was first performed
by Andrei and Gross \cite{Andrei:xg}, see also \cite{Nason:1993ak}.
Disconnected correlation function were first considered in 
\cite{Geshkenbein:vb} and a more recent study can be found
in \cite{Dorokhov:2003kf}.

 In order to make contact with our calculation of the vector 
and axial-vector three-point functions in the next section, we 
briefly review the calculation of Andrei and Gross, and then 
compute the disconnected contribution. Using the expansion in 
powers of the quark mass, equ.~(\ref{prop_expan}), we can write 
\be
\label{vec_dec}
 (P^{con}_V)_\pm^{\mu\nu}(x,y)
  = (P^{con}_V)^{\mu\nu}_{0} + A^{\mu\nu}_{\pm} + B^{\mu\nu}_\pm 
\ee
with 
\bea 
(P_V^{con})^{\mu\nu}_{0} &=&  
  -{\rm Tr}\left[\gamma^{\mu}S_0(x,y)\gamma^{\nu}S_0(y,x)\right],\\
\label{A}
A^{\mu\nu}_\pm  &=&
  -{\rm Tr}\left[\gamma^{\mu}S^{NZ}_\pm(x,y)
                 \gamma^{\nu}S^{NZ}_\pm(y,x)\right]
  - (P^{con}_V)_0^{\mu\nu}(x,y), \\
\label{B}
B^{\mu\nu}_\pm  &=&
  -2 {\rm Tr}\left[\gamma^\mu\Psi_{0\pm}(x)\Psi_{0\pm}^+(y)
                   \gamma^\nu \Delta_{\pm}(y,x)\gpm\right].
\eea
Using the explicit expression for the propagators given in 
the previous section we find
\bea
\label{A_2}
A^{\mu\nu}_{\pm}&=&
  \frac{\rho^2 h_x h_y}{2\pi^4\Delta^4}
  \left\{
    S^{\mu\alpha\nu\beta}
    \Big[\rho^2 h_x h_y(2\Delta_\alpha\Delta_\beta - 
      \Delta^2\delta_{\alpha\beta}) +
      h_y(y_\beta \Delta_\alpha + y_\alpha \Delta_\beta) 
  \right. \nonumber \\
 & & \hspace{1cm}\mbox{} \left. 
   - h_x (x_\beta 
      \Delta_\alpha + x_\alpha \Delta_\beta) \Big]
     \mp 2\epsilon^{\mu\nu\alpha\beta}(h_y\Delta_\alpha y_\beta 
  - h_x   \Delta_\beta x_\alpha)
  \right\}
\eea
and 
\be
\label{B_2}
B^{\mu\nu}_\pm = -\frac{\rho^2}{\pi^4\Delta^2}h_x^2 h_y^2
 \left\{
   (x\cdot y +\rho^2)\delta^{\mu\nu} - (y^\mu x^\nu - x^\mu y^\nu) 
   \mp \epsilon^{\mu\alpha\nu\beta}x_\alpha y_\beta
 \right\}
\ee
with $h_x=1/(x^2+\rho^2)$, $\Delta=x-y$, and $S^{\mu\alpha\nu\beta}
=g^{\mu\alpha}g^{\nu\beta} - g^{\mu\nu}g^{\alpha\beta} + g^{\mu\beta}
g^{\alpha\nu}$. Our result agrees with \cite{Andrei:xg} up to a color 
factor of $3/2$, first noticed in \cite{Dubovikov:bf}, a '-' sign in 
front of the epsilon terms, which cancels after adding instantons and 
anti-instantons, and a '-' sign in front of the 2nd term in $B^{\mu\nu}$. 
This sign is important in order to have a conserved current, but it
does not affect the trace $P^{\mu\mu}$. Summing up the contributions 
from instantons and anti-instantons we obtain
\bea
(P^{con}_V)^{\mu\nu} (x,y)&=&
 2\frac{12 S^{\mu\alpha\nu\beta} \Delta_\alpha 
    \Delta_\beta}{(2\pi^2)^2\Delta^8}
  + \frac{1}{2\pi^4}(h_x h_y)^2\rho^2 
    \left[
     -\frac{2\Sigma^2}{\Delta^4} \Delta^\mu\Delta^\nu
    \right . \nonumber\\
& &     +\frac{2\Sigma\cdot\Delta}{\Delta^4}
  (\Sigma^\mu\Delta^\nu + \Delta^\mu 
       \Sigma^\nu  
   - \Sigma\cdot \Delta g^{\mu\nu})
  \nonumber \\
& & \left.
 + \frac{2}{\Delta^2}  (\Delta^2g^{\mu\nu} - \Delta^\mu\Delta^\nu - 
   \Delta^\mu\Sigma^\nu + \Delta^\nu \Sigma^\mu)
   \right] .
\eea 
This result has to be averaged over the position of the instanton.
We find
\bea
 2a^{\mu\nu} &=& \sum_\pm \int d^4 z A^{\mu\nu}_\pm (x-z,y-z) 
   \nonumber\\
 &=& -\frac{1}{\pi^2} \left[ 
    \frac{\partial^2}{\partial\Delta_\mu\partial\Delta_\nu} 
    G(\Delta^2,\rho) + 2G'(\Delta^2,\rho)g^{\mu\nu} \right], \\
 2b^{\mu\nu} &=& \sum_\pm \int d^4 z B^{\mu\nu}_\pm (x-z,y-z) 
 =\frac{1}{\pi^2}\left[ 
   \frac{\partial^2}{\partial\Delta^2} 
    G(\Delta^2,\rho) + 2G'(\Delta^2,\rho)\right]g^{\mu\nu},
	\nonumber\\
\eea
with 
\be
 \label{G}
 G'(\Delta^2,\rho) = 
  \frac{\partial G(\Delta^2,\rho)}{\partial \Delta^2}
  = \frac{\rho^2}{\Delta^4}
    \left[-\frac{2\rho^2}{\Delta^2}\xi
        \log\frac{1-\xi}{1+\xi}-1 \right]
\ee
and $\xi^2= \Delta^2/(\Delta^2+4\rho^2)$. The final result 
for the single instanton contribution to the connected part 
of the vector current correlation function is
\bea
\delta (P_V^{con})^{\mu\mu}&=& 
      (P_V^{con})^{\mu\mu} - 2 (P^{con}_V)^{\mu\mu}_0
  =\frac{24}{\pi^2}\frac{\rho^4}{\Delta^2} \;
  \frac{\partial}{\partial \Delta^2}
  \left(\frac{\xi}{\Delta^2}\log\frac{1+\xi}{1-\xi}
  \right)\nonumber\\
& \equiv &\frac{24}{\pi^2}\frac{\rho^4}{\Delta^2} F(\Delta,\xi),
\eea
where we defined 
\be
F(\Delta,\xi) = \frac{\partial}{\partial \Delta^2}
\left(\frac{\xi}{\Delta^2}\log\frac{1+\xi}{1-\xi}
\right) .
\ee
The computation of the connected part of the axial-vector 
correlator is very similar. Using equs.~(\ref{A},\ref{B})
we observe that the only difference is the sign in front 
of $B^{\mu\nu}$. We find
\be
\delta (P^{con}_A)^{\mu\mu}= 
  (P^{con}_A)^{\mu\mu} - 2 (P^{con}_A)^{\mu\mu}_{0} 
=-\frac{1}{\pi^2}\left[ 20\Delta^2G'' + 56 G' \right],
\ee
with $G'$ given in equ.~(\ref{G})

\begin{figure}
\begin{center}
\leavevmode
\includegraphics[width=13.8cm,angle=0,clip=true]{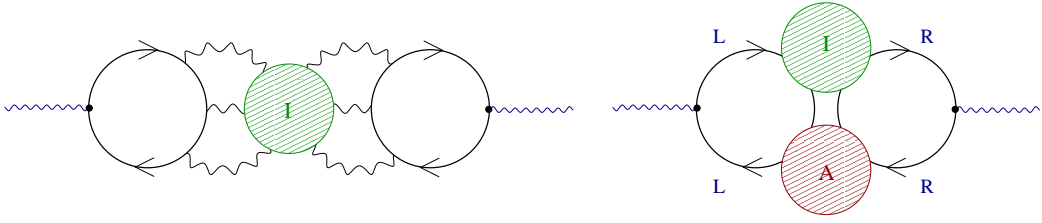}
\end{center}
\caption{\label{fig_f1_inst}
Instanton contributions to the disconnected axial-vector
correlation function. The left panel shows the single-instanton
(non-zero mode) contribution. The right panel shows the
instanton-anti-instanton (fermion zero mode) contribution.}
\end{figure}

\begin{figure}
\begin{center}
\includegraphics[width=8cm,angle=0,clip=true]{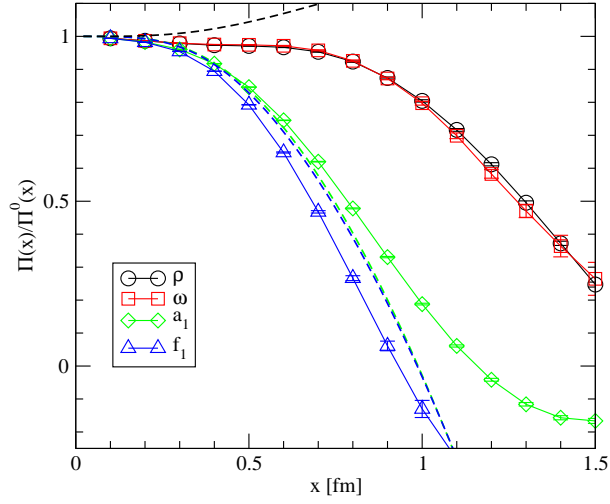}
\end{center}
\caption{\label{fig_vec_cor_num}
correlation functions in the $\rho,\omega,a_1$ and
$f_1$ channel. All correlation functions are normalized
to free field behavior. The data points show results from
a numerical simulation of the random instanton liquid. The
dashed lines show the single instanton approximation.}
\end{figure}

 We now come to the disconnected part, see Fig.~\ref{fig_f1_inst}. 
In the vector channel the single instanton contribution to 
the disconnected correlator vanishes \cite{Geshkenbein:vb}. 
In the axial-vector channel we can use the result for 
${\rm Tr}[\gf\gamma^\mu S(x,x)]$ derived in the previous 
section. The correlation function is
\be
(P^{dis}_A)^{\mu\nu}(x,y)=\frac{4\rho^4(x-z)^\mu(y-z)^\nu}
 {\pi^4((x-z)^2+\rho^2)^3((y-z)^2+\rho^2)^3} .
\ee
Summing over instantons and anti-instantons and integrating 
over the center of the instanton gives
\be
(P^{dis}_A)^{\mu\nu} = 2\frac{\rho^4}{2\pi^2}
 \frac{\partial^2}{\partial \Delta_\mu\partial \Delta_\nu}
   F(\Delta,\xi)
\ee
and 
\be
 (P^{dis}_A)^{\mu\mu} = \frac{4\rho^4}{\pi^2}
 \left\{2\frac{d}{d\Delta^2} + 
       \Delta^2\left(\frac{d}{d\Delta^2}\right)^2\right\}
        F(\Delta,\xi). 
\ee
We can now summarize the results in the vector singlet 
($\omega$) and triplet ($\rho$), as well as axial-vector
singlet ($f_1$) and triplet ($a_1$) channel. The result 
in the $\rho$ and $\omega$ channel is 
\be
(\Pi^3_V)^{\mu\mu} = (\Pi^0_V)^{\mu\mu}=
 -\frac{12}{\pi^4\Delta^6} +
  2\int d\rho n(\rho) \frac{24}{\pi^2}\frac{\rho^4}{\Delta^2}
  F(\Delta,\xi) .
\ee
In the $a_1,f_1$ channel we have
\bea
(\Pi^3_A)^{\mu\mu} &=& 
   -\frac{12}{\pi^4\Delta^6} +
   \int d\rho n(\rho) \left[ - \frac{2}{\pi^2}
     \left(20\Delta^2G'' + 56 G'\right)  \right], \\
(\Pi^0_A)^{\mu\mu} &=& 
 -\frac{12}{\pi^4\Delta^6} 
 +
 \int d\rho n(\rho) 
   \left[ 
	- \frac{2}{\pi^2}\left(20\Delta^2G'' + 56 G'\right)
   \right .
 \nonumber\\
& & \hspace{2.5cm} 
  \left .
	+ \frac{16\rho^4}{\pi^2} \left( 2F'+\Delta^2F''\right)  
  \right].
\eea
In order to obtain a numerical estimate of the instanton 
contribution we use a very simple model for the instanton
size distribution, $n(\rho)=n_0\delta(\rho-\bar{\rho})$,
with $\bar{\rho}=0.3$ fm and $n_0=0.5\, {\rm fm}^{-4}$. The 
results are shown in Fig.~\ref{fig_vec_cor_num}. 

 We observe that the OZI rule violating difference 
between the singlet and triplet axial-vector correlation 
functions is very small and repulsive. We can also see
this by studying the short distance behavior of the 
correlation functions. The non-singlet correlators 
satisfy 
\bea
(\Pi_V^3)^{\mu\mu} &=& (\Pi_V^3)^{\mu\mu}_{0} 
  \left\{ 1 + \frac{\pi^2x^4}{3} 
  \int d\rho n(\rho) + \ldots \right\}, \\
(\Pi_A^3)^{\mu\mu} &=& (\Pi_A^3)^{\mu\mu}_{0} 
  \left\{ 1 - \pi^2x^4
  \int d\rho n(\rho) + \ldots \right\}.
\eea
As explained by Dubovikov and Smilga, this result 
can be understood in terms of the contribution of the 
dimension $d=4$ operators $\langle g^2G^2\rangle$ and 
$\langle m\bar{q}q\rangle$ in the operator product 
expansion (OPE). The OZI violating contribution 
\be
(\Pi^{OZI}_A)^{\mu\mu} =  
(\Pi^0_A)^{\mu\mu} - (\Pi^3_A)^{\mu\mu}
 =  -(\Pi^3_A)^{\mu\mu}_0
   \left(\frac{4\pi^2}{45}\frac{x^6}{\rho^2}\right)
  \int d\rho n(\rho)
\ee
is of $O(x^6)$ and not singular at short distance. Our 
results show that it remains small and repulsive even if 
$x>\rho$. We also note that the sign of the OZI-violating 
term at short distance is model independent. The quark 
propagator in euclidean space satisfies the Weingarten 
relation
\be
\label{wein}
S(x,y)^\dagger  = \gamma_5 S(y,x) \gamma_5 .
\ee
This relation implies that ${\rm Tr}[S(x,x)\gamma_\mu\gamma_5]$ 
is purely real. As a consequence we have
\be 
\label{ozi_ineq}
\lim_{x\to y}  \Big\{ {\rm Tr}[S(x,x)\gamma_\mu\gamma_5]
  {\rm Tr}[S(y,y)\gamma_\mu\gamma_5] \Big\} > 0.
\ee
Since the path integral measure in euclidean space is 
positive this inequality translates into an inequality
for the correlation functions. In our convention the 
trace of the free correlation function is negative, and
equ.~(\ref{ozi_ineq}) implies that the interaction is 
repulsive at short distance. The result is in agreement 
with the single instanton calculation. 

 We can also study higher order corrections to the single 
instanton result. The two-instanton (anti-instanton) 
contributions is of the same form as the one-instanton
result. An interesting contribution arises from 
instanton-anti-instanton pairs, see Fig.~\ref{fig_f1_inst}. 
This effect was studied in \cite{Schafer:1994nv}. It was 
shown that the instanton-anti-instanton contribution to 
the disconnected meson channels can be described in terms 
of an effective lagrangian 
\be 
{\cal L} = \frac{2G}{N_c^2}(\bar{\psi}\gamma_\mu\gamma_5\psi)^2
\ee
with 
\be 
G = \int d\rho_1d\rho_2 (2\pi\rho_1)^2(2\pi\rho_2)^2
 \frac{n(\rho_1,\rho_2)}{8T_{IA}^2}
\ee
where $n(\rho_1,\rho_2)$ is the tunneling rate for an 
instanton-anti-instanton pair and $T_{IA}$ is the 
matrix element of the Dirac between the two (approximate)
zero modes. We note that this interaction is also repulsive,
and that there is no contribution to the $\omega$ channel. 
 
\begin{figure}
\begin{center}
\includegraphics[width=8cm,angle=0,clip=true]{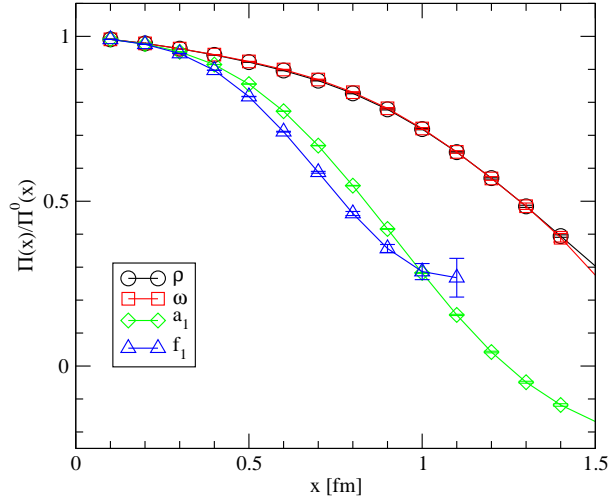}
\end{center}
\caption{\label{fig_vec_cor_unq}
Correlation functions in the $\rho,\omega,a_1$ and
$f_1$ channel. All correlation functions are normalized
to free field behavior. The data points show results from
unquenched simulations of the instanton liquid model.}
\end{figure}

 Numerical results for the vector meson correlation functions 
are shown in Figs.~\ref{fig_vec_cor_num} and \ref{fig_vec_cor_unq}.
The correlation functions are obtained from Monte Carlo 
simulations of the instanton liquid as described in 
\cite{Schafer:1995pz,Schafer:1995uz}. We observe that OZI 
violation in the vector channel is extremely small, both in 
quenched and unquenched simulations. The OZI violating 
contribution to the $f_1$ channel is repulsive. In quenched 
simulations this contribution becomes sizable at large distance. 
Most likely this is due to mixing with an $\eta'$ ghost pole. 
We observe that the effect disappears in unquenched simulations. 
The pion contribution to the $a_1$ correlator is of course
present in both quenched and unquenched simulations.

 Experimentally we know that the $\rho$ and $\omega$, as 
well as the $a_1$ and $f_1$ meson, are indeed almost 
degenerate. Both iso-singlet states are slightly heavier
than their iso-vector partners. To the best of our knowledge
there has been only one attempt to measure OZI violating 
correlation functions in the vector and axial-vector channel 
on the lattice, see \cite{Isgur:2000ts}. Isgur and Thacker
concluded that OZI violation in both channels was too 
small to be reliably measurable in their simulation.

\section{Axial Vector Coupling of a Quark}
\label{sec_gaq}

 In this section we wish to study the iso-vector and iso-singlet 
axial coupling of a single quark. Our purpose is twofold. One
reason is that the calculation of the axial-vector three-point 
function involving a single quark is much simpler than that of 
the nucleon, and that it is closely connected to the axial-vector 
two-point function studied in the previous section. The second, 
more important, reason is the success of the constituent quark 
model in describing many properties of the nucleon. It is clear 
that constituent quarks have an intrinsic structure, and that the 
axial decay constant of a constituent quark need not be close to 
one. Indeed, Weinberg argued that the axial coupling of a quark 
is $(g_{A}^3)_Q \simeq 0.8$ \cite{Weinberg:gf}. Using this value 
of $(g_{A}^3)_Q$ together with the naive $SU(6)$ wave function of 
the nucleon gives the nucleon axial coupling $g_A^3=0.8\cdot 5/3\simeq 
1.3$, which is a significant improvement over the naive quark model
result $5/3$. It is interesting to study whether, in a similar fashion, 
the suppression of the flavor singlet axial charge takes place on the 
level of a constituent quark. 

\begin{figure}
\begin{center}
\includegraphics[width=6.5cm,angle=0,clip=true]{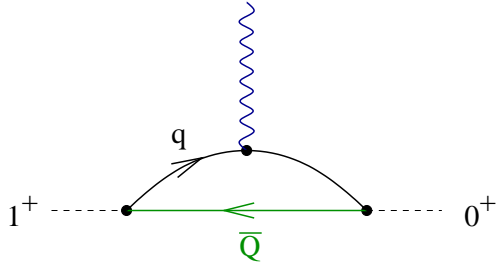}
\end{center}
\caption{\label{fig_hl}
Physical interpretation of the gauge invariant
axial-vector three-point function of a quark in
terms of a weak light-quark transition in a heavy-light
$\bar{Q}q$ meson.}
\end{figure}

 In order to address this question we study three-point functions
involving both singlet and triplet vector and axial-vector 
currents. The vector three-point function is
\be
\label{VQQ}
(\Pi^a_{VQQ})^{\alpha\beta}_\mu(x,z,y) = 
\langle q^\alpha(x)V^a_\mu(z)\bar{q}^\beta(y)\rangle .
\ee
The axial-vector function $(\Pi^a_{AQQ})$ is defined analogously.
We should note that equ.~(\ref{VQQ}) is not gauge invariant.
We can define a gauge invariant correlation function by 
including a gauge string. The gauge string can be interpreted
as the propagator of a heavy anti-quark, see Fig.~\ref{fig_hl}. 
This implies that the gauge invariant quark axial-vector 
three-point function is related to light quark weak transitions 
in heavy-light mesons. 
 
  The spectral representation of vector and axial-vector 
three-point functions is studied in some detail in the 
appendix \ref{app_spect_rep}. The main result is that in the limit that 
$y_4\gg z_4 \gg x_4$ the ratio 
\be
  \frac{{\rm Tr}[(\Pi^a_{AQQ})_3 \gamma_5\gamma_3]}
                {{\rm Tr}[(\Pi^a_{VQQ})_4 \gamma_4]}
 \to \frac{g_A}{g_V}
\ee
tends to the ratio of axial-vector and vector coupling
constants, $g_A/g_V$. We therefore define the following 
Dirac traces
\bea
\label{correl_def}
(\Pi^{a}_{VQQ})^{\mu\nu}(x,z,y)
  &=& {\rm Tr}[(\Pi^{a}_{VQQ})^{\mu} \gamma^\nu], \\
(\Pi^{a}_{AQQ})^{\mu\nu}(x,z,y)
  &=& {\rm Tr}[(\Pi^{a}_{AQQ})^{\mu} \gf \gamma^\nu] .
\label{correl_ax_def} 
\eea

\begin{figure}
\begin{center}
\leavevmode
\includegraphics[width=12cm,angle=0,clip=true]{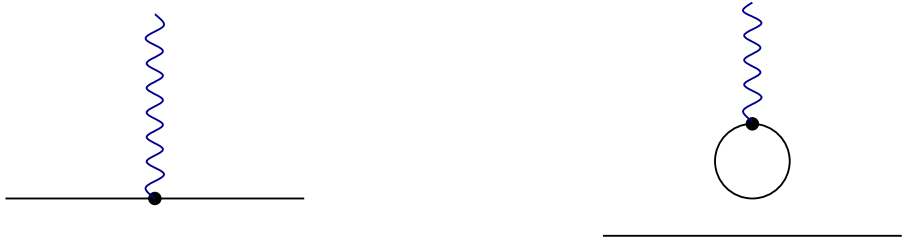}
\end{center}
\caption{\label{fig_VQQ}
Quark line diagrams that contribute to the
vector and axial-vector three-point function of a
constituent quark. The solid lines denote quark
propagators in a gluonic background field.
The two diagrams show the connected and disconnected
contribution.}
\end{figure}

As in the case of the two-point function the iso-triplet 
correlator only receives quark-line connected contributions,
whereas the iso-singlet correlation function has a disconnected 
contribution, see Fig.~\ref{fig_VQQ}. We find
\be 
\label{Pi_tripl}
(\Pi^3_{VQQ})^{\mu\nu} (x,z,y) =  (P^{con}_{VQQ})^{\mu\nu}(x,z,y) 
 = \langle {\rm Tr} [S(x,z)\gamma^\mu  S(z,y)\gamma^\nu] \rangle 
\ee
and 
\be
\label{Pi_singl}
(\Pi^0_{VQQ})^{\mu\nu}(x,z,y) = (P^{con}_{VQQ})^{\mu\nu}(x,z,y)
   -2(P^{dis}_{VQQ})^{\mu\nu}(x,z,y)
\ee
with
\be
(P^{dis}_{VQQ})^{\mu\nu}(x,z,y) = 
 \langle {\rm Tr} [S(x,y)\gamma^\mu]{\rm Tr}[S(z,z)\gamma^\nu]
 \rangle ,
\ee
as well as the analogous result for the axial-vector correlator.

 In the following we compute the single instanton contribution
to these correlation functions. We begin with the connected 
part. We again write the propagator in the field of the 
instanton as $S_{ZM}+S_{NZ}+S_m$ where $S_{ZM}$ is the 
zero-mode term, $S_{NZ}$ is the non-zero mode term, and
$S_m$ is the mass correction. In the three-point correlation
function we get contribution of the type $S_{NZ}S_{NZ}$, 
$S_{ZM}S_{m}$ and $S_{m}S_{ZM}$
\bea
\label{pi_3pt_z/nz}
 (P^{con}_{A/VQQ})^{\mu\nu} &=& 
  P_{NZNZ}^{\mu\nu} + c_{A/V}
   \left(P^{\mu\nu}_{ZMm} + P^{\mu\nu}_{mZM}
   \right)\nonumber\\
  &=& {\rm Tr} [S^{NZ}(x,z)\gamma^{\mu}S^{NZ}(z,y)\gamma^{\nu}]
\nonumber\\ 
& & +
c_{A/V}{\rm Tr} [-\Psi_0(x)\Psi_0^{+}(z)\gamma^{\mu}
(-\Delta_{\pm}(z,y)\gamma_{\pm}\gamma^{\nu}]\nonumber\\
& & +
c_{A/V }{\rm Tr} [(-\Delta_{\pm}(x,z)\gamma_{\pm})\gamma^{\mu}
  (-\Psi_0(z)\Psi_0^{+}(y))\gamma^{\nu}] ,
\eea
where the only difference between the vector and axial-vector 
case is the sign of 'ZMm' and 'mZM' terms. We have  $c_{A/V}
=\pm 1$ for vector (axial vector) current insertions. The 
detailed evaluation of the traces is quite tedious and we 
relegate the results to the appendix \ref{sec_app5}.

\begin{figure}
\begin{center}
\includegraphics[width=12cm,angle=0,clip=true]{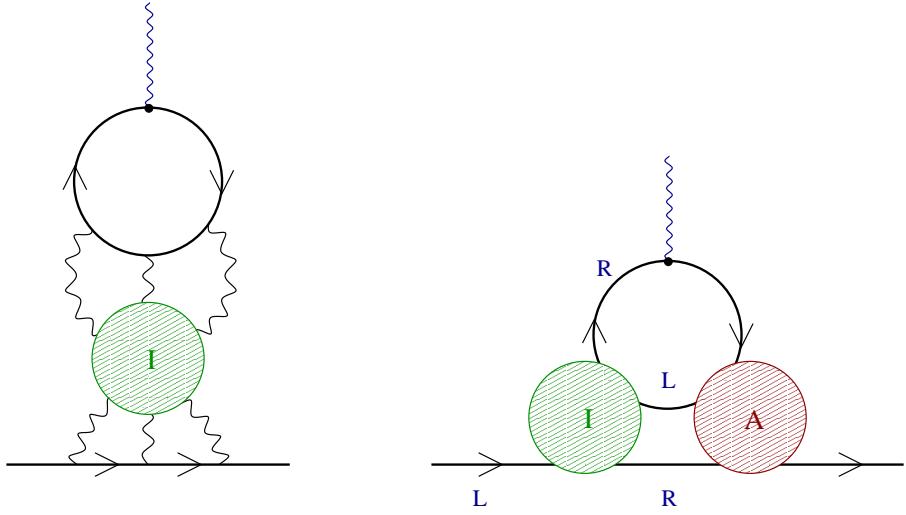}
\end{center}
\caption{\label{fig_gaq_inst}
Instanton contributions to the disconnected axial-vector
three-point correlation function of a quark. The left panel
shows the single-instanton (non-zero mode) contribution. The
right panel shows the instanton-anti-instanton (fermion zero
mode) contribution.}
\end{figure}

 Our main goal is the calculation of the disconnected 
correlation function, which is related to OZI violation. 
In the single instanton approximation only the axial-vector
correlator receives a non-zero disconnected contribution
\be
(P^{dis}_{AQQ})^{\mu\nu}={\rm Tr}[S(x,y)\gf\gamma^\nu] 
  {\rm Tr}[S(z,z)\gf \gamma^\mu], 
\ee
see Fig.~\ref{fig_gaq_inst}. We observe that the second trace 
is the axial-vector current in the field of an instanton, see 
equ.~(\ref{ax_cur_inst}). As for the first trace, it is easy 
to see that neither the zero mode part of the propagator nor 
the part of $S_{NZ}$ proportional to the free propagator can
contribute. A straight-forward computation gives
\be
{\rm Tr} [S(x,y)\gf\gamma^\nu]=\mp\frac{\rho^2}{\pi^2(x-y)^2x^2y^2}
 \frac{x^\alpha y^\beta (x-y)^\sigma}
 {\sqrt{(1+\frac{\rho^2}{x^2})(1+\frac{\rho^2}{y^2})}}
 \left(\frac{S^{\alpha\nu\sigma\beta}}{\rho^2+x^2} -
       \frac{S^{\alpha\sigma\nu\beta}}{\rho^2+y^2} \right).
\ee
Combined with equ.~(\ref{ax_cur_inst}) we obtain
\bea
(P^{dis}_{AQQ})^{\mu\nu}&=& - 
  \frac{2\rho^4 z^\mu x^\alpha y^\beta (x-y)^\sigma}
  {\pi^4(x-y)^2x^2y^2(z^2 + \rho^2)^3}
   \frac{1}{\sqrt{(1+\frac{\rho^2}{x^2})(1+\frac{\rho^2}{y^2})}}
	\nonumber\\
 & & \hspace{.3cm}\x
  \left(\frac{S^{\alpha\nu\sigma\beta}}{\rho^2+x^2} -
        \frac{S^{\alpha\sigma\nu\beta}}{\rho^2+y^2}\right),
\eea
which has to be multiplied by a factor 2 in order to take 
into account both instantons and anti-instantons. 

\begin{figure}
\begin{center}
\leavevmode
\includegraphics[width=8cm,angle=0,clip=true]{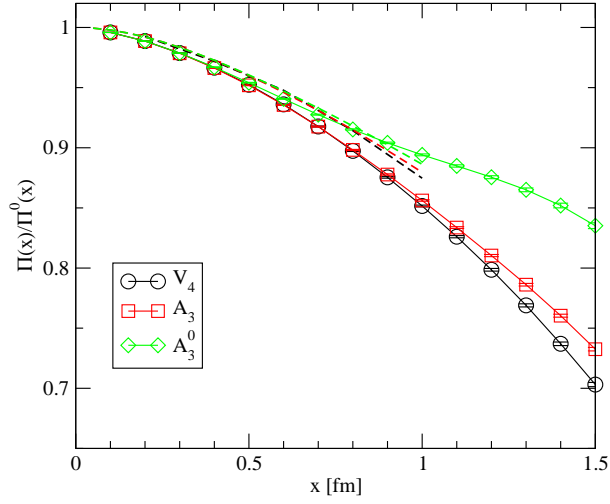}
\end{center}
\caption{\label{fig_qu_va}
Axial and vector three-point functions of a quark as a
function of the separation between the two quark sources.
The correlation functions are normalized to free field
behavior. The data points show results from numerical
simulations of the instanton liquid and the dashed lines
show the single instanton approximation.}
\end{figure}

  Results for the vector three-point function 
$(\Pi_{VQQ}^3)^{44}(\tau,\tau/2,0)$ 
and axial-vector three-point function 
$(\Pi_{AQQ}^{0,3})^{33} (\tau,\tau/2,0)$
 are shown in Fig.~\ref{fig_qu_va}. We observe 
that the vector and axial-vector correlation functions are very 
close to one another. We also note that the disconnected
contribution adds to the connected part of the axial-vector 
three-point function. This can be understood from the short 
distance behavior of the correlation function. The disconnected 
part of the gauge invariant three-point function satisfies
\be 
\lim_{y,z\to x} \Big\{ (\Pi_{AQQ}^{dis})^{33}(x,z,y) \Big\}
 = \lim_{x\to z} \Big\{ {\rm Tr}[S(x,x)\gamma_3\gamma_5]
  {\rm Tr}[S(z,z)\gamma_3\gamma_5] \Big\} >0.
\ee
\begin{figure}
\begin{center}
\leavevmode
\includegraphics[width=8cm,angle=0,clip=true]{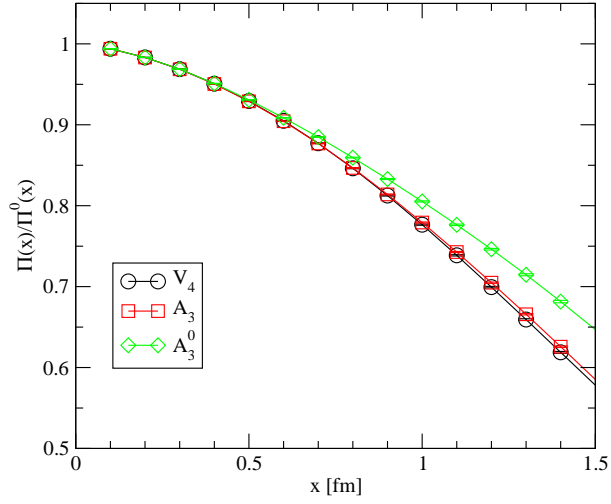}
\end{center}
\caption{\label{fig_qu_va_unq}
Axial and vector three-point functions of a quark 
as a function of the separation between the two quark
sources. The data points show results from an unquenched
instanton simulation.}
\end{figure}
This expression is exactly equal to the short distance term in the 
disconnected $f_1$ meson correlation function. The short distance 
behavior of the connected three-point function, on the other hand, 
is opposite in sign to the two-point function. This is related to the 
fact that the two-point function involves one propagator in the forward 
direction and one in the backward direction, whereas the three-point 
function involves two forward propagating quarks. A similar connection 
between the interaction in the $f_1$ meson channel and the flavor 
singlet coupling of a constituent quark was found in a Nambu-Jona-Lasinio 
model \cite{Vogl:1991qt,Steininger:ed}. It was observed, in particular, 
that an attractive coupling in the $f_1$ channel is needed in 
order to suppress the flavor singlet $(g_A^0)_Q$.   

\begin{figure}
\begin{center}
\leavevmode
\includegraphics[width=8cm,angle=0,clip=true]{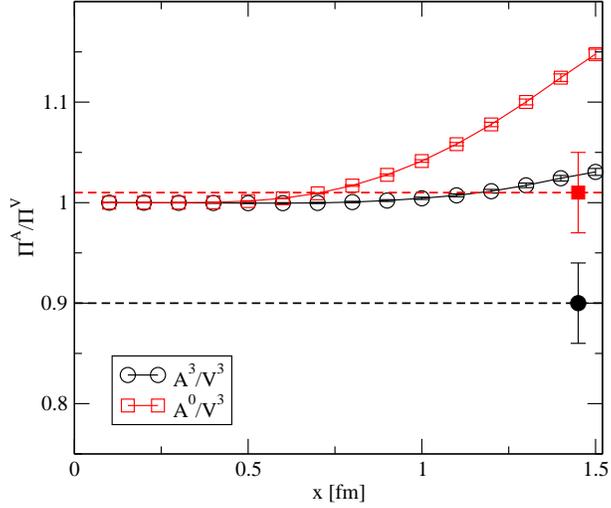}
\end{center}
\caption{\label{fig_qu_ga}
Ratio of axial-vector to vector correlation functions
of a constituent quark calculated in the instanton
liquid model. The open points show point-to-point
correlation functions while the solid point is the
zero momentum (point-to-plane) limit. The figure
shows the iso-vector and iso-singlet correlation functions.}
\end{figure}

 The same general arguments apply to the short distance contribution 
from instanton-anti-instanton pairs. At long distance, on the other 
hand, we expect that IA pairs reduce the flavor singlet axial current 
correlation function. The idea can be understood from Fig.~\ref{fig_gaq_inst}, 
see \cite{Dorokhov:ym,Dorokhov:2001pz,Kochelev:1997ux}. In an IA 
transition a left-handed valence up quark emits a right handed 
down quark which acts to shield its axial charge. We have studied 
this problem numerically, see Figs.~\ref{fig_qu_va}-\ref{fig_qu_ga}.
We find that in quenched simulations the flavor singlet axial
three-point function is significantly enhanced. This effect is 
analogous to what we observed in the $f_1$ channel and disappears
in unquenched simulations. We have also studied the axial three-point 
function at zero three-momentum $\vec{q}=0$. This correlation function 
is directly related to the coupling constant, see App.~\ref{app_spect_rep}. 
We find that the iso-vector coupling is smaller than one, $(g_A^3)_Q
\simeq 0.9$, in agreement with Weinberg's idea. The flavor singlet 
coupling, on the other hand, is close to one. We observe no suppression of 
the singlet charge of a constituent quark. We have also checked 
that this result remains unchanged in unquenched simulations. 

\section{Axial Structure of the Nucleon}
\label{sec_gan}
 
  In this section we shall study the axial charge of the 
nucleon in the instanton model. We consider the same
correlation functions as in the previous section, but 
with the quark field replaced by a nucleon current. The
vector three-point function is given by
\be
(\Pi^a_{VNN})^{\alpha\beta}_\mu(x,y) = 
\langle \eta^\alpha(0)V^a_\mu(y)\bar{\eta}^\beta(x)\rangle .
\ee
Here, $\eta^\alpha$ is a current with the quantum numbers
of the nucleon. Three-quark currents with the nucleon
quantum numbers were introduced by Ioffe \cite{Ioffe:kw}.
He showed that there are two independent currents with 
no derivatives and the minimum number of quark fields
that have positive parity and spin $1/2$. In the case 
of the proton, the two currents are
\bea
\label{ioffe}
\eta_1 = \epsilon_{abc} (u^a C\gamma_\mu u^b) 
            \gamma_5 \gamma_\mu d^c, \hspace{1cm}
\eta_2 = \epsilon_{abc} (u^a C\sigma_{\mu\nu} u^b) 
            \gamma_5 \sigma_{\mu\nu} d^c .
\eea
It is sometimes useful to rewrite these currents in terms of scalar
and pseudo-scalar diquark currents. We find
\bea
\label{ioffe_ps}
\eta_{1} &=& 2 \left\{ \epsilon_{abc} (u^a C d^b)\gamma_5 u^c 
  - \epsilon_{abc} (u^a C\gamma_5 d^b) u^c \right\}, \\
\eta_{2} &=& 4 \left\{ \epsilon_{abc} (u^a C d^b)\gamma_5 u^c 
  + \epsilon_{abc} (u^a C\gamma_5 d^b) u^c \right\}.
\eea
Instantons induce a strongly attractive interaction in the scalar 
diquark channel $\epsilon^{abc}(u^bC\gamma_5d^c)$ 
\cite{Schafer:1993ra,Rapp:1997zu}. As a consequence, the nucleon 
mainly couples to the scalar diquark component of the Ioffe currents 
$\eta_{1,2}$. This phenomenon was also observed on the lattice 
\cite{Leinweber:1994nm}. This result is suggestive of a model 
of the spin structure that is quite different from the naive 
quark model. In this picture the nucleon consists of a tightly 
bound scalar-isoscalar diquark, loosely coupled to the third 
quark \cite{Anselmino:1992vg}. The quark-diquark model suggests 
that the spin and isospin of the nucleon are mostly carried by a 
single constituent quark, and that $g_A^N\simeq g_A^Q$.

\begin{figure}
\begin{center}
\includegraphics[width=8cm,angle=0,clip=true]{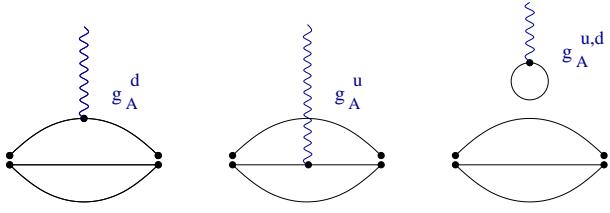}
\end{center}
\caption{\label{fig_VNN}
Quark line diagrams that contribute to the
axial-vector three-point function of the proton.
The solid lines denote quark propagators in a gluonic
background field. The lines are connected in the same
way that the Dirac indices of the propagators are
contracted. The iso-vector and iso-singlet correlation functions
correspond to $g_A^u=-g_A^d=1$ and $g_A^u=g_A^d=1$,
respectively. The disconnected diagram only contributes
to the iso-scalar three-point function.}
\end{figure}
 Nucleon correlation functions are defined by $\Pi^N_{\alpha\beta}
(x) = \langle \eta_\alpha(0)\bar\eta_\beta(x) \rangle$, where 
$\alpha,\beta$ are Dirac indices. The correlation function of the 
first Ioffe current is
\be
\label{nn_cor}
 \Pi_{\alpha\beta}(x)=2\epsilon_{abc}\epsilon_{a'b'c'}
    \,\langle \left( \gamma_\mu\gamma_5S_d^{cc'}(0,x)
       \gamma_\nu\gamma_5 \right)_{\alpha\beta}\,
 {\rm Tr}\left[
     \gamma_\mu S_u^{aa'}(0,x)\gamma_\nu C(S_u^{bb'}(0,x))^TC
        \right]\rangle .
\ee
The vector and axial-vector three-point functions can be 
constructed in terms of vector and axial-vector insertions
into the quark propagator, 
\bea
\label{SV}
(\Gamma^V_\mu)_f^{aa'}(x,y) &=& 
          S_f^{ab}(0,y)\gamma_\mu S_f^{ba'}(y,x), \\
\label{SA}
(\Gamma^A_\mu)_f^{aa'}(x,y) &=& 
         S_f^{ab}(0,y)\gamma_\mu\gamma_5 S_f^{ba'}(y,x). 
\eea
The three-point function is given by all possible 
substitutions of equ.~(\ref{SV}) and (\ref{SA}) into
the two-point function. We have 
\bea
&\hspace{-4cm}
(\Pi^a_{VNN})^{\alpha\beta}_\mu(x,y) =
 2\epsilon_{abc}\epsilon_{a'b'c'}\x\hspace{-3.2cm}   & \nonumber \\
  \x\langle \Bigg\{ & 
    g_V^d \left( \gamma_\rho\gamma_5 (\Gamma^V_\mu)_d^{cc'}(x,y)
       \gamma_\sigma\gamma_5 \right)_{\alpha\beta} & 
 {\rm Tr}\left[
     \gamma_\rho S_u^{aa'}(0,x)\gamma_\sigma C(S_u^{bb'}(0,x))^TC
        \right]  \nonumber  \\
  + & 2g_V^u  \left( \gamma_\rho\gamma_5 S_d^{cc'}(0,x)
       \gamma_\sigma\gamma_5 \right)_{\alpha\beta} & 
 {\rm Tr}\left[
     \gamma_\rho (\Gamma^V_\mu)_u^{aa'}(x,y)\gamma_\sigma C(S_u^{bb'}(0,x))^TC
        \right] \nonumber  \\
 - &  \left( \gamma_\rho\gamma_5S_d^{cc'}(0,x)
       \gamma_\sigma\gamma_5 \right)_{\alpha\beta} & 
 {\rm Tr}\left[
     \gamma_\rho S_u^{aa'}(0,x)\gamma_\sigma C(S_u^{bb'}(0,x))^TC
        \right]  \nonumber   \\
 &  &\label{VNN_con} 
 \hspace{-5.25cm}
 \x
 {\rm Tr}\left[g_V^u\gamma_\mu S_u^{dd}(y,y)
              +g_V^d\gamma_\mu S_d^{dd}(y,y)\right]
    \Bigg\}\rangle ,
\eea
where the first term is the vector insertion 
into the $d$ quark propagator in the proton, the second
term is the insertion into the $uu$ diquark, and the third 
term is the disconnected contribution, see Fig.~\ref{fig_VNN}. 
The vector charges of the quarks are denoted by $g_V^f$. In 
the case of the iso-vector three-point function we have $g_V^u=1$,
$g_V^d=-1$ and in the iso-scalar case $g_V^u=g_V^d=1$.

\begin{figure}
\begin{center}
\leavevmode
\includegraphics[width=8cm,angle=0,clip=true]{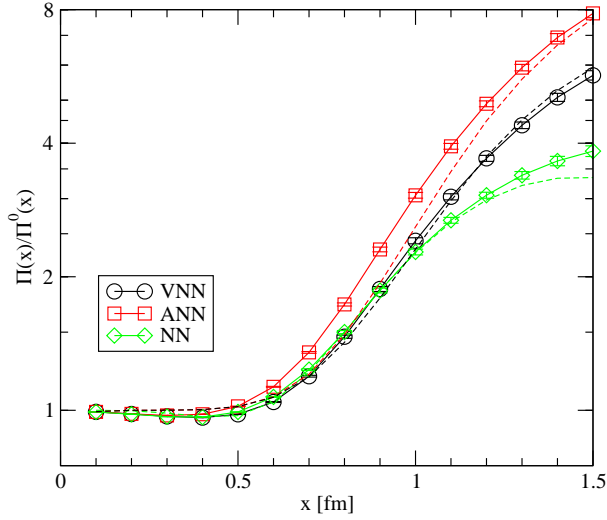}
\end{center}
\caption{\label{fig_nn_3pt}
Vector, axial-vector three-point functions
of the nucleon and nucleon two-point function calculated
in the instanton liquid model. All correlation functions
are normalized to free field behavior. The results are
compared to a simple pole fit of the type discussed
in the appendix.}
\end{figure}

\begin{figure}
\begin{center}
\leavevmode
\includegraphics[width=8cm,angle=0,clip=true]{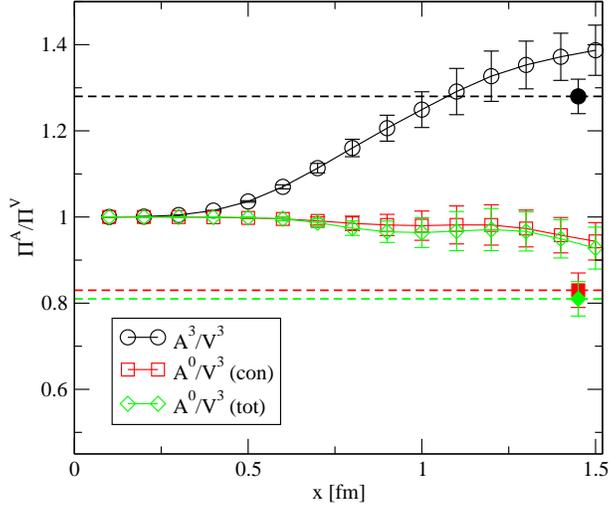}
\end{center}
\caption{\label{fig_gan}
Ratio of axial-vector to vector correlation
functions of the nucleon calculated in the instanton
liquid model. The open points show point-to-point
correlation functions while the solid point is the
zero momentum (point-to-plane) limit. The figure
shows the iso-vector, connected iso-singlet, and
full iso-singlet axial-vector correlation functions.}
\end{figure}

 Vector and axial-vector three-point functions of the nucleon
are shown in Figs.~\ref{fig_nn_3pt} and \ref{fig_gan}. In order to 
verify that the correlation functions are dominated by the nucleon 
pole contribution we have compared our results to the spectral 
representation discussed in the appendix \ref{app_spect_rep},
 see Fig.~\ref{fig_nn_3pt}.
The nucleon coupling constant was determined from the nucleon
two-point function. The figure shows that we can describe 
the three-point functions using the phenomenological values
of the vector and axial-vector coupling constants. We have 
also checked that the ratio of axial-vector and vector 
current three-point functions is independent of the nucleon
interpolating field for $x>1$ fm. The only exception is 
a pure pseudo-scalar diquark current, which has essentially 
no overlap with the nucleon wave function. 

 The main result is that the iso-vector axial-vector correlation
function is larger than the vector correlator. The corresponding 
ratio is shown in Fig.~\ref{fig_gan}, together with the ratio 
of $\vec{q}=0$ correlation functions. We find that the iso-vector 
axial coupling constant is $g_A^3=1.28$, in good agreement with 
the experimental value. We also observe that the ratio of point-to-point
correlation functions is larger than this value. As explained in 
the appendix, this shows that the axial radius of the nucleon is 
smaller than the vector radius. Taking into account only the connected 
part of the correlation function we find a singlet coupling $g_A^0=
0.79$. The disconnected part is very small, $g_A^0(dis)=-(0.02
\pm 0.02)$. Assuming that $\Delta s\simeq \Delta u(dis)=
\Delta d (dis)$ this implies that the OZI violating difference 
$g_A^8-g_A^0$ is small. This result does not change in going
from the quenched approximation to full QCD.

\begin{figure}
\begin{center}
\leavevmode
\includegraphics[width=8cm,angle=0,clip=true]{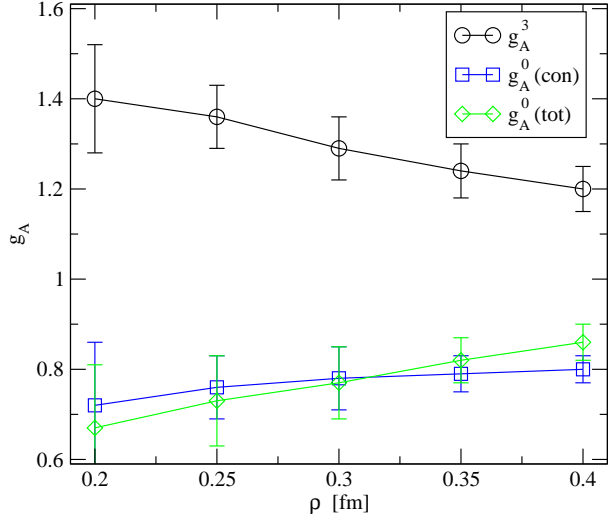}
\end{center}
\caption{\label{fig_gan_rho}
Axial coupling constants of the nucleon as a function of
the instanton size $\rho$ with the instanton density
fixed at $(N/V)=1\,{\rm fm}^{-4}$. We show the iso-vector,
connected iso-singlet, and full iso-singlet axial coupling
constant.}
\end{figure}

 We have also studied the dependence of the results on the 
average instanton size, see Fig.~\ref{fig_gan_rho}. We observe
that there is a slight decrease in the iso-singlet coupling
and a small increase in the iso-vector coupling as the instanton
size is decreased. What is surprising is that the disconnected
term changes sign between $\rho= 0.3$ fm and $\rho=0.35$ fm.
The small value $g_A^0(dis)=-(0.02\pm 0.02)$ obtained above 
is related to the fact that the phenomenological value of 
the instanton size is close to the value where $g_A^0(dis)$ 
changes sign. However, even for $\rho$ as small as 0.2 fm
the disconnected contribution to the axial coupling $g_A^0(dis)
=-(0.05\pm 0.02)$ is smaller in magnitude than phenomenology 
requires. 

\section{Conclusions}
\label{sec_sum}

 The main issue raised by the EMC measurement of the
flavor singlet axial coupling is not so much why $g_A^0$ 
is much smaller than one -  except for the naive quark model
there is no particular reason to expect $g_A^0$ to be 
close to one - but why the OZI violating observable $g_A^0
-g_A^8$ is large. Motivated by this question we have studied 
the contribution of instantons to OZI violation in the axial-vector 
channel. We considered the $f_1-a_1$ meson splitting, the flavor 
singlet and triplet axial coupling of a constituent quark, and the 
axial coupling constant of the nucleon. We found that instantons 
provide a short distance contribution which is repulsive in the 
$f_1$ meson channel and adds to the gauge invariant flavor singlet 
three-point function of a constituent quark. We showed that 
the sign of this term is fixed by positivity arguments. 

 We computed the axial coupling constants of the constituent 
quark and the nucleon using numerical simulations of the 
instanton liquid. We find that the iso-vector axial coupling 
constant of a constituent quark is $(g_A^3)_Q=0.9$ and that of 
a nucleon is $g_A^3=1.28$, in good agreement with experiment. 
The result is also in qualitative agreement with the 
constituent quark model relation $g_A^3=5/3\cdot (g_A^3)_Q$.
The flavor singlet coupling of quark is close to one, while 
that of a nucleon is suppressed, $g_A^0=0.77$. However, this
value is still significantly larger than the experimental 
value $g_A^0=(0.28-0.41)$. In addition to that, we find very
little OZI violation, $\Delta s\simeq \Delta u(dis) \simeq
-0.01$. We observed, however, that larger values 
of the disconnected contribution can be obtained if the 
average instanton size is smaller than the phenomenological
value of $\rho \simeq 1/3$ fm.

 There are many questions that remain to be addressed. In
order to understand what is missing in our calculation it
would clearly be useful to perform a systematic study of 
OZI violation in the axial-vector channel on the lattice. 
The main question is whether the small value of $g_A^0$ 
is a property of the nucleon, or whether large OZI violation 
is also seen in other channels. A study of the connected
contributions to the axial coupling constant in cooled 
as well as quenched quantum QCD configurations was performed 
in \cite{Dolgov:1998js}. These authors find $g_A^0(con)=\Delta 
u(con)+\Delta d (con) \simeq 0.6$ in both cooled and full 
configurations. The disconnected term was computed by 
Dong et al.~\cite{Dong:1995rx}. They find $\Delta u(dis)
+\Delta d(dis)\simeq -0.24$. 

 In the context of the instanton model it is important to 
study whether the results for $g_A^0$ obtained from the 
axial-vector current three-point function are consistent
with calculations of $g_A^0$ based on the matrix element 
of the topological charge density $G\tilde{G}$
\cite{Forte:1990xb,Hutter:1995cs,Kacir:1996qn,Diakonov:1995qy}. 
It would also be useful to further clarify the connection of 
the instanton liquid model to soliton models of the nucleon 
\cite{Diakonov:1987ty}. In soliton models the spin of the 
nucleon is mainly due to the collective rotation of the 
pion cloud, and a small value for $g_A^0$ is natural 
\cite{Brodsky:1988ip,Blotz:1993am}. The natural parameter
that can be used in order to study whether this picture 
is applicable is the number of colors, $N_c$. Unfortunately, 
a direct calculations of nucleon properties for $N_c>3$ 
would be quite involved. Finally it would be useful to study 
axial form factors of the nucleon. It would be interesting to 
see whether there is a significant difference between the 
iso-vector and iso-singlet axial radius of the nucleon. A 
similar study of the vector form factors was recently presented 
in \cite{Faccioli:2003yy}.

\chapter{Group integration}

\section{General remarks}
Group integration became an important tool in physics with the 
de\-ve\-lop\-ment 
of lattice gauge theories\cite{Creutz1}. Since then it enjoyed more 
attention in other areas as well, like instantons 
\cite{Shifman2,Zahed} or random matrix theory, e.g.\cite{Prosen}. 
In this chapter we elaborate on an idea of Michael Creutz and present an 
algorithm for computing integrals over elements of a compact group.

A lot of progress has been made in the direction of solving the generating 
functional 
\beq\label{gen_func}
Z(J,J^\dagger)=\int \!\! du\;\exp\{J^\dagger u + u^\dagger J\} \:\:,
\eeq
which enables one to express any other integral of a general function 
$f(u,u^\dagger)$ as:
\beq
\int\!\!du\;f(u,u^\dagger)=
\left.
 f(\delta_{J^\dagger},\delta_J)\;\;Z(J,J^\dagger)
\right|_{J,J^\dagger=0}.
\eeq
Different methods have been used to compute (\ref{gen_func}). Direct 
integration over parametrized $SU(2)$ and $SU(3)$ was given in 
\cite{Eriksson}, and polar decomposition of $U(N)$ and integration over 
angular variables in \cite{Fateev}. Another approach was to 
construct and solve a partial differential equation which $Z(J,J^\dagger)$ 
sa\-tis\-fies due to its invariance w.r.t. left and right actions of the 
group\cite{Brower,Creutz2}. The author of \cite{Jaap} expanded 
$Z(J,J^\dagger)$ in powers of invariants and computed numerically the 
coefficients up to some order using again a differential equation. 
Character expansion (for recent work see \cite{Balantekin}) proved to be 
a powerful method, used for $U(N)$ in \cite{Bars}. More 
recently\cite{Prosen} a 
$\frac{1}{N}$ expansion of integrals over monomials in $U(N)$ and $O(N)$ 
matrix 
elements was 
developed, with the leading and next-to-leading order terms computed using 
a standard integral over $\mathbf{C}^{N^2}$ with a modified Gaussian 
measure.

The method we propose lacks the elegance of character expansion of 
generating functional.
Its simplicity, however, makes it a handy tool for evaluating group 
integrals over any compact group and any representation. Moreover, it is easily 
implementable as a computer algorithm once the group invariants are  
known. The same idea may also be applied to evaluating integrals of 
tensorial structure over any manifold with a measure possessing some kind 
of symmetry. 

The flow of ideas has the following structure:
 in section \ref{Haar} we introduce the 
main 
properties of Haar measure. We state the algorithm in section 
\ref{algorithm} and 
exemplify it on 
fundamental and adjoint representations of $SU(N)$ in section \ref{fund} 
and \ref{adjoint} respectively.

\section{ Properties of Haar measure}\label{Haar}
The algorithm being proposed relies heavily on the properties of the Haar 
measure
\footnote{for a more detailed 
discussion see e.g. \cite{Creutz1} and references therein.}:

For any compact group there exists a unique left-right invariant measure, normalized to unity, 
such that:
\begin{eqnarray}
\int\!\! du &=& 1 \label{normal}\\
\int\!\! du\; f(gu) &=& \int\!\! du\; f(ug) = \int\!\! du\; f(u) 
\label{left-right} \\
\int\!\! du\; f(u) &=& \int\!\! du\; f(u^{-1})\:\:,\label{inverse}
\end{eqnarray}
where $u,g$ are elements of the group and $du$ is the Haar measure.

The left-right invariance of the measure is the analog of 
translational invariance of the $\mathbf{R}^n$ integral and for finite 
groups represents the invariability of the number of elements w.r.t. 
multiplication by a fixed element of the group.
The examples of group measure \cite{Creutz1}:

for $ U(1)=\{e^{i\phi}\:|\:-\pi<\phi\leq\pi\}$:
\beq
 \int\!\! du\; 
f(u)=\frac{1}{2\pi}\int_{-\pi}^{\pi}\!\! d\phi\; f(e^{i\phi})
\label{exmpl_measure}
\eeq

for
$ SU(2)=\{a_0+i\vec{a}\cdot\vec{\tau}\: |\: a_0^2+\vec{a}^2=1\}$:
\beq\label{exmpl_measure2}
\int\!\! du\; f(u) = \frac{1}{\pi^2} \int\!\! d^4a\: \delta(a^2-1)\: 
f(u(a))
\eeq

\section{The algorithm for group integration}
\label{algorithm}
We will present here the main idea, mentioned before in 
\cite{Creutz2} but, to our knowledge, never carried out 
completely. 

For the purpose of integration we will not need an exact expression 
for the measure, like (\ref{exmpl_measure}),(\ref{exmpl_measure2}). We 
will only use the general 
properties of the Haar measure, making the algorithm applicable to any 
compact group. 

We are interested in the most general integral of a monomial in group 
elements:
\beq
I_{\underbrace{i_1\ldots i_n}_{'left'},\underbrace{j_1\ldots 
j_n}_{'right'\;indices}}^
{\overbrace{l_1\ldots 
l_m}\:,\:\overbrace{k_1\ldots k_m}}=\int\!\!du\: 
u_{i_1 j_1}\ldots u_{i_n\ldots j_n}\: \bar{u}^{k_1 l_1}\ldots 
\bar{u}^{k_m 
l_m}\:\:, \label{gen_int}
\eeq
where $\bar{u}$ represents $u^\dag$ for SU(N) or $u^+$ for SO(N) or  
$u^{-1}$ for a general compact group. 

In order to exemplify the idea, we will consider first the simple 
example of
\beq\label{example}
I_{i,j}^{l,k} = \int\!\! du\: u_{ij}\: \bar{u}^{kl} \:\:.
\eeq
The right invariance (\ref{left-right}) implies:
\beq 
I_{i,j}^{l,k}=\bar{g}_{kk'}\: I_{i,j'}^{l,k'}\: g_{j'j} \:\:.
\eeq
This shows that $I_{i,j}^{l,k}$ has to be an invariant of the group in 
'right' indices $j,k$. The same argument with left invariance of the 
measure compels $I_{i,j}^{l,k}$ to be separately an invariant in 'left' 
indices $i,l$ 
as well. There is only one second rank invariant tensor with upper and 
lower 
index for SU(N): $\delta_i^l$. 
Therefore, the integral (\ref{example}) has to be of the form:
$$
I_{i,j}^{l,k}=a_1\:\delta_i^l\:\delta_j^k\:\: .
$$
The constant $a_1$ is found easily by contracting with $\delta_j^k$ and 
using $u_{ij}\bar{u}^{jl}=\delta_i^l$ and
the unit normalization (\ref{normal}):
$$\delta_i^l=I_{i,j}^{l,j}=a_1\:N\:\delta_i^l\:\: .$$
So the integral then is:
\beq\label{result_1}
I_{i,j}^{l,k}=\int\!\!du\:u_{ij}\:\bar{u}^{kl}=
\frac{1}{N} \: \delta_{i}^{l} \: \delta_{j}^{k}\:\:,
\eeq
which is nothing else but the orthogonality relation for fundamental 
representation.
The non-existence of one-index invariant leads directly to
\beq
I_{i,j}=\int\!\!du\:u_{ij}=0\:\: .
\eeq
The same steps as in the above examples can be followed for any integral 
of the type (\ref{gen_int}). 
Let us summarize them in the form of the following algorithm:
\begin{itemize}
\item Generate all the independent invariant tensors with the index 
structure of (\ref{gen_int}). The result of integration will be take the form:
$$
\sum_{all\:invariants}const\times 
(left\:\:indices\:\:invariant)\times(right\:\:indices\:\:invariant).
$$ 
\item Use index pairs exchange symmetry(e.g. $i_1j_1 \leftrightarrow 
i_2j_2$) 
to identify the tensors with the 
same numerical coefficients.
\item Compute the independent coefficients by multiplying (\ref{gen_int}) 
with the corresponding invariant tensors (e.g. contracting indices with 
$\delta$ tensor). 
\end{itemize}
Throughout the this work we deal with SU(N). However, the same philosophy is 
applicable to any compact group.
\section{Integration over fundamental representation of SU(N)}
\label{fund}
The invariant tensors depend on the group (e.g. vary with N in SU(N)) as 
well as representation. For fundamental representation of SU(N) we have 
two 
basic invariants, out of which the rest is constructed: $\delta_i^l$ and 
$\epsilon^{i_1 \ldots i_N}$. The invariance can be seen easily from the 
unitarity and unimodularity.

The result of a particular integral will depend on N, e.g.\footnote{the 
coefficient of $\frac{1}{2}$ is determined by multiplication with the 
invariant tensor itself: $\epsilon_{i_1i_2}\epsilon_{j_1j_2}$} 
\beq
	\int\!\!du\:u_{i_1j_1}\:u_{i_2j_2}  =  
	\left\{
		\begin{array}{ll}
			\frac{1}{2}\epsilon_{i_1i_2}\epsilon_{j_1j_2} 
			& \mbox{for $SU(2)$}\\
			 0 & \mbox{for $SU(N>2)$}
		\end{array}
 	\right.\:\: .
\eeq
Despite of having different invariants for different N, there are many 
cases the result has the same form, but coefficients depend on N. An 
example is the integral with the same number of $u$ and $\bar{u}$'s, 
schematically $I_n \equiv \int\!\!du (u\:\bar{u})^n$,on which we will focus 
in this paragraph. 

Let us then follow the proposed algorithm and compute $I_n$. 
First we generate 
all left-right invariant terms. Using symmetries of the measure 
we identify the terms with the same 
numeric coefficients. Finally we multiply
 both sides with the very same set of invariants and solve the obtained
 system of linear equations for the unknown coefficients.  

So what are the possible invariants we can construct? The building blocks 
are Kronecker's $\delta$ and $\epsilon$ tensors.

First of all, there are $(n!)^2$ terms of the form
\beq\label{deltas}
(\underbrace{\delta \times \ldots \times \delta}_{n\times left\: indices}) 
\times 
(\underbrace{\delta \times \ldots \times \delta}_{n\times right\: 
indices})\:\:, 
\eeq
i.e. a $\delta$ for each up-down pair both in left and right indices.

Then, for $n>N$ we can use $\epsilon_{i_1\ldots i_N}$ or 
$\epsilon^{l_1\ldots l_N}$ to form terms like 
\beq
(\underbrace{\delta\times\ldots \delta}_{n-N}\times \epsilon_{lower\:
indices}\times 
\epsilon^{upper\: indices})_{left}\times (\ldots)_{right}\:\:.
\eeq
Notice that \emph{$\epsilon$ tensors must come in pairs}, both with 
upper and 
lower indices, left or right(or both). The reason is 
obvious: $\delta$'s require both an upper and a lower index; if an 
$\epsilon$ with lower indices is used, then the only invariant left for 
upper extra indices is 
another $\epsilon$. One could thus have pairs of $\epsilon$ in left and 
right indices, having different structure for different N. 
Fortunately, this is not the case, as products of $\epsilon$'s
 can be expanded in products of $\delta$'s.

The important observation is, that any pair of $\epsilon$'s in upper and 
lower indices can be written as a completely antisymmetric sum of products 
of Kronecker $\delta$ tensors:
\beq\label{epsilon}
\epsilon_{i_1i_2\ldots i_n}\epsilon^{l_1 l_2 \ldots 
l_n}=\sum_{permutations\: P}(-1)^P \delta_{i_1}^{P(l_1)}\ldots 
\delta_{i_n}^{P(l_n)}\:\: .
\eeq

Therefore, independently on N, the final result will only contain products 
of $\delta$'s, it will be of the form of (\ref{deltas}).

Having noticed this, we only need to generate all the possible 
($\delta)^n\times (\delta)^n$ combinations, identify common coefficients 
and 
compute them.(the actual values of coefficients will depend on N).

We will present the calculation up to $n=3$, for the higher order 
integrals one would use a computer program to generate terms and compute
the coefficients.
 \subsection{Case n=2}
We will treat the integral
\beq\label{2-integral}
I_{i_1i_2,j_1j_2}^{l_1l_2,k_1k_2}=\int\!\!du
\:u_{i_1j_1}\:\bar{u}^{k_1l_1}\:
u_{i_2j_2}\:\bar{u}^{k_2l_2}
\eeq
in some detail in order to unveil the ideas that are useful in the 
more involving case of higher n.

The integral will contain $(2!)\times(2!)$ terms:
%
%
\beq\label{4-diagrams}
\raisebox{-0.75cm}{
\psfig{file=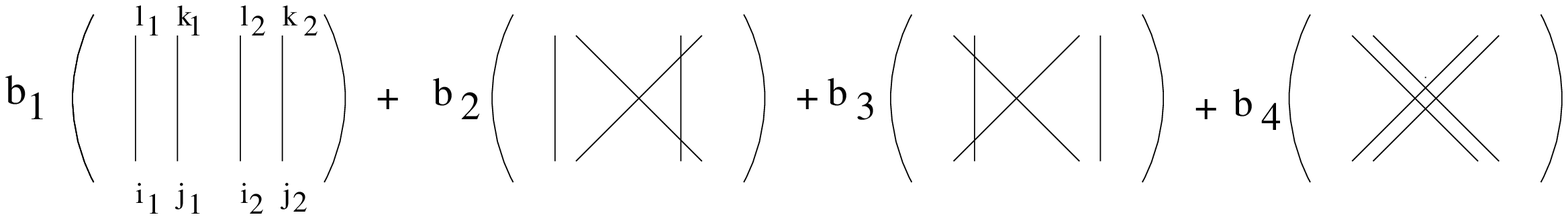,width=12cm,height=1.5cm}
},
\eeq
%
%
where we represent $\delta$ by a line joining the indices.\footnote{In 
all the subsequent diagrams the 
places of 
indices are the same as in the first diagram, so we omit them for 
the sake of simplicity} Each line has 
to join an upper left(right) index with a lower left(right) index. No 
connections between left and right indices are allowed. The indices in the above graphs have 
always the same position as in the first parenthesis of (\ref{4-diagrams}) 

The $(i_1j_1) \leftrightarrow (i_2j_2)$ and $(k_1l_1) \leftrightarrow 
(k_2l_2)$ 
symmetry of  (\ref{2-integral}) allows us to identify the equal 
coefficients among $b_1\ldots b_4$. Under $(i_1j_1)\leftrightarrow 
(i_2j_2)$ 
the diagrams (1 and 4) as well as (2 and 3) interchange. Therefore 
$b_1=b_4$ and $b_2=b_3$.

It is easy to identify the diagrams that have the same coefficients by the 
topology of diagrams. Take pairs of 1 left and 1 right index, e.g. lower 
$(i_1j_1), (i_2j_2)$ or upper $(l_1k_1),(l_2k_2)$. What matters is how 
many of these pairs of indices are interconnected:  in $b_1$ and $b_4$ 
terms one only has 
(1 upper pair - 1 lower pair) $\delta$ connections. These two diagrams are 
pair-exchange equivalent. The $b_2$ diagram is topologically different: 
there is a (2 upper pairs - 2 lower pairs) 
$\delta$-interconnection: both pairs of upper indices (lk) are connected 
to both pairs of lower (ij) indices. This diagram is pair-exchange 
equivalent to 
$b_3$ and falls in the same topological set of diagrams. This easy way to 
identify diagrams with the same coefficients will 
prove very useful in calculating integrals with higher n.

Computing the coefficients is straight-forward: contract with 'right' 
$\delta_{j_1}^{k_1}$ to get on one hand, from (\ref{result_1}):
$$
\frac{1}{N}\delta_{i_1}^{l_1}\delta_{i_2}^{l_2}\delta_{j_2}^{k_2}
$$
and, on the other hand (\ref{4-diagrams}):
$$(N\:b_1+b_2)
\delta_{i_1}^{l_1}\delta_{i_2}^{l_2}\delta_{j_2}^{k_2} +
(N\:b_2 + b_1) 
\delta_{i_1}^{l_2}\delta_{i_2}^{l_1}\delta_{j_2}^{k_2}\:\: .
$$
Matching the two sides we get the coefficients. The result then is:
\beq
\begin{split}
&I_{i_1i_2,j_1j_2}^{l_1l_2,k_1k_2}
=\int\!\!du\:u_{i_1j_1}\:\bar{u}_{k_1l_1}\:
u_{i_2j_2}\:\bar{u}_{k_2l_2}=
\\ 
&=
\frac{1}{N^2-1}\left[
\raisebox{-0.4cm}{
\psfig{file=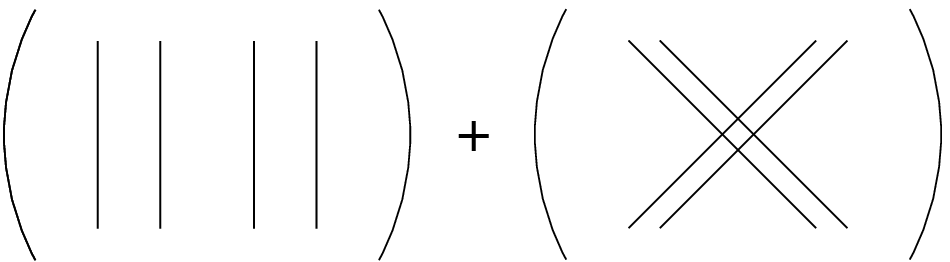,width=3cm,height=1cm}
}\right]
-\frac{1}{N(N^2-1)}
\left[
\raisebox{-0.4cm}{
\psfig{file=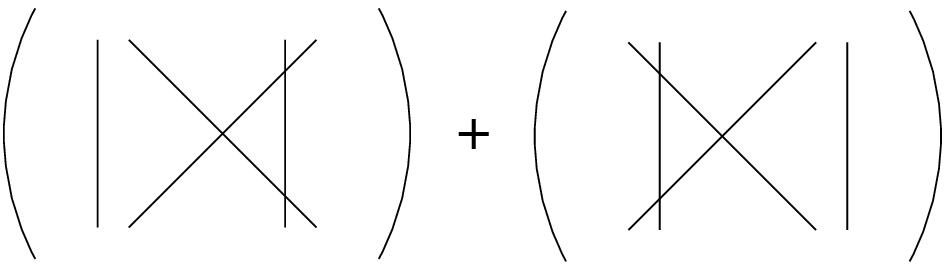,width=3cm,height=1cm}
}
\right]
\\  
&=
\frac{1}{N^2-1}[
\delta_{i_1}^{l_1}\delta_{j_1}^{k_1}\delta_{i_2}^{l_2}\delta_{j_2}^{k_2} +
\delta_{i_1}^{l_2}\delta_{j_1}^{k_2}\delta_{i_2}^{l_1}\delta_{j_2}^{k_1} 
]
- \frac{1}{N(N^2-1)}
[
\delta_{i_1}^{l_1}\delta_{j_1}^{k_2}\delta_{i_2}^{l_2}\delta_{j_2}^{k_1} +
\delta_{i_1}^{l_2}\delta_{j_1}^{k_1}\delta_{i_2}^{l_1}\delta_{j_2}^{k_2}
]\\
\end{split}
\label{result_2}
\eeq

\subsection{Case n=3}
The number of terms in 
\beq
I_{i_1i_2i_3,j_1j_2j_3}^{l_1l_2l_3,k_1k_2k_3}=
\int\!\!du \: 
u_{i_1j_1}\:\bar{u}^{k_1l_1}\:
u_{i_2j_2}\:\bar{u}^{k_2l_2}\:
u_{i_3j_3}\:\bar{u}^{k_3l_3}
\eeq
increases to $(3!)\times(3!)=36$. They will be grouped in 3 topologically 
different classes having the same coefficients in the front:
\begin{itemize}
\item 6 terms of type (1-1,1-1,1-1), like e.g.
$$
\psfig{file=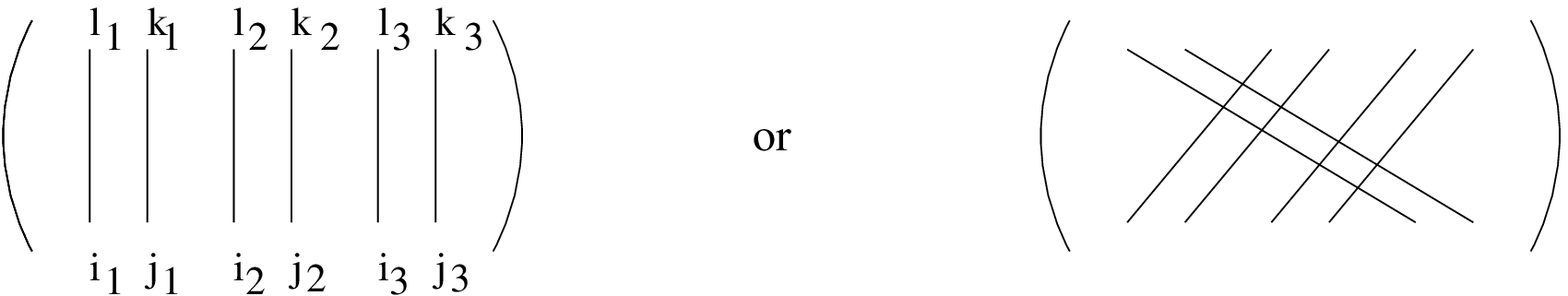,width=12cm,height=1.5cm}
$$

linking 1 pair of upper to 1 pair of lower indices.
\item 18 terms of type (2-2, 1-1) having 2 pairs of lower and 2 pairs of 
upper indices interconnected in addition to one upper to one lower pair 
$\delta$ connection. Examples of diagrams from this category:
$$
\psfig{file=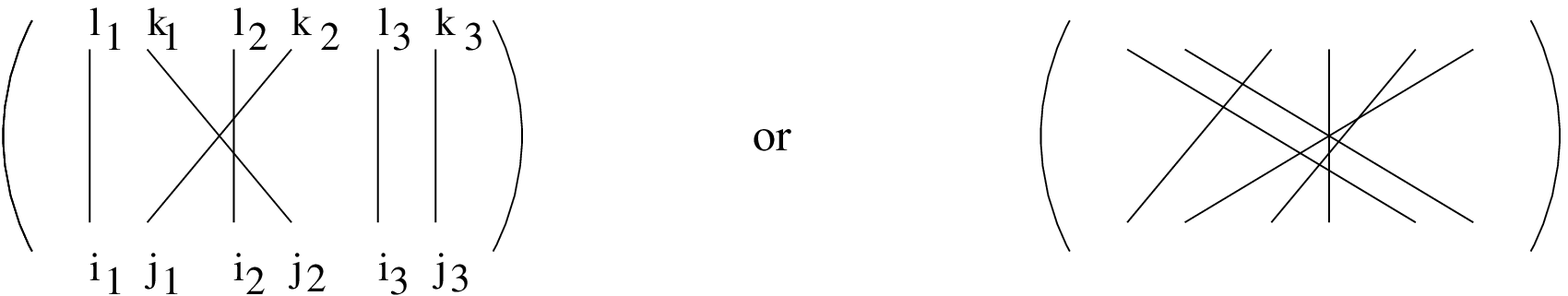,width=12cm,height=1.5cm}
$$

\item 12 terms of type(3-3) having all 3 pairs of upper indices and 3 
pairs of lower indices interconnected, e.g. diagrams:
$$ 
\psfig{file=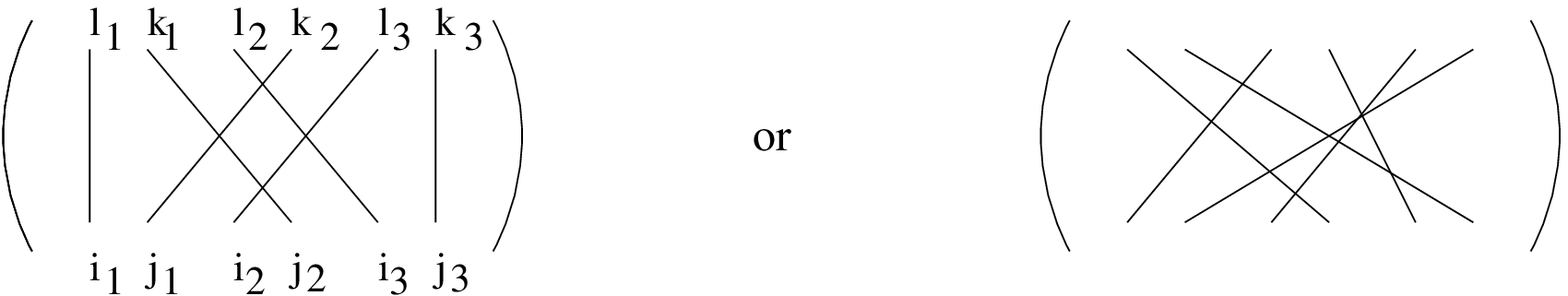,width=12cm,height=1.5cm}
$$
\end{itemize}

The coefficients can be found in a similar way to the case of $n=2$. The 
result is:
\begin{eqnarray}\label{result_3} 
\nonumber
& &\int\!\!du \: 
u_{i_1j_1}\:\bar{u}^{k_1l_1}\:
u_{i_2j_2}\:\bar{u}^{k_2l_2}\:
u_{i_3j_3}\:\bar{u}^{k_3l_3} = \\ \nonumber
& &c_1\left[
\raisebox{-.6cm}{
\psfig{file=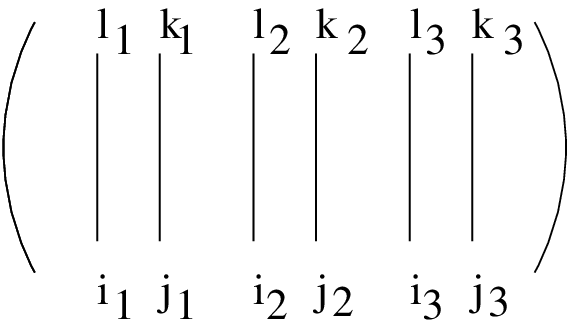,width=4cm,height=1.25cm}
}+\ldots altogether\;\;6\;\;terms
\right] + \\ \nonumber
&+& c_2\left[
\raisebox{-.6cm}{
\psfig{file=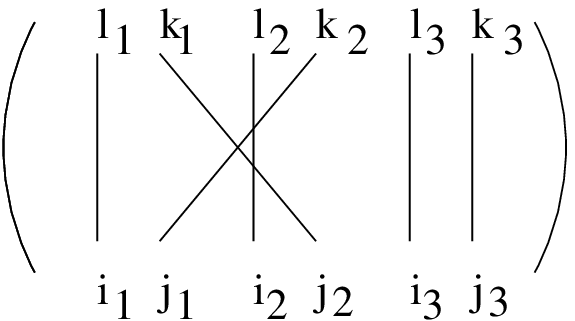,width=4cm,height=1.25cm}
}+\ldots altogether\;\;18\;\;terms
\right] + \\ 
&+& c_3\left[
\raisebox{-.6cm}{
\psfig{file=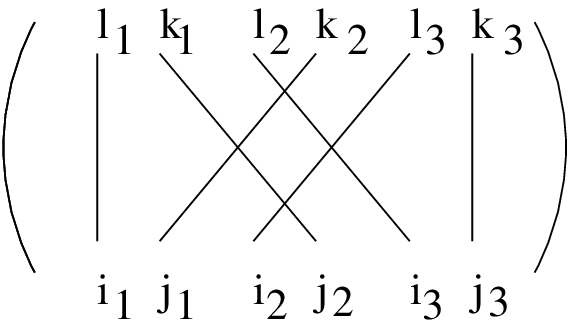,width=4cm,height=1.25cm}
}+\ldots altogether\;\;12\;\;terms
\right],  \\  \nonumber
\end{eqnarray}

where different terms in $\ldots$ are generated by interchanges of $(lk)$ 
pairs or $(ij)$ pairs, and have the same topology as the first 
term in parenthesis.

For $SU(N\ge3)$ the coefficients are:
\begin{eqnarray}
c_1 &=& \frac{(N^2-2)}{N(N^2-1)(N^2-4)}\nonumber\\
c_2 &=& -\frac{1}{(N^2-1)(N^2-4)}\nonumber\\
c_3 &=& \frac{2}{N(N^2-1)(N^2-4)}
\end{eqnarray}

An important requirement is, that the invariant tensors are independent. 
The 3! terms of type 
$\delta_{i_1}^{l_1}\delta_{i_2}^{l_2}\delta_{j_2}^{k_2}$ and permutations 
are indeed independent for $SU(N), N\ge 3$. However, for $N=2$ there are 
'too many' indices and only 2 possible values for each index. This leads 
to the following relationship:
\beq
0=\epsilon_{i_1i_2i_3}\epsilon^{l_1l_2l_3}=
\left[
\raisebox{-.45cm}{
\psfig{file=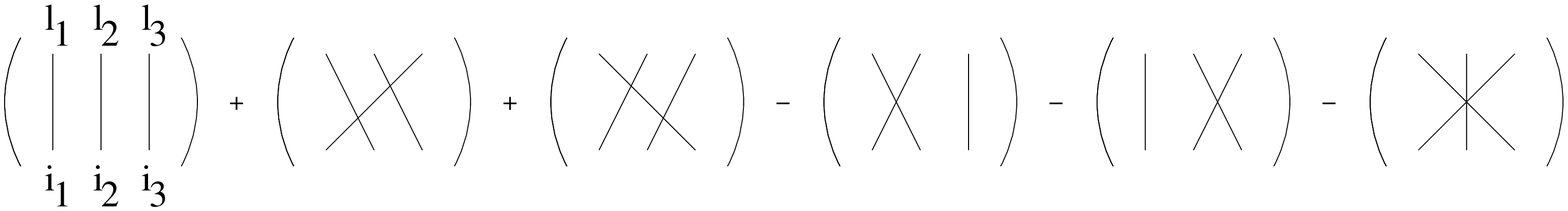,width=9cm,height=1cm}
}
\right].
\eeq
In this case of having dependent tensors one can follow the same strategy 
up to one point: after 
contracting the indices in order to figure out the coefficients, one has 
to express everything in terms of a basis of independent tensors, and only 
then compare the coefficients. For SU(2) this leads to the same form of 
result (\ref{result_3}) with the parameter-dependent coefficients:
\begin{eqnarray}
c_1 &=& t+\frac{1}{6}\nonumber\\
c_2 &=& -t-\frac{1}{24} \nonumber\\
c_3 &=& t
\end{eqnarray}
The parameter t cancels if one goes to the basis of 5 independent 
$\d\d\d$ tensors so we can as well set it to zero.

For higher $n=4,5\ldots$ one has to take this subtlety into account also 
for $SU(3),SU(4)\ldots$ The same idea will prove useful in the next 
paragraph, when 
we treat the integral over adjoint representation.
\section{The integral over adjoint representation of SU(N)}
\label{adjoint}
In a number of cases one encounters integrals of type
\beq \label{adjoint_gen}
I_{\underbrace{a_1a_2\ldots a_n}_{left},\underbrace{b_1b_2\ldots 
b_n}_{right}}=
\int\!\! du\: 
R^{a_1b_1} 
R^{a_2b_2} 
\ldots R^{a_nb_n}\:\: , 
\eeq
where 
$$
R^{ab}=\frac{1}{2}Tr(\lambda^b u \lambda^a \bar{u})
$$
is the adjoint representation of $SU(N)$ satisfying
 $R^{ab}\lambda^{b}=u\lambda^a \bar{u}$, with $\lambda^{a}$ the 
generators of SU(N) normalized to $Tr(\lambda^a\lambda^b)=2\delta_b^a$.

To solve this integral, one can straight-forwardly integrate over u's and 
then contract the result with $\lambda$'s and perform the trace. For 
example:
\begin{eqnarray}
\int\!\!du\: R^{ab} &=& 
\frac{1}{2}\lambda^b_{li}\lambda^a_{jk}\int\!\!du\:u_{ij}\:\bar{u}_{kl}
\nonumber \\
&=&\frac{1}{2}\lambda^b_{li}\lambda^a_{jk}\frac{1}{N}\delta_{jk}\delta_{il}
=\frac{1}{2N}Tr\lambda^b\:Tr\lambda^a = 0\:\: .
\label{int_R}
\end{eqnarray}
For higher n the number of terms of $\int\!\!du\:(u\bar{u})^n$ grows as 
$(n!)^2$ and the e\-va\-lu\-a\-tion of the integral, although 
straight-forward, 
becomes tedious. For this reason, we present here a more manageable way of 
computation, based on the above algorithm.

First of all, let's note some properties of $R^{ab}$ matrices:
\begin{itemize}
\item $R^{ab}$ is the adjoint representation of $SU(N)$ 
group\footnote{We will denote by $R^{ab}_u$ the adjoint matrix 
corresponding to the group element $u$. When there is no ambiguity on what 
element the R corresponds to, we will skip writing u's, usually it will 
mean that all the R matrices correspond to the same group element }: 
\beq 
R^{ab}_{uu'}=R^{ac}_{u'}\: 
R^{cb}_u \eeq \item The matrices form a subgroup of $SO(N^2-1)$:
\beq
R^{ab}R^{cb}=R^{ab}(R^+)^{bc}=\delta^{ac}
\eeq
\item
Under the left-right action of the group, it rotates as follows:
\beq
R^{ab}\stackrel{u\rightarrow g\:u\:h}{\longrightarrow}
\tilde{R}^{ab}=R^{bc}_{g^{-1}}\:R^{ad}_h\:R^{dc} 
=R^{ad}_h \: R^{dc} \: R^{cb}_g 
\eeq
\end{itemize}
The main idea of the algorithm applies here as well: the left-right 
invariance of the measure in (\ref{adjoint_gen}) implies that
\beq
I_{a_1\ldots a_n,b_1\ldots b_n}=
R^{a_1a_1'}_h\ldots R^{a_na_n'}_h
R^{b_1b_1'}_{g^{-1}}\ldots R^{b_nb_n'}_{g^{-1}}
\times
I_{a_1'\ldots a_n',b_1'\ldots b_n'}\:\: .
\eeq
In other words, $I_{a_1\ldots a_n,b_1\ldots b_n}$ has to be of the form
\beq
\sum_{all\:invariants}const\times(invariant\:\: in\:\: left\:\: 
indices)\times(invariant\:\: in\:\: right\:\: indices)
\eeq
Up to now, everything is similar to what we had for integration over 
fundamental representation. However, the construction of invariants will 
be 
more involved, due to two extra 'building blocks': $d^{abc}$ and 
$f^{abc}$. The SU(3) invariants up to the 5th rank can be found e.g. in 
\cite{CORE,Invar}.
One might also find useful some $SU(N)$ relations in \cite{General}.

The symmetries used for determining the common coefficients of 
in\-va\-riants 
are:
\begin{itemize}
\item $(a_ib_i)\leftrightarrow (a_jb_j)$ exchange symmetry
\item $a\leftrightarrow b$ symmetry stemming from inversion invariance of 
the measure, as in (\ref{inverse})
\end{itemize}

Following the algorithm presented above, we computed the following 
integrals:
\subsection{Case n=1}The non-existence of 1 index invariant leads to 
\beq
\int\! du\; R^{ab} = 0
\eeq 
as we explicitly computed in (\ref{int_R}).

\subsection{Case n=2} 
There is only one\footnote{we need separately an invariant for left and 
one for right indices} second rank invariant:
$\delta$, and one easily gets the orthogonality relation for the adjoint 
representation: 
\beq
\int\!\!du\:R^{a_1b_1}\:R^{a_2b_2}=\frac{1}{N^2-1}\delta^{a_1a_2}
\delta^{b_1b_2}\:\:.
\eeq 

\subsection{Case n=3} There are 2 third rank invariants:
$d^{abc}$ and $f^{abc}$ and so 

\begin{eqnarray}
\int\!\!du\;\;R^{a_1b_1}\:R^{a_2b_2}\:R^{a_3b_3}&=&
\frac{1}{N(N^2-1)}f^{a_1a_2a_3}f^{b_1b_2b_3}\nonumber \\ 
&+&
\frac{N}{(N^2-4)(N^2-1)}d^{a_1a_2a_3}d^{b_1b_2b_3}\:\: . 
\end{eqnarray}
For SU(2) we have
$d^{a_1a_2a_3}=0$ and one just disregards the second term in the result
above. 
\subsection{Case n=4} 
We will use the following diagrammatic representation: 
$$ \raisebox{-.65cm}{
\psfig{file=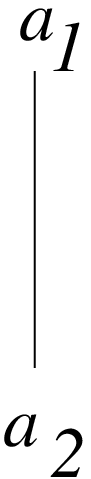,width=0.3cm,height=1.25cm} } =
\delta_{a_2}^{a_1}\hspace{1.5cm} 
\raisebox{-.65cm}{
\psfig{file=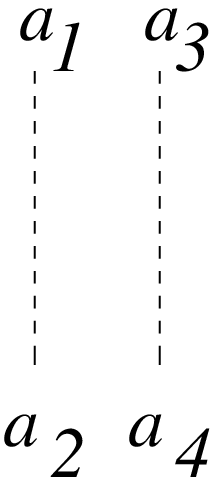,width=.7cm,height=1.25cm} 
} =
d_{a_1a_2a}d_{a_3a_4a}\hspace{1.5cm} 
\raisebox{-.65cm}{
\psfig{file=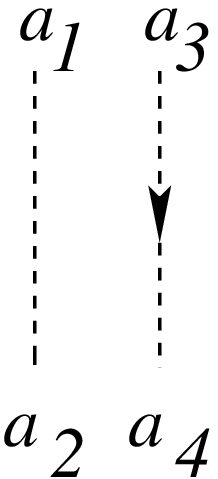,width=.7cm,height=1.25cm} 
} = d_{a_1a_2a}f_{a_3a_4a}\:\: . 
$$
The bracketed terms will mean invariants both in left ('a') and right
('b') indices. The places of indices is always the same as in the
following examples: 
$$ \raisebox{-.5cm}{
\psfig{file=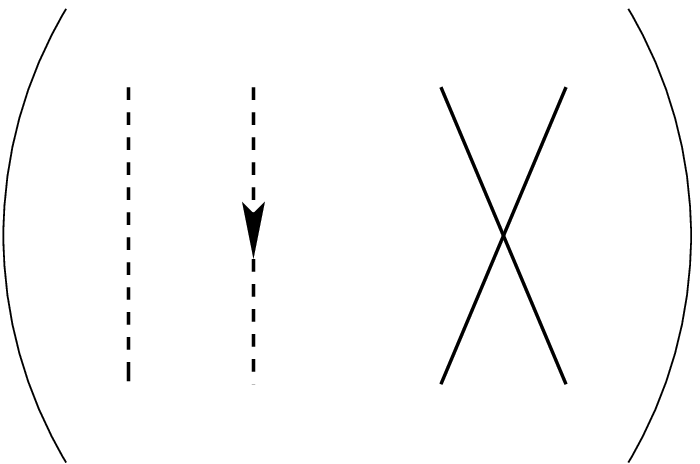,width=2cm,height=1.25cm} 
} \equiv
\raisebox{-.5cm}{ 
\psfig{file=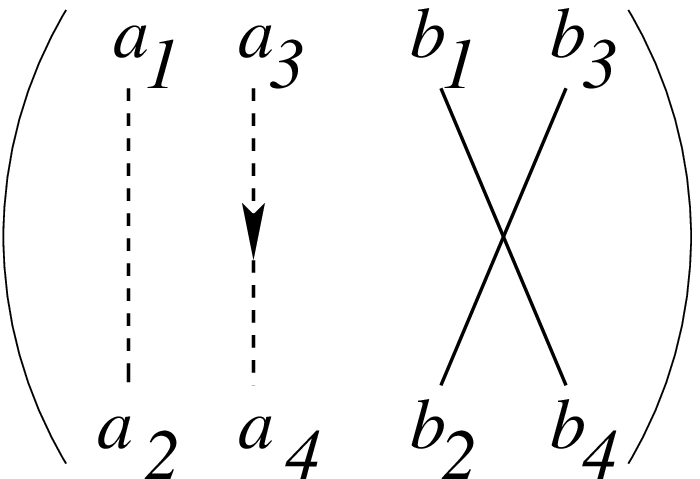,width=2cm,height=1.25cm} 
}
=d_{a_1a_2a}d_{a_3a_4a}\;\delta_{b_4}^{b_1}\delta_{b_2}^{b_3} $$ 
or 
$$
\raisebox{-.5cm}{ 
\psfig{file=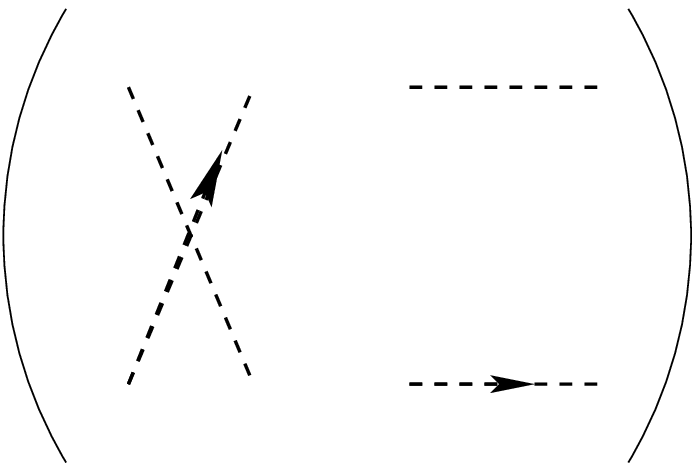,width=2cm,height=1.25cm} 
} \equiv
\raisebox{-.5cm}{ 
\psfig{file=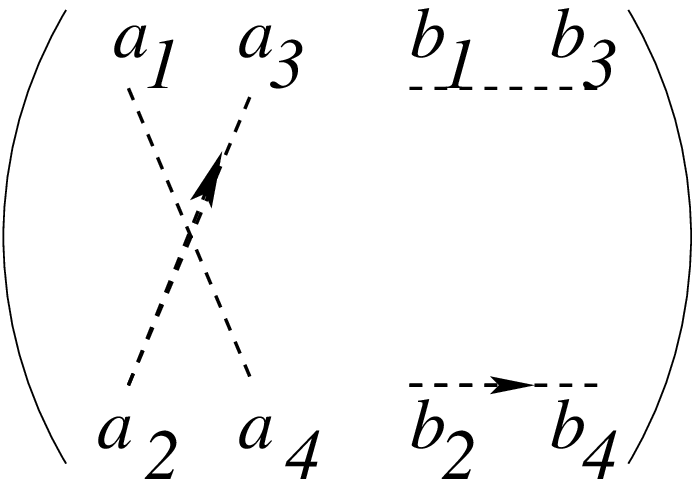,width=2cm,height=1.25cm} 
}
=d_{a_1a_4a}f_{a_2a_3a}\;\; d_{b_1b_3b}f_{b_2b_4b} 
$$ 
and so on.  
For
SU(2) this is still a simple integral, since the only independent
invariant tensors are again products of $\delta$(The structure constants
tensors reduce to product of $\delta$'s as well, due to (\ref{epsilon})).
The straight-forward application of the algorithm leads for SU(2) to:
\begin{eqnarray}
& &
\int\!\!du\:R^{a_1b_1}\:R^{a_2b_2}\:R^{a_3b_3}\:R^{a_4b_4}= 
\\ 
\nonumber
& &
\frac{2}{15}\left[
\raisebox{-.5cm}{ 
\psfig{file=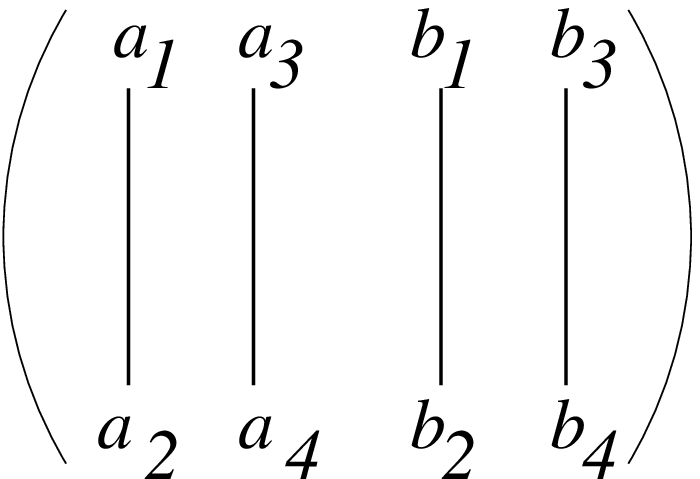,width=1.8cm,height=1.25cm} 
}+
\raisebox{-.5cm}{ 
\psfig{file=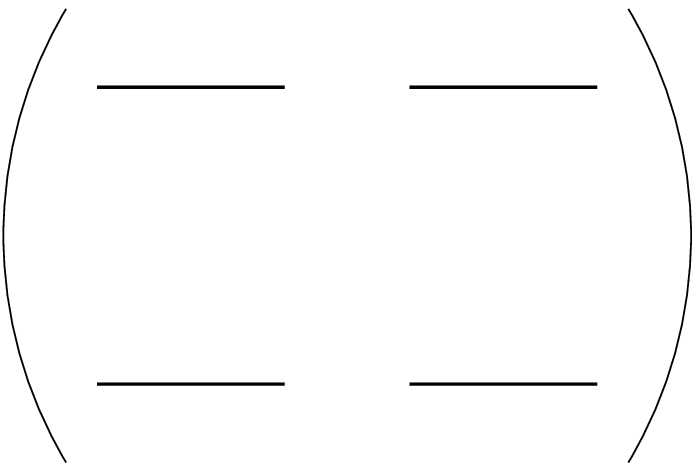,width=1.8cm,height=1.25cm} 
}
+
\raisebox{-.5cm}{ 
\psfig{file=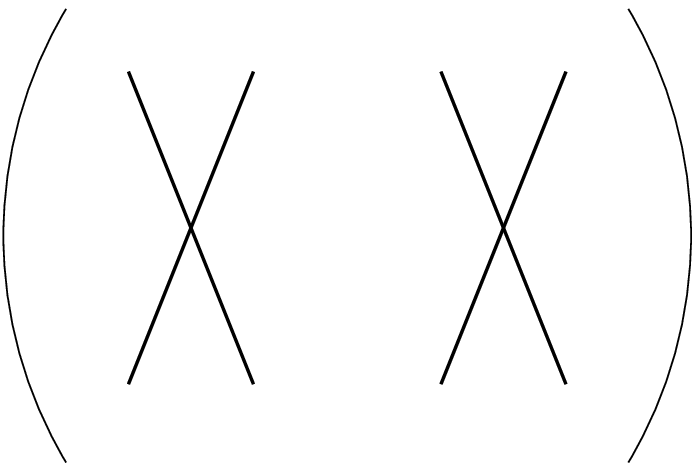,width=1.8cm,height=1.25cm} 
}
\right] 
-\frac{1}{30} \left[
\raisebox{-.5cm}{ 
\psfig{file=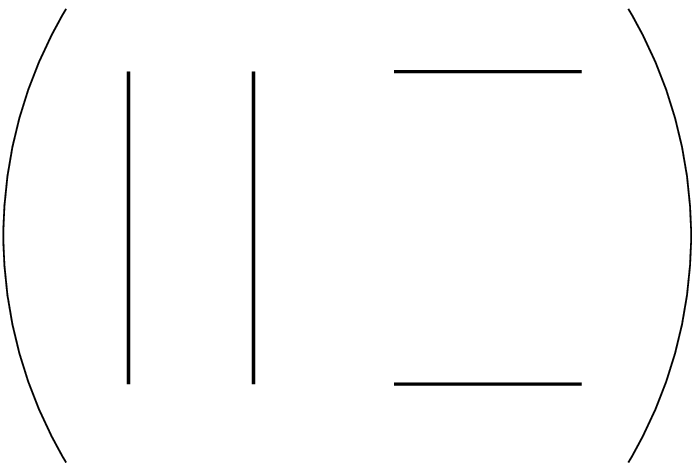,width=1.8cm,height=1.25cm} 
}+
\right.
\\ \nonumber
& &
\left.
\raisebox{-.5cm}{ 
\psfig{file=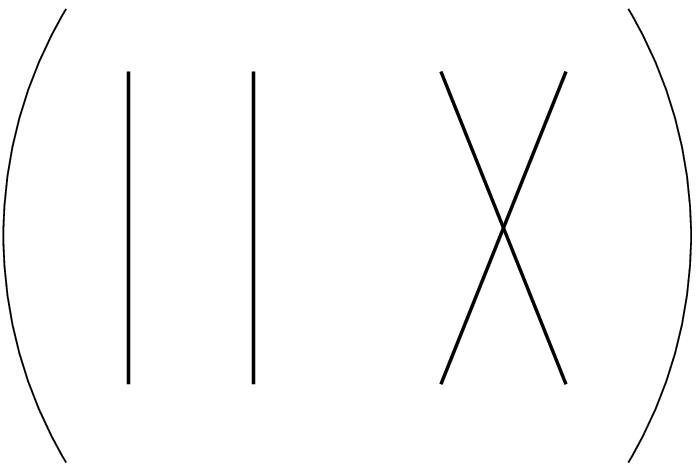,width=1.8cm,height=1.25cm} 
}+
\raisebox{-.5cm}{ 
\psfig{file=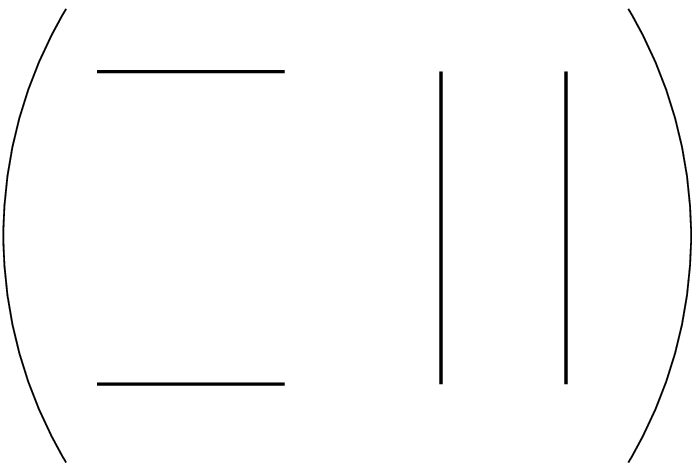,width=1.8cm,height=1.25cm} 
}+
\raisebox{-.5cm}{ 
\psfig{file=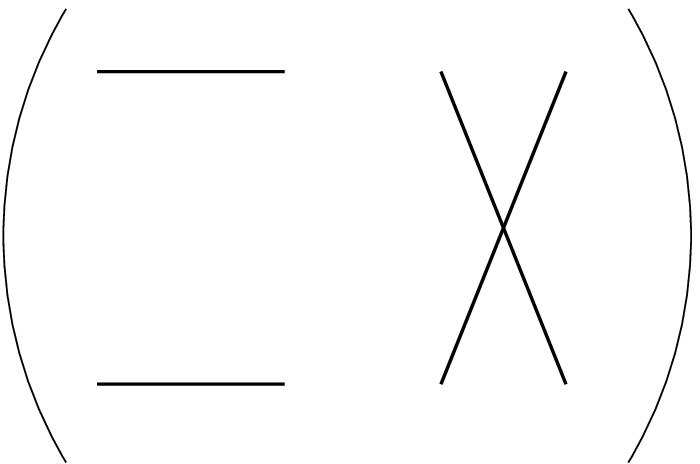,width=1.8cm,height=1.25cm} 
}+
\raisebox{-.5cm}{ 
\psfig{file=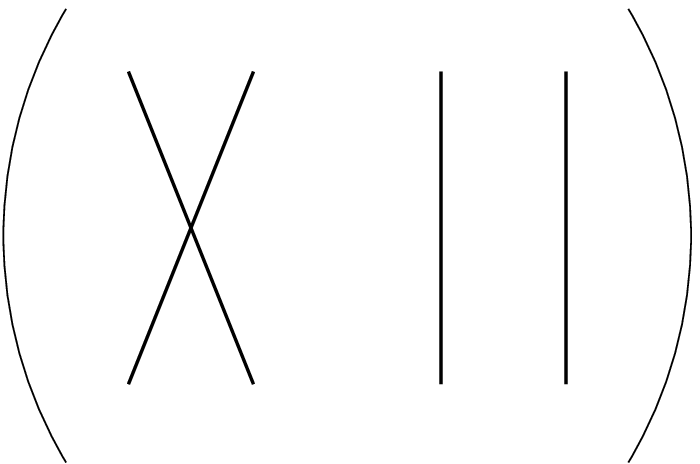,width=1.8cm,height=1.25cm} 
}+
\raisebox{-.5cm}{ 
\psfig{file=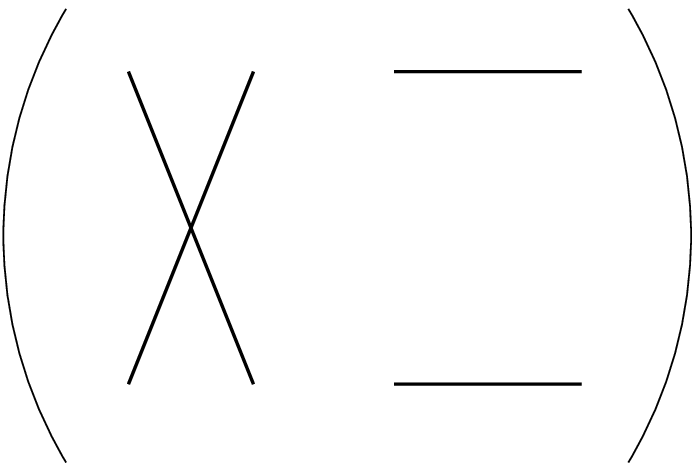,width=1.8cm,height=1.25cm} 
}
\right]\\ 
\nonumber
\end{eqnarray} 
Following \cite{Invar}, one can construct 9 independent invariants of 
4th rank for adjoint 
$SU(N>3)$ 
and 8 for SU(3). 

In terms of diagrams, the $n=4$ result for $SU(N\ge 3)$ is:
\beq
\begin{split}
	&\int\!\!du\:R^{a_1b_1}\:R^{a_2b_2}\:R^{a_3b_3}\:R^{a_4b_4}=
	K_{11}\left[
		\raisebox{-.4cm}{ 
		\psfig{file=gr_int_files/dldl_dldl_ind.eps,width=1.2cm,height=1cm} 
		}+
		\raisebox{-.4cm}{ 
		\psfig{file=gr_int_files/dldl_x_dldl_x.eps,width=1.2cm,height=1cm} 
		}+
		\raisebox{-.4cm}{ 
		\psfig{file=gr_int_files/dldl_hor_dldl_hor.eps,width=1.2cm,height=1cm} 
		}
	\right] \\
	&+K_{44}\left[
		\raisebox{-.4cm}{ 
		\psfig{file=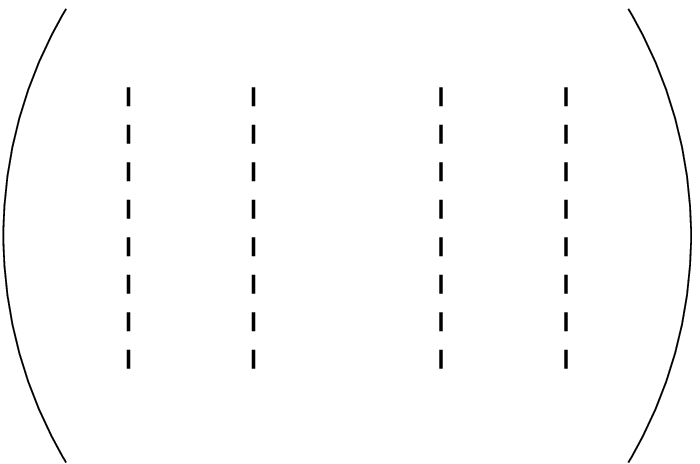,width=1.2cm,height=1cm} 
		}+
		\raisebox{-.4cm}{ 
		\psfig{file=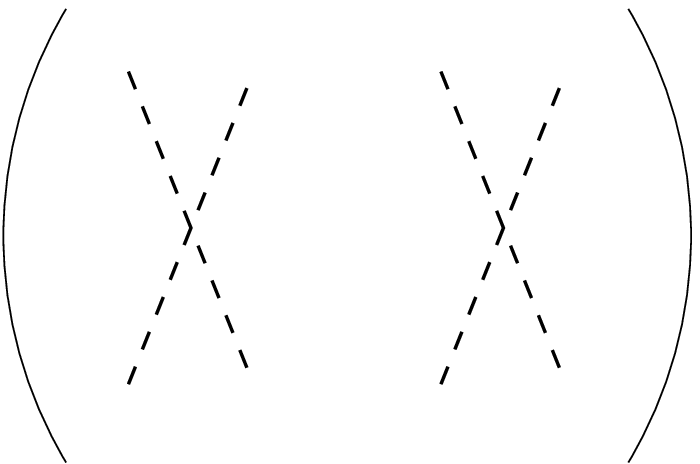,width=1.2cm,height=1cm} 
		}+
		\raisebox{-.4cm}{ 
		\psfig{file=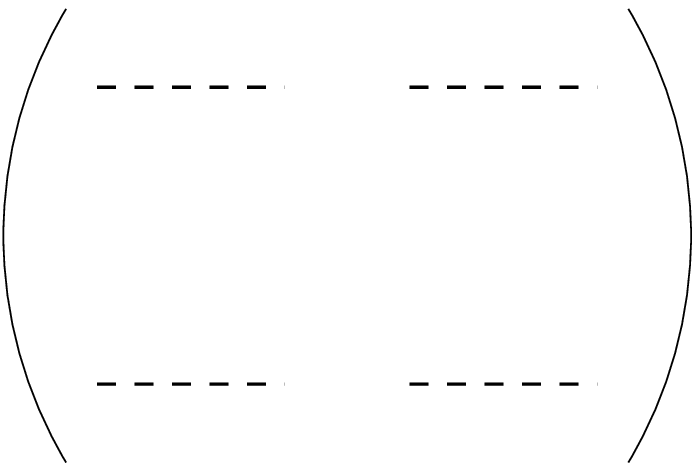,width=1.2cm,height=1cm} 
		}
	\right]  \\
	&+K_{41}\left[
		\raisebox{-.4cm}{ 
		\psfig{file=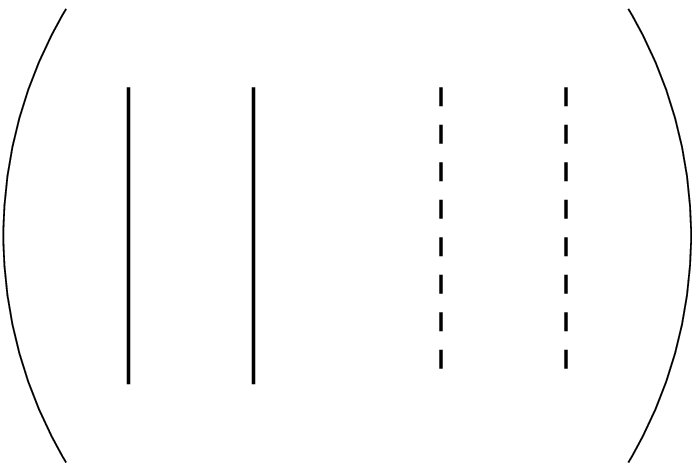,width=1.2cm,height=1cm} 
		}+
		\raisebox{-.4cm}{ 
		\psfig{file=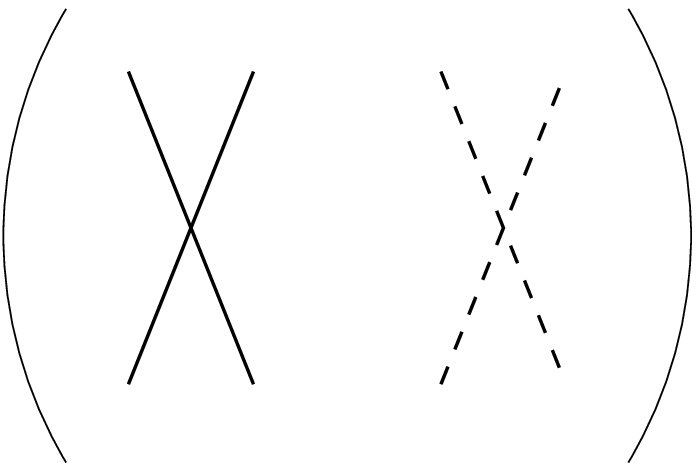,width=1.2cm,height=1cm} 
		}+
		\raisebox{-.4cm}{ 
		\psfig{file=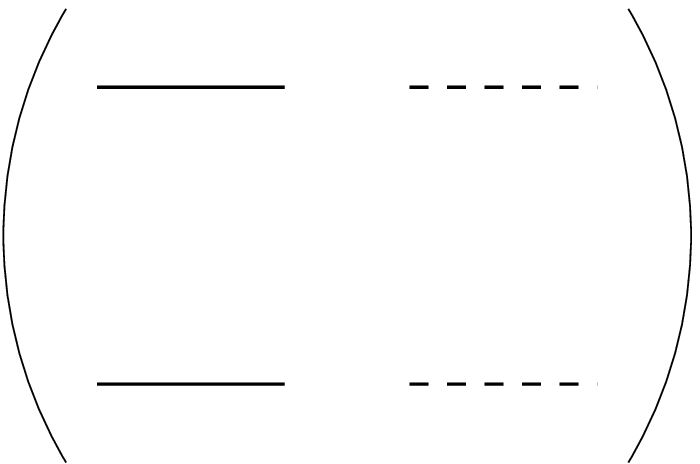,width=1.2cm,height=1cm} 
		}+
		\raisebox{-.4cm}{ 
		\psfig{file=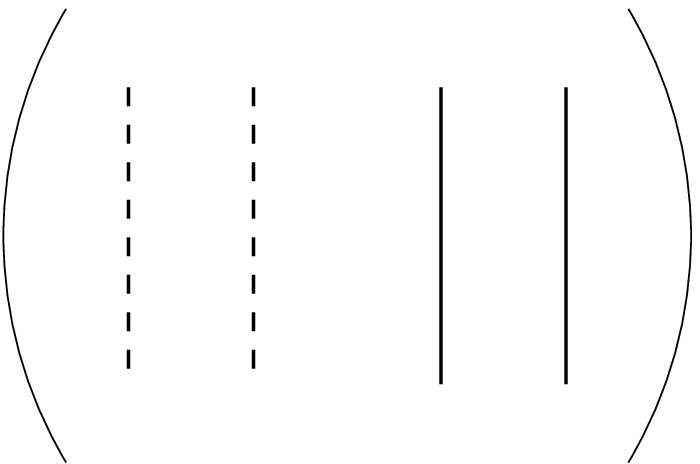,width=1.2cm,height=1cm} 
		}+
		\raisebox{-.4cm}{ 
		\psfig{file=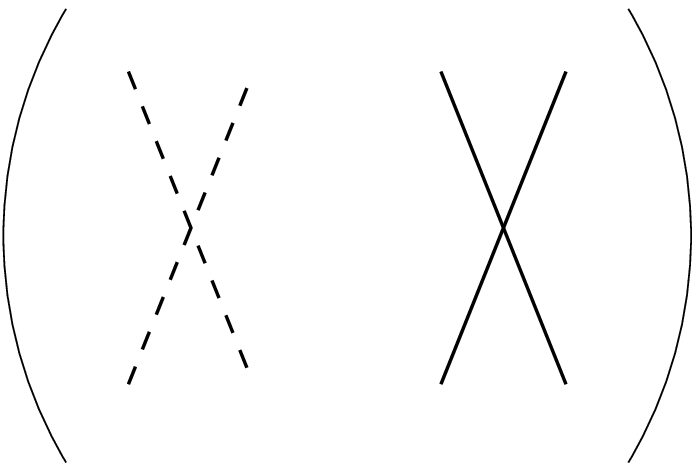,width=1.2cm,height=1cm} 
		}+
		\raisebox{-.4cm}{ 
		\psfig{file=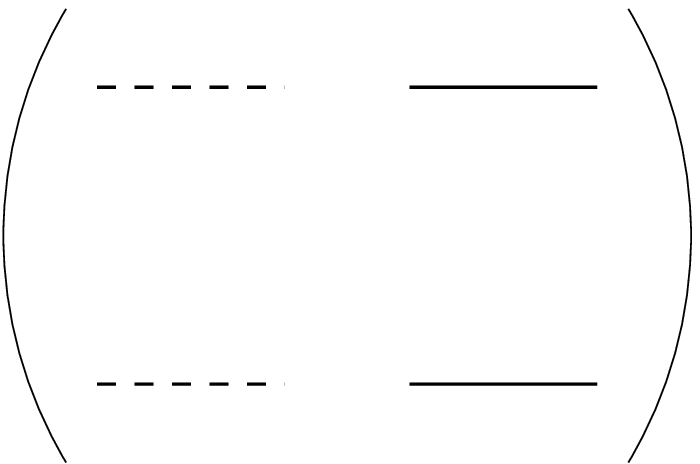,width=1.2cm,height=1cm} 
		}
	\right]  \\
	&+K_{12}\left[
		\raisebox{-.4cm}{ 
		\psfig{file=gr_int_files/dldl_v_dldl_x.eps,width=1.2cm,height=1cm} 
		}+
		\raisebox{-.4cm}{ 
		\psfig{file=gr_int_files/dldl_v_dldl_h.eps,width=1.2cm,height=1cm} 
		}+
		\raisebox{-.4cm}{ 
		\psfig{file=gr_int_files/dldl_x_dldl_v.eps,width=1.2cm,height=1cm} 
		}+
		\raisebox{-.4cm}{ 
		\psfig{file=gr_int_files/dldl_x_dldl_h.eps,width=1.2cm,height=1cm} 
		}+
		\raisebox{-.4cm}{ 
		\psfig{file=gr_int_files/dldl_h_dldl_v.eps,width=1.2cm,height=1cm} 
		}+
		\raisebox{-.4cm}{ 
		\psfig{file=gr_int_files/dldl_h_dldl_x.eps,width=1.2cm,height=1cm} 
		}
	\right] \\
	&+K_{42}\left[
		\raisebox{-.4cm}{ 
		\psfig{file=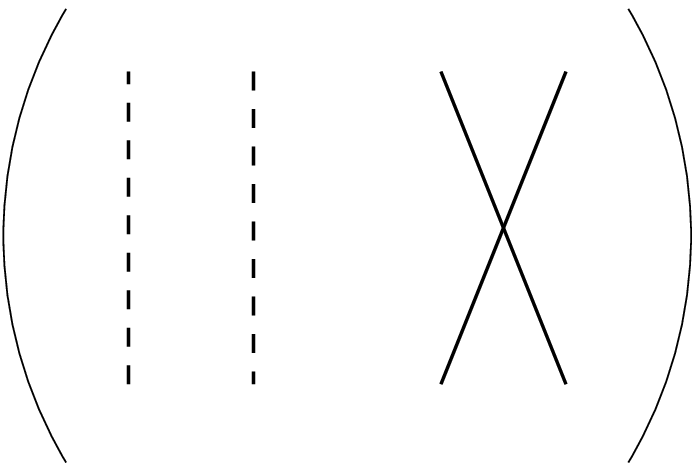,width=1.2cm,height=1cm} 
		}+
		\raisebox{-.4cm}{ 
		\psfig{file=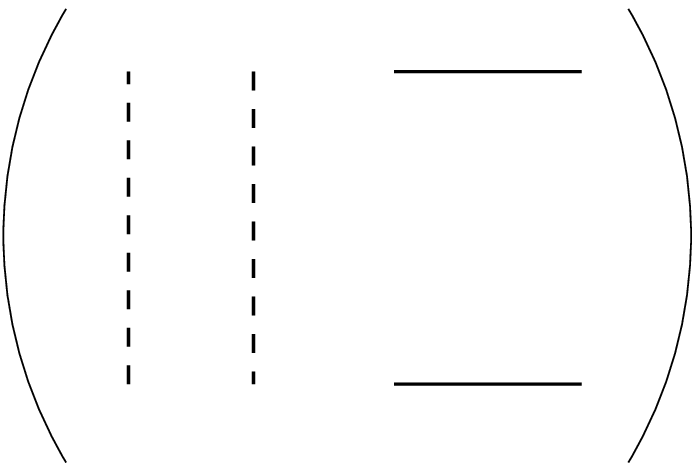,width=1.2cm,height=1cm} 
		}+
		\raisebox{-.4cm}{ 
		\psfig{file=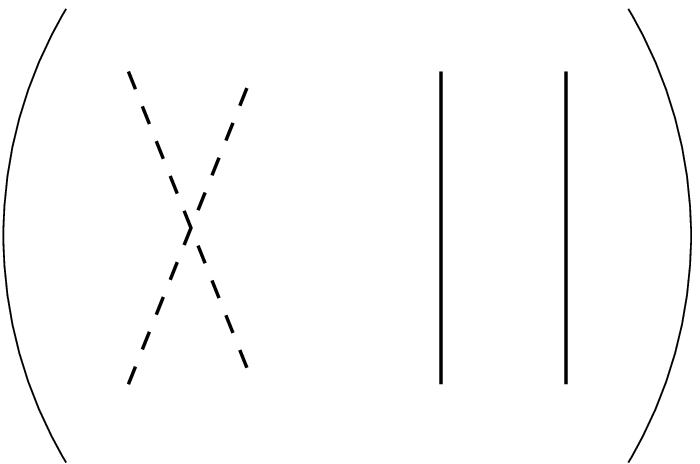,width=1.2cm,height=1cm} 
		}+
		\raisebox{-.4cm}{ 
		\psfig{file=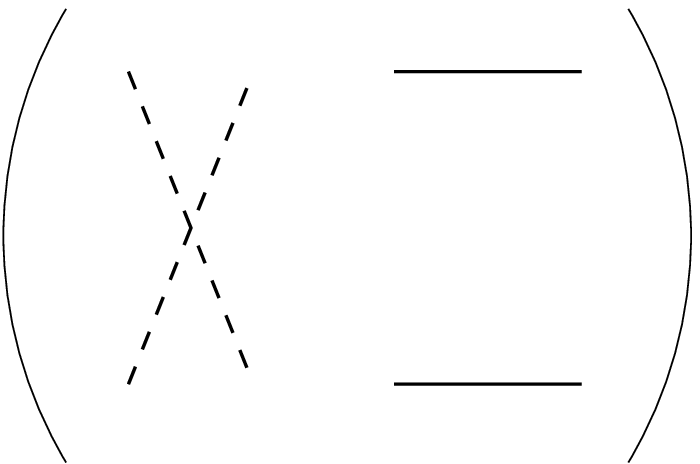,width=1.2cm,height=1cm} 
		}+
		\raisebox{-.4cm}{ 
		\psfig{file=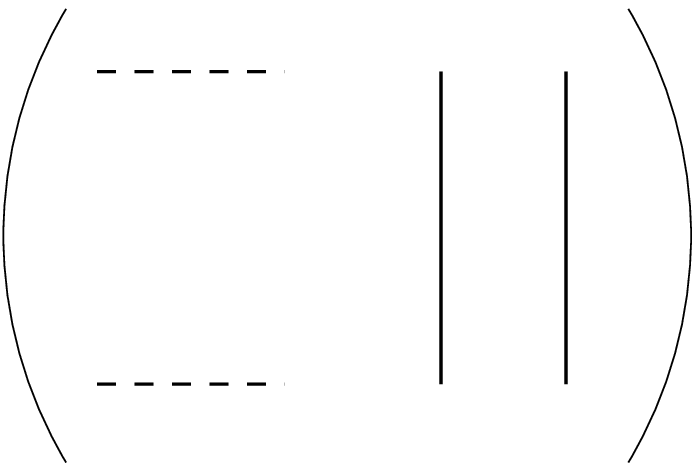,width=1.2cm,height=1cm} 
		}+
		\raisebox{-.4cm}{ 
		\psfig{file=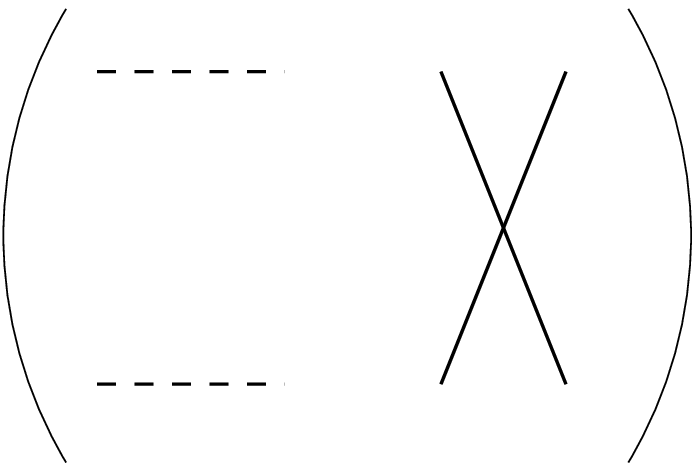,width=1.2cm,height=1cm} 
		}
	\right] \\
	&+K_{42}\left[
		\raisebox{-.4cm}{ 
		\psfig{file=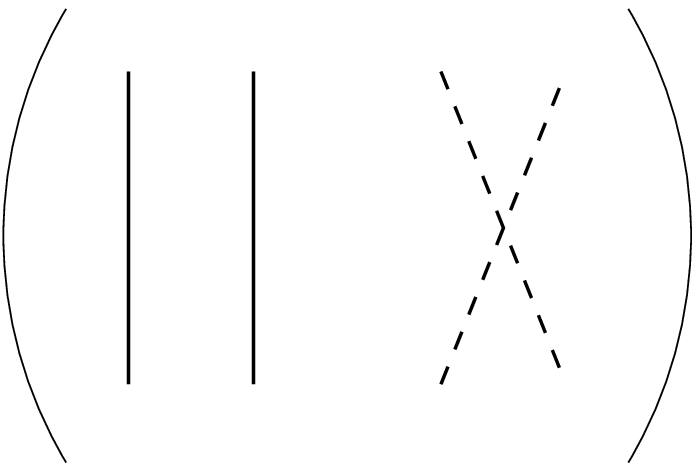,width=1.2cm,height=1cm} 
		}+
		\raisebox{-.4cm}{ 
		\psfig{file=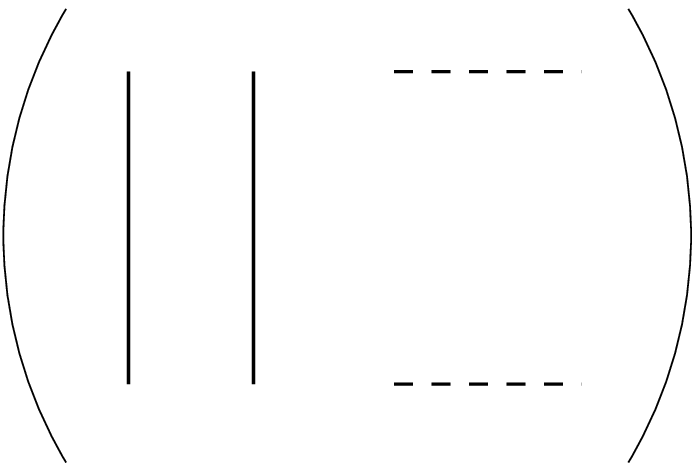,width=1.2cm,height=1cm} 
		}+
		\raisebox{-.4cm}{ 
		\psfig{file=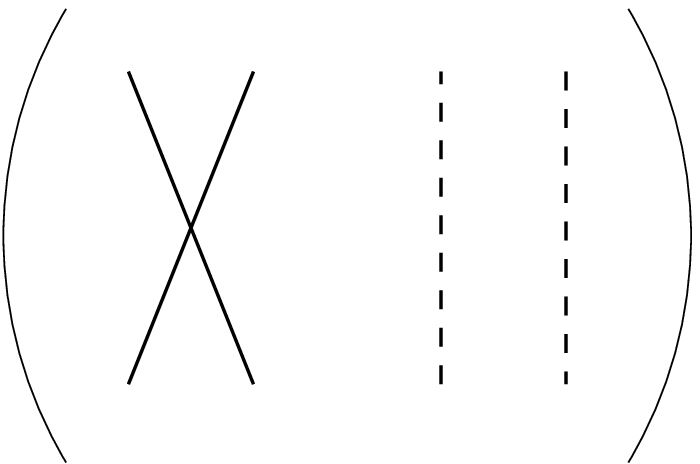,width=1.2cm,height=1cm} 
		}+
		\raisebox{-.4cm}{ 
		\psfig{file=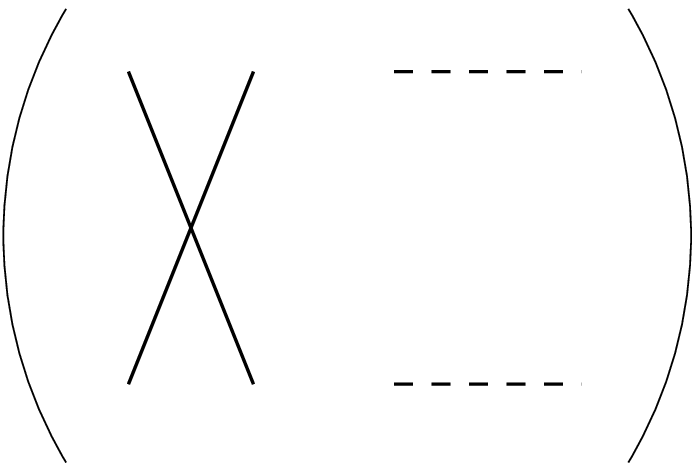,width=1.2cm,height=1cm} 
		}+
		\raisebox{-.4cm}{ 
		\psfig{file=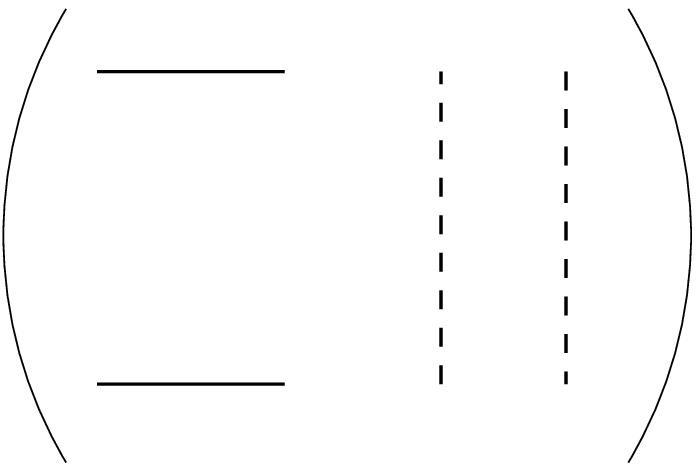,width=1.2cm,height=1cm} 
		}+
		\raisebox{-.4cm}{ 
		\psfig{file=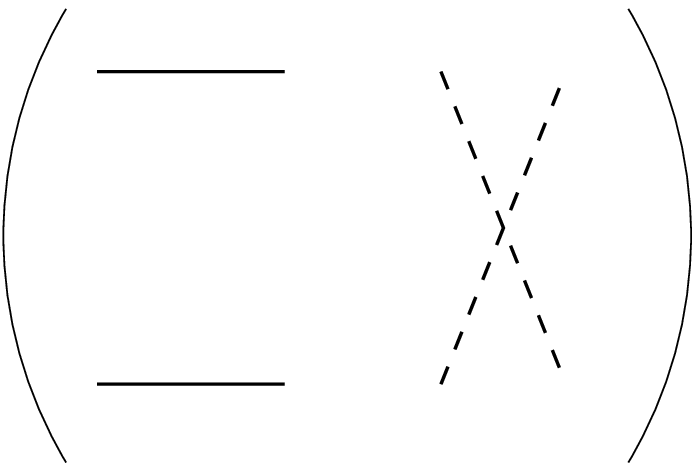,width=1.2cm,height=1cm} 
		}
	\right]\\
	&+K_{45}\left[
		\raisebox{-.4cm}{ 
		\psfig{file=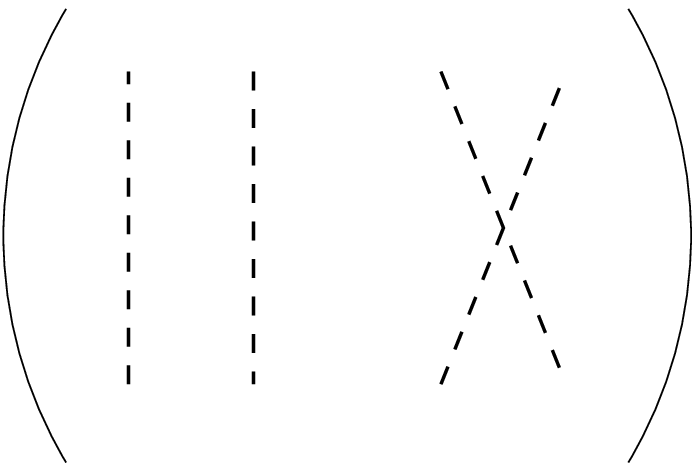,width=1.2cm,height=1cm} 
		}+
		\raisebox{-.4cm}{ 
		\psfig{file=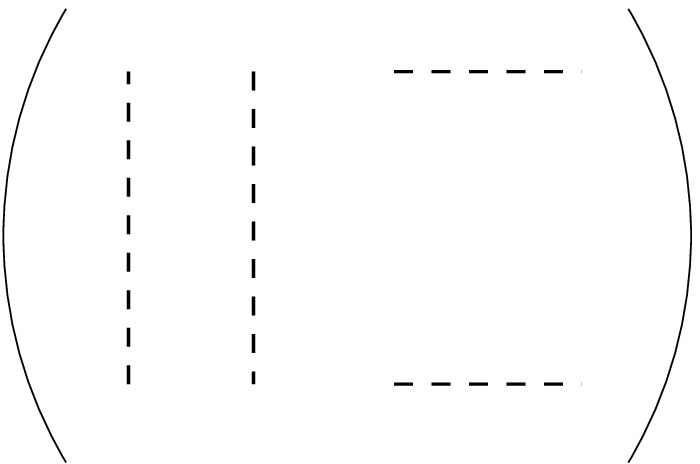,width=1.2cm,height=1cm} 
		}+
		\raisebox{-.4cm}{ 
		\psfig{file=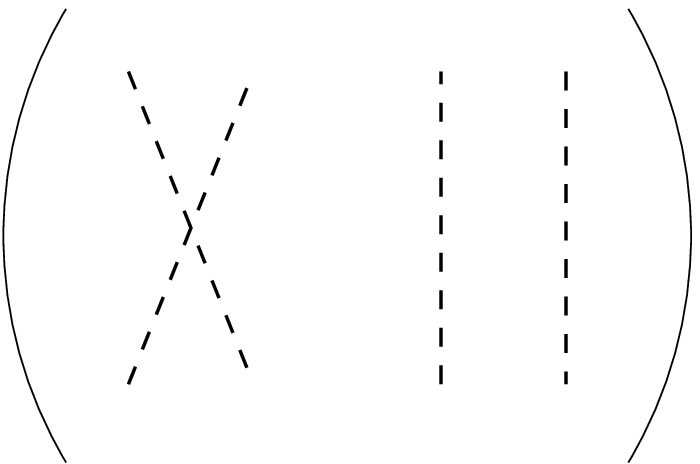,width=1.2cm,height=1cm} 
		}+
		\raisebox{-.4cm}{ 
		\psfig{file=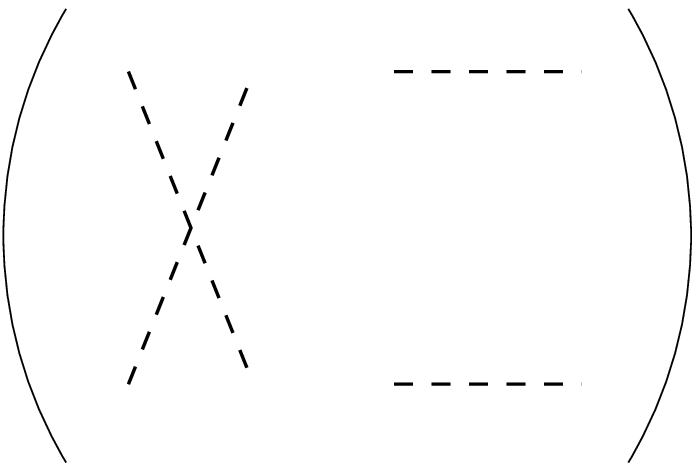,width=1.2cm,height=1cm} 
		}+
		\raisebox{-.4cm}{ 
		\psfig{file=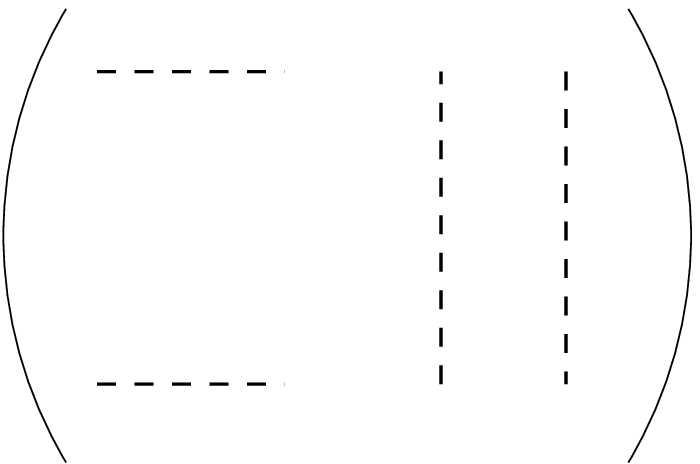,width=1.2cm,height=1cm} 
		}+
		\raisebox{-.4cm}{ 
		\psfig{file=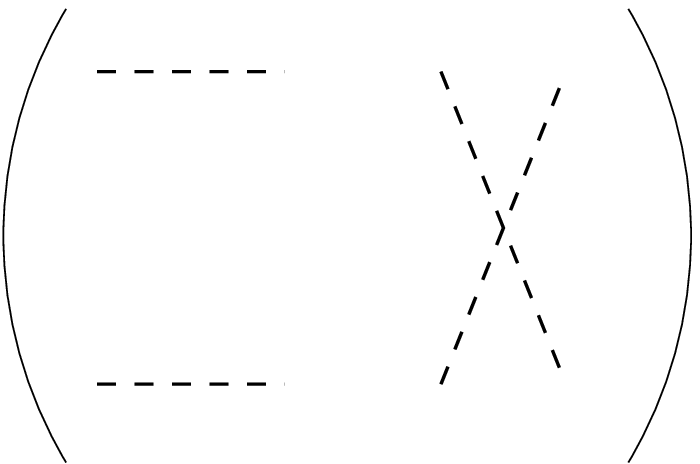,width=1.2cm,height=1cm} 
		}
	\right]\\
	&+K_6\left[
		3 \raisebox{-.4cm}{ 
		\psfig{file=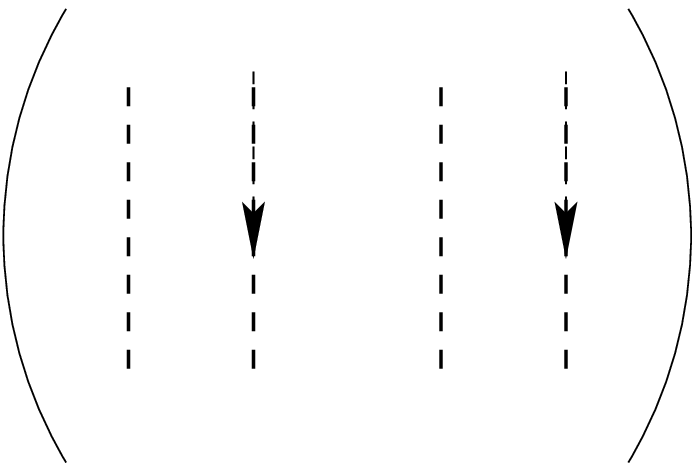,width=1.2cm,height=1cm} 
		}+
		3 \raisebox{-.4cm}{ 
		\psfig{file=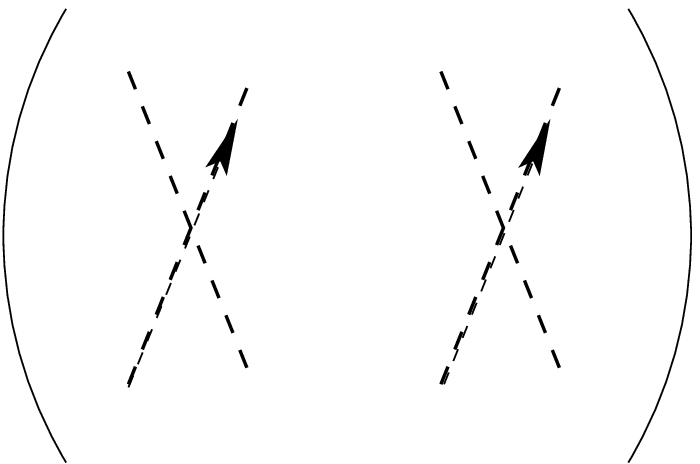,width=1.2cm,height=1cm} 
		}+
		3 \raisebox{-.4cm}{ 
		\psfig{file=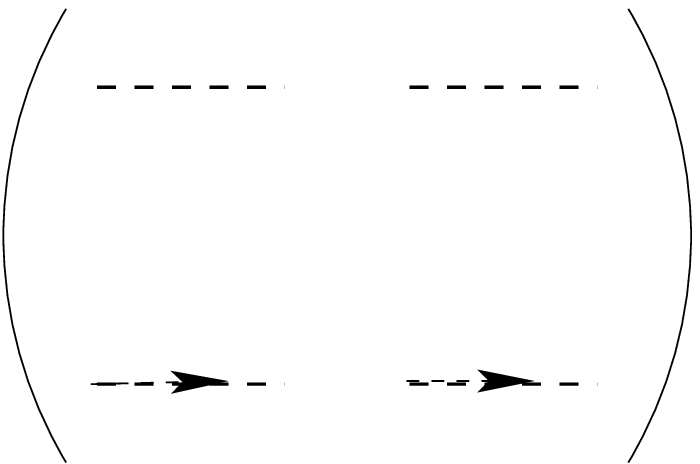,width=1.2cm,height=1cm} 
		}-
		\raisebox{-.4cm}{ 
		\psfig{file=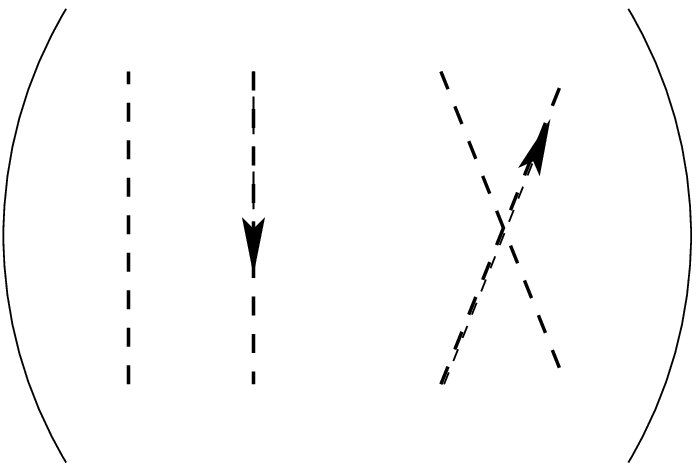,width=1.2cm,height=1cm} 
		}+
		\raisebox{-.4cm}{ 
		\psfig{file=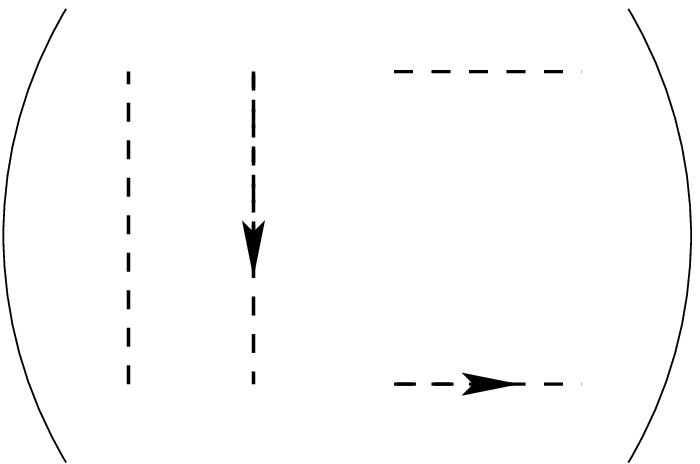,width=1.2cm,height=1cm} 
		}-
		\raisebox{-.4cm}{ 
		\psfig{file=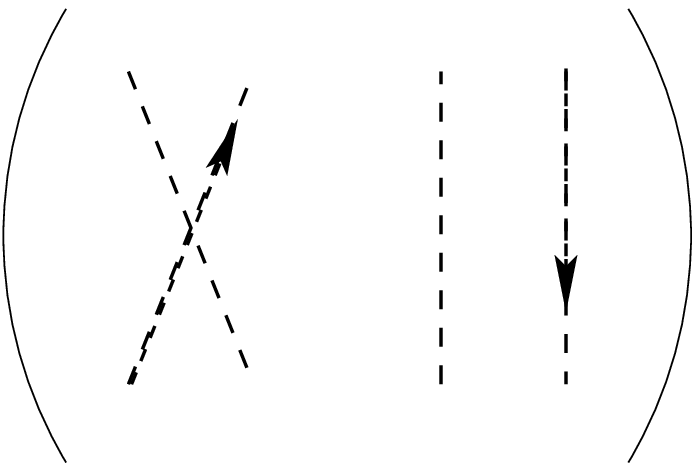,width=1.2cm,height=1cm} 
		} \right. \\
	&+\left.
		\raisebox{-.4cm}{ 
		\psfig{file=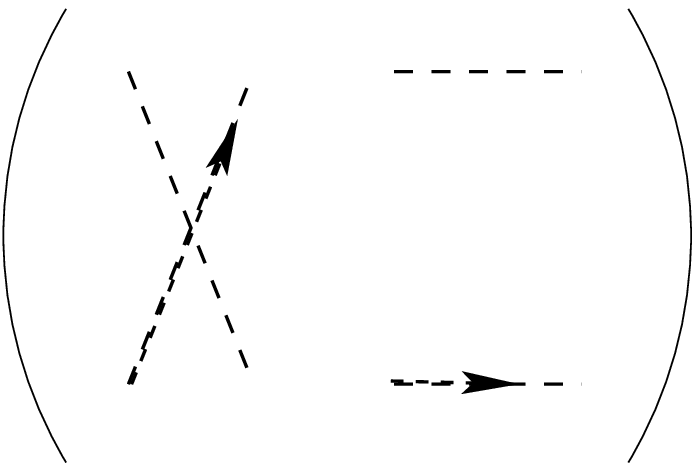,width=1.2cm,height=1cm} 
		}+
		\raisebox{-.4cm}{ 
		\psfig{file=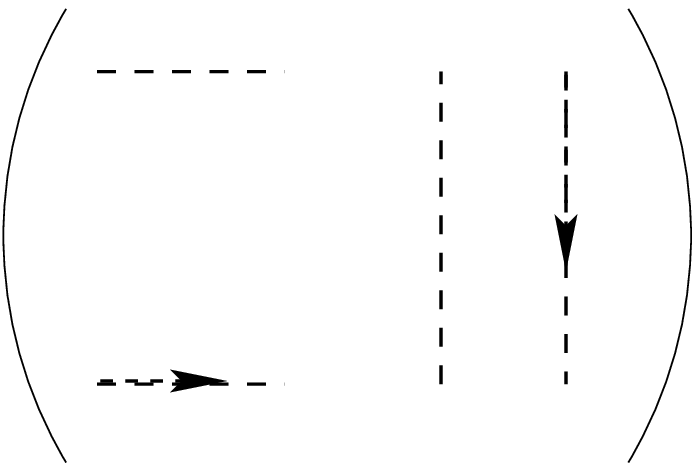,width=1.2cm,height=1cm} 
		}+
		\raisebox{-.4cm}{ 
		\psfig{file=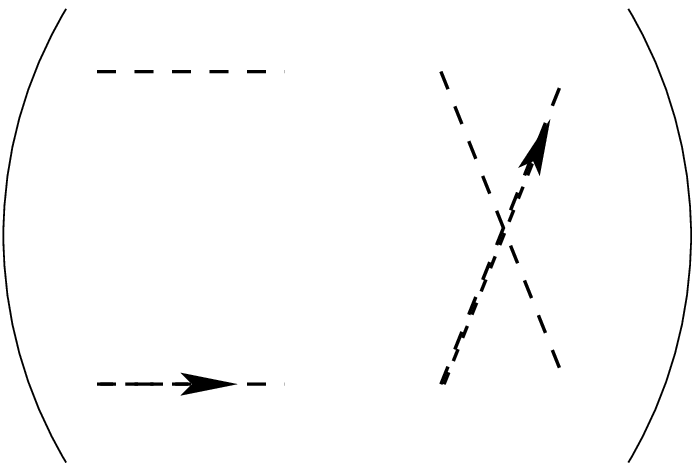,width=1.2cm,height=1cm} 
		}
	\right]\\
\end{split}
\eeq
where the coefficients for the case of $SU(N>3)$ are:
\beq
\begin{split}
K_{11} &= \frac{N^4-6N^2-24}{N^4(N^2-9)(N^2-1)}
\nonumber\\
K_{12} &= - \frac{ N^2 - 12}{N^4(N^2-9)(N^2-1)}\nonumber\\
K_{41} &= \frac{ N^2 - 12}{N^3(N^2-9)(N^2-1)}
\nonumber\\
K_{42} &= -  \frac{ N^2 - 6}{N^3(N^2-9)(N^2-1)}
\nonumber\\
\end{split}
\eeq
\beq
\begin{split}
K_{44} &= \frac{3N^4-29N^2+48}{2N^2(N^2-1)(N^2-4)(N^2-9)}
\\
K_{45} &= -\frac{N^4-15N^2+24}{2N^2(N^2-1)(N^2-4)(N^2-9)}
\\
K_6   &= \frac{1}{2(N^2-4)(N^2-1)}\\
\end{split}
\eeq
The subtlety for SU(3) relies on an extra, SU(3)-specific relation 
between d tensor components\cite{General}. Diagrammatically: 
\beq\label{d_rel_su3}
	\raisebox{-.4cm}{ 
	\psfig{file=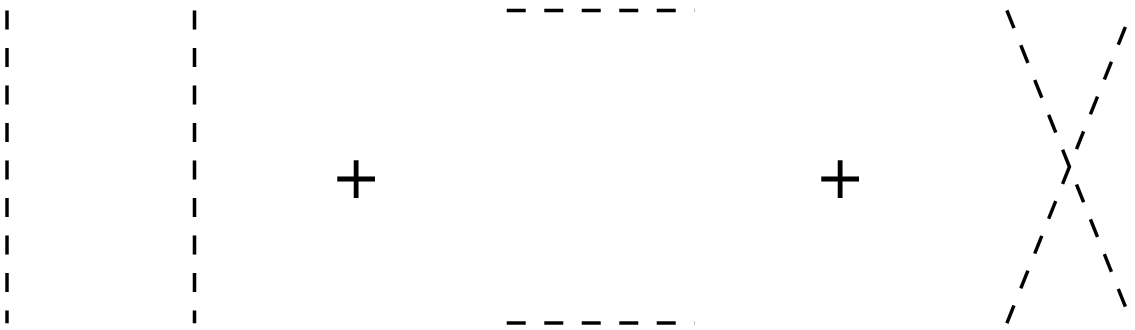,width=4cm,height=1cm} 
	} = 
	\frac{1}{3}\left(
	\raisebox{-.4cm}{ 
	\psfig{file=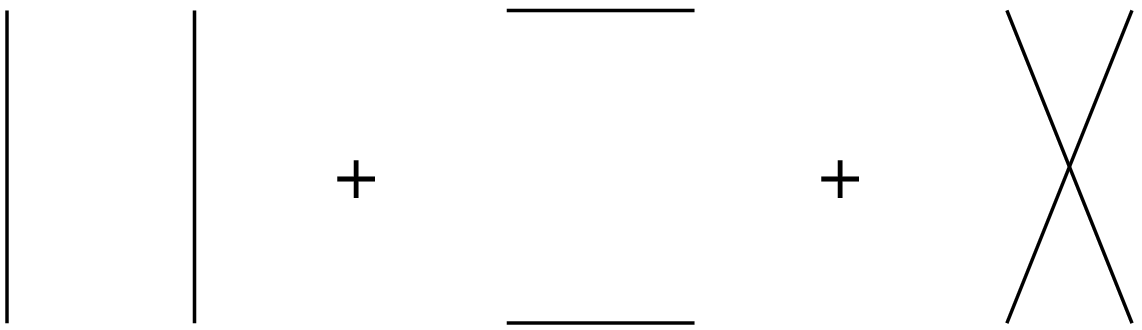,width=4cm,height=1cm} 
	}
	\right)\:\: .
\eeq

This reduces the basis of independent tensors to 8. One has to keep this 
in mind when making contractions to get the coefficients. The system of 
equations will then be under-determined and the parametric\footnote{Since 
$K_{42}$, $K_{45}$ cancel when one expresses all in terms 
of 
independent basis, we can as well choose them to be zero.}
solution for $SU(3)$ is:
\begin{eqnarray} 
K_{11} &=& \frac{91}{6480} -\frac{2}{3}K_{42} - \frac{1}{9}K_{45} 
\nonumber \\
K_{12} &=& -\frac{13}{2160} -\frac{2}{3}K_{42} - \frac{1}{9}K_{45} 
\nonumber \\
K_{41} &=& \frac{1}{108} + K_{42}
\nonumber \\
K_{44} &=& \frac{1}{180} + K_{45}
\nonumber \\
K_6&=&\frac{1}{80}
\end{eqnarray}

For higher $n=5,6\ldots$ one needs to take into consideration the 
analogous relations of the equation (\ref{d_rel_su3}) for SU(4), SU(5)... 
\section{Conclusions}

We have presented a 'down to earth' algorithm for evaluating group 
integrals. While lacking the elegance of more group-theoretic 
approaches(like character expansion), the strength of our method is in its 
simplicity, allowing one to reach the result in an easy and short way. We 
illustrated the method on examples of integrals over $SU(N)$ fundamental 
and adjoint representation, but the algorithm is not restricted to these 
particular groups. While some of the results for $\int du (u\bar{u})^n$ we 
derived have already been known, the results for $\int du (R^{ab})^4$, to 
our knowledge, have not been published.

\chapter{Summary and outlook}

We have studied instanton contributions to processes with unusual spin-flavor
structure. The first part of thesis focused on the decay of scalar and 
pseudoscalar charmonium and glueball. We found an internally consistent picture 
that agrees well with instanton phenomenology and lattice results.
The decay rates of studied processes are reproduced by instanton computation
with the average size of instanton $\bar{\rho}\cong 0.3\;{\rm fm}$. The ratios
of decay rates do not depend on the average size of instantons and agree, with
one exception, with the experimental values.

A better understanding of $\eta_c$ decay would come with more experimental data.
For example, it would be useful to compare our predictions with the data on
smaller decay channels $\eta_c\to K\bar{K}\eta,\; K\bar{K}\eta'$.
 The distribution of final state mesons would also be of interest, as we
could test the prediction of instanton-based isotropic production.

In the second part of our work we turned our attention to the problem of 
nucleon spin. The experimental findings point to a large OZI violation in the
flavor singlet axial vector channel. Therefore we studied the OZI violation 
in axial-vector two-point functions and axial vector coupling of quark
and nucleon. We found little reduction in the value of the flavor singlet
coupling of quark, $g^0_Q\cong 1$, while some suppression was present
in the coupling of the nucleon: $g_A^0 = 0.77$. However, the experimental 
value $g_A^0=(0.28-0.41)$ is significantly lower. Since the studied quark
two and three point functions do not show large OZI violation we conclude that
the structure of nucleon spin is given by another, possibly more complicated
mechanism which does not show on the quark level, but is specific to nucleon.

It would be interesting to study the connection to calculations of $g_A^0$
 based on the matrix element of the topological charge density $G\tilde{G}$.
One should also be able to get some insight from the lattice calculations of
OZI violation in the axial vector channel.

In the last part of present work we elaborated an algorithm for group 
integration and exemplified it on $SU(N)$ fundamental and adjoint 
representations. Although straight-forward, we hope the reader will
find it a useful gadget, handy but inexpensive, with no reason
 not to have it in one's toolbox.

\chapter{Appendix}

\begin{Appendix}{}
\section{Spectral Representation}
\label{app_spect_rep}
\subsection{Nucleon Two-Point Function}
\label{sec_app1}

 Consider the euclidean correlation function
\be
\Pi^{\alpha\beta}_N(x)= \langle \eta^\alpha(0)\bar{\eta}^\beta(x)\rangle,
\ee
where $\eta^\alpha(x)$ is a nucleon current and $\alpha$ is a 
Dirac index. We can write 
\be 
\Pi^{\alpha\beta}_N(x)=  \Pi_1(x) (\hat{x}\cdot\gamma)^{\alpha\beta}
 +\Pi_2(x)\delta^{\alpha\beta}.
\ee
The functions $\Pi_{1,2}(x)$ have spectral representations
\bea
\Pi_1(x) &=& \int_0^\infty ds\,\rho_1(s) D'(\sqrt{s},x),\\
\Pi_2(x) &=& \int_0^\infty ds\,\rho_2(s) D(\sqrt{s},x),
\eea
where $\rho_{1,2}(s)$ are spectral functions and
\bea
D(m,x) &=& \;\frac{m}{4\pi^2x}K_1(mx) ,\\
D'(m,x)&=& - \frac{m^2}{4\pi^2x} K_2(mx) ,
\eea
are the euclidean coordinate space propagator of a scalar 
particle with mass $m$ and its derivative with respect to
$x$. The contribution to the spectral function arising from 
a nucleon pole is 
\be
\rho_1(s) = |\lambda^2_N| \delta(s-m_N^2), \hspace{1cm}
\rho_2(s) = |\lambda^2_N| m_N\delta(s-m_N^2),
\ee
where $\lambda_N$ is the coupling of the nucleon to the current,
$\langle 0 |\eta|N(p)\rangle = \lambda_N u(p)$, and $m_N$ is the 
mass of the nucleon. It is often useful to consider the point-to-plane 
correlation function
\be
K^{\alpha\beta}_N(\tau) = \int d^3x\, \Pi^{\alpha\beta}_N(\tau,
 \vec{x}).
\ee
The integral over the transverse plane insures that all intermediate 
states have zero three-momentum. The nucleon pole contribution 
to the point-to-plane correlation function is 
\be
K^{\alpha\beta}_N(\tau) = \frac{1}{2}(1+\gamma_4)^{\alpha\beta}
 |\lambda_N|^2 \exp(-m_N\tau).
\ee

\subsection{Scalar Three-Point Functions}
\label{sec_app2}

 Next we consider three-point functions. Before we get to 
three-point functions of spinor and vector currents we consider 
a simpler case in which the spin structure is absent. We study 
the three-point function of two scalar fields $\phi$ and a scalar 
current $j$. We define
\be 
\label{3pt_scal}
\Pi(x,y) = \langle \phi(0) j(y)\phi(x) \rangle .
\ee
The spectral representation of the three-point function is 
complicated and in the following we will concentrate on the 
contribution from the lowest pole in the two-point function
of the field $\phi$. We define the coupling of this state
to the field $\phi$ and the current $j$ as
\bea
\langle 0|\phi(0)|\Phi(p)\rangle &=& \lambda, \\
\langle \Phi(p')|j(0)|\Phi(p)\rangle &=& F(q^2),
\eea
where $F(q^2)$ with $q=p-p'$ is the scalar form factor. The 
pole contribution to the three-point function is
\be
\label{3pts_spec}
\Pi(x,y) = \lambda^2 \int d^4z\, D(m,y+z)F(z)D(m,x-y-z),
\ee
where $D(m,x)$ is the scalar propagator and $F(z)$ is 
the Fourier transform of the form factor. In order to 
study the momentum space form factor directly it is 
convenient to integrate over the location of the endpoint 
in the transverse plane and Fourier transform with 
respect to the midpoint 
\be 
\int d^3x\,\int d^3y\, e^{iqy}
 \langle \phi(0) j(\tau/2,\vec{y})\phi(\tau,\vec{x}) \rangle 
 =\frac{\lambda^2}{(2m)^2}\exp(-m\tau)F(q^2).
\ee
The correlation function directly provides the form factor
for space-like momenta. Maiani and Testa showed that there 
is no simple procedure to obtain the time-like form factor
from euclidean correlation functions \cite{Maiani:ca}.
 
 Form factors are often parametrized in terms of monopole, dipole, 
or monopole-dipole functions
\bea 
F_M(q^2)    &=& F_M(0)\frac{m_V^2}{Q^2+m_V^2}, \\
F_D(q^2)    &=& F_D(0)\left(\frac{m_V^2}{Q^2+m_V^2}\right)^2, \\
F_{MD}(q^2) &=& F_{MD}(0)\frac{m_1^2}{Q^2+m_1^2}
 \left(\frac{m_2^2}{Q^2+m_2^2}\right)^2 ,
\eea
with $Q^2=-q^2$. For these parametrization the Fourier transform 
to euclidean coordinate space can be performed analytically. We find
\bea
 F_M(x) &=& m_V^2 D(x,m_V) \\
 F_D(x) &=& m_V^2\left(
    -\frac{x}{2} D'(x,m_V)-D(x,m_V)\right)\\
 F_{MD}(x) &=& \frac{m_1^2m_2^4}{m_2^2-m_1^2}
 \left\{ \frac{1}{m_2^2-m_1^2} 
      \left( D(x,m_1)-D(x,m_2) \right)\right. \nonumber \\
 & & \left.\hspace{2cm}\mbox{}+ \frac{1}{m_2^2} 
  \left( \frac{x}{2} D'(x,m_2) + D(x,m_2) \right) \right\}.
\eea
We also consider three-point functions involving a vector 
current $j_\mu$. The matrix element is 
\be
\langle \Phi(p')|j_\mu(0)|\Phi(p)\rangle = q_\mu F(q^2).
\ee
The pole contribution to the vector current three-point 
function is 
\be
\label{3pts_v_spec}
\Pi_\mu(x,y) = \lambda^2 \int d^4z\, D(m,y+z)\hat{z}_\mu 
     F'(z)D(m,x-y-z),
\ee
with $F'(z)=dF(z)/dz$ and $\hat{z}_\mu=z_\mu/|z|$. For the 
parameterizations given above the derivative of the coordinate 
space form factor can be computed analytically. We get 
\bea
 F_M'(x) &=& m_V^2 D'(x,m_V) \\
 F_D'(x) &=& -\frac{m_V^4}{2} D(x,m_V) \\
 F_{MD}'(x) &=& \frac{m_1^2m_2^4}{m_2^2-m_1^2}
 \left\{ \frac{1}{m_2^2-m_1^2} 
      \left( D'(x,m_1)-D'(x,m_2) \right)\right. \nonumber \\
 & & \left.\hspace{2cm}\mbox{}
    + \frac{x}{2} D(x,m_2)  \right\}.
\eea

\subsection{Nucleon three-point functions}
\label{sec_app3}

 Next we consider three-point functions of the nucleon involving vector 
and axial-vector currents. The vector three-point function is 
\be
(\Pi^a_{VNN})^{\alpha\beta}_\mu(x,y) = 
\langle \eta^\alpha(0)V^a_\mu(y)\bar{\eta}^\beta(x)\rangle .
\ee
The axial-vector three-point function is defined analogously. 
The nucleon pole contribution involves the nucleon coupling 
to the current $\eta$ and the nucleon matrix element of the 
vector and axial vector currents. The vector current matrix 
element is 
\be
\langle N(p')|V_\mu^a|N(p)\rangle = 
 \bar{u}(p')\left[ F_1(q^2)\gamma_\mu
 +\frac{i}{2M}F_2(q^2) \sigma_{\mu\nu}q^\nu \right]
  \frac{\tau^a}{2}u(p),
\ee
where the form factors $F_{1,2}$ are related to the 
electric and magnetic form factors via
\bea
G_E(q^2) &=& F_1(q^2) +\frac{q^2}{4M^2}F_2(q^2), \\
G_M(q^2) &=& F_1(q^2) + F_2(q^2).
\eea
The axial-vector current matrix element is 
\be
\langle N(p')|A_\mu^a|N(p)\rangle = 
 \bar{u}(p')\left[ G_A(q^2)\gamma_\mu
 +\frac{1}{2M}G_P(q^2)(p'-p)_\mu \right]\gamma_5
  \frac{\tau^a}{2}u(p)
\ee
where $G_{A,P}$ are the axial and induced pseudo-scalar
form factors. 

 We are interested in extracting the vector and 
axial-vector coupling constants $g_V=F_1(0)$ and
$g_A=G_A(0)$. In order to determine the vector coupling
$g_V$ we study the three-point function involving 
the four-component of the vector current in the 
euclidean time direction. For simplicity we take
$y=x/2$. We find that 
\be
(\Pi_{VNN})^{\alpha\beta}_4(x,x/2) = 
   \Pi_{VNN}^{1}(\tau) \delta^{\alpha\beta}
 + \Pi_{VNN}^{2}(\tau) (\gamma_4)^{\alpha\beta},
\ee
where $x_\mu=(\vec{0},\tau)$. The two independent 
structures $\Pi_{VNN}^{1,2}$ are given by 
\bea 
\Pi_{VNN}^{1}(\tau) &=& |\lambda_N|^2\int d^4y\,
 \left\{ 
     \frac{\tau+2y_4}{2x_1}m D'(x_1)D(x_2)F_1(y)
      \right. 
    \nonumber\\
    &+&\left. 
    \frac{\tau-2y_4}{2x_2}m D(x_1)D'(x_2)F_1(y) 
     +\frac{\tau|\vec{y}|}{x_1x_2}D'(x_1)D'(x_2)
     \frac{F'_2(y)}{2m} \right\},\nonumber \\
\Pi_{VNN}^{2}(\tau) &=&  |\lambda_N|^2\int d^4y\,
 \left\{  \left[
   \frac{\tau^2-4y_4^2+4\vec{y}^2}{4x_1x_2} D'(x_1)D'(x_2) 
     +m^2D(x_1)D(x_2)  \right]
      \right. \nonumber \\
 & & \x F_1(y) +\left.
     |\vec{y}| \left[ 
     \frac{m}{x_1}D'(x_1)D(x_2) 
    +\frac{m}{x_2}D'(x_2)D(x_1) 
       \right] \frac{F'_2(y)}{2m} \right\},\nonumber
\eea
where $F_{1,2}(y)$ are the Fourier transforms of the 
the Dirac form factors $F_{1,2}(q^2)$, and we have 
defined $x_{1}=(\vec{y},\tau/2+y_4)$ and $x_2=
(-\vec{y},\tau/2-y_4)$. 

 In order to extract the axial-vector coupling we study 
three-point functions involving spatial components of the 
axial-vector current. We choose the three-component of the 
current and again take $y=x/2$ with $x_\mu=(\vec{0},\tau)$. 
We find
\be
(\Pi_{ANN})^{\alpha\beta}_3(x,x/2) = 
   \Pi_{ANN}^{1}(\tau) (\gamma_5)^{\alpha\beta}
 + \Pi_{ANN}^{2}(\tau) (\gamma_3\gamma_5)^{\alpha\beta}
 + \Pi_{ANN}^{3}(\tau) (\gamma_3\gamma_4\gamma_5)^{\alpha\beta},
\ee
with 
\bea
\Pi_{ANN}^{1}(\tau) &=& |\lambda_N|^2\int d^4y \,
    my_3\left[ 
     \frac{\tau+2y_4}{2x_1} D'(x_1)D(x_2)G_A(y)
	\right. \nonumber\\
& & \hspace{1cm}
	\left.
    -\frac{\tau-2y_4}{2x_2} D(x_1)D'(x_2)G_A(y) \right],\;\;\;\;\;\;\;\;\; \\
\Pi_{ANN}^{2}(\tau) &=&  |\lambda_N|^2\int d^4y\,
 \left\{  
   \frac{\tau^2+8y_3^2-4y^2}{4x_1x_2} D'(x_1)D'(x_2)G_A(y) 
 \right.
  \nonumber\\
& & \hspace{-2.3cm} \left.
     +m^2D(x_1)D(x_2)G_A(y) 
     +\frac{y_3^2}{|\vec{y}|} \left[ 
     \frac{m}{x_1}D'(x_1)D(x_2) 
    +\frac{m}{x_2}D'(x_2)D(x_1) 
       \right] \frac{G'_P(y)}{2m} \right\},\nonumber\\
\eea
where $G_{A,P}(y)$ are the Fourier transforms of the 
nucleon axial and induced pseudo-scalar form factors.

 These results are quite complicated. The situation 
simplifies if we consider three-point functions 
in which we integrate all points over their location 
in the transverse plane. The vector three-point function
is 
\be
\int d^3x\int d^3y\, (\Pi_{VNN})^{\alpha\beta}_4
 (\tau,\vec{x};\tau/2,\vec{y}) = 
    \frac{g_V}{2}(1+\gamma_4)^{\alpha\beta}
    |\lambda_N|^2 \exp(-m_N\tau), 
\ee
where $g_V=F_1(0)$ is the vector coupling. Note that 
the three-point function of the spatial components 
of the current vanishes when integrated over the 
transverse plane. The axial-vector three-point function
is 
\be
\int d^3x\int d^3y\, (\Pi_{ANN})^{\alpha\beta}_3
 (\tau,\vec{x};\tau/2,\vec{y}) = \frac{g_A}{2}
    \left((1+\gamma_4)\gamma_3\gamma_5\right)^{\alpha\beta}
    |\lambda_N|^2 \exp(-m_N\tau), 
\ee
where $g_A=G_A(0)$ is the axial-vector coupling. In
the case of the axial-vector current the three-point 
function of the time component of the current vanishes
when integrated over the transverse plane. This is why 
we consider three-point function involving the spatial 
components of the axial current. 

\subsection{Phenomenology}
\label{sec_app4}

\begin{figure}
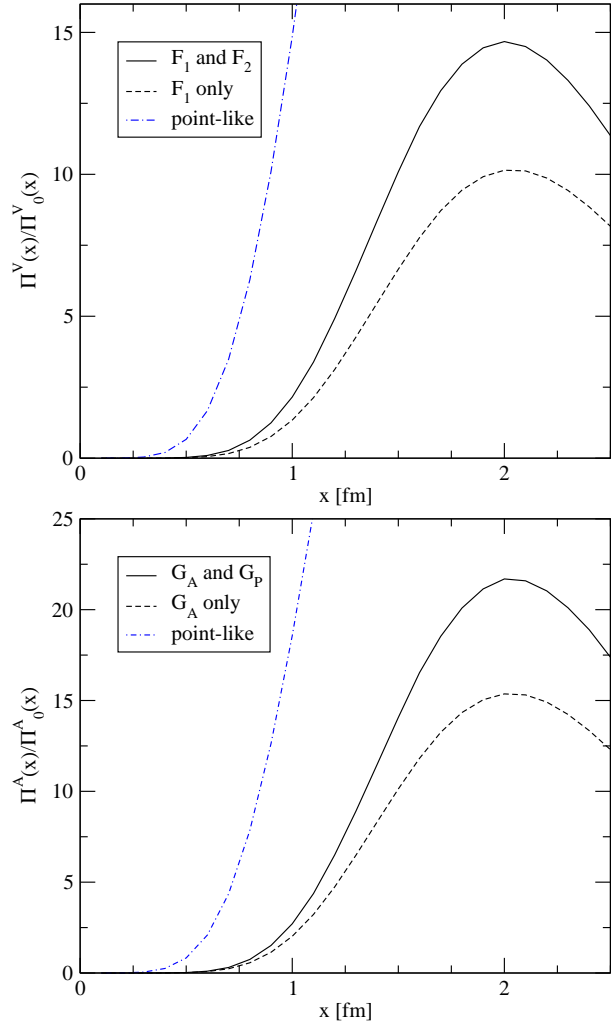

\begin{center}
\leavevmode
\includegraphics[width=8cm,angle=0,clip=true]{spin/pheno_gv.eps}
\includegraphics[width=8cm,angle=0,clip=true]{spin/pheno_ga.eps}
\end{center}
\caption{\label{fig_pheno}
Nucleon pole contribution to the vector (upper panel) and
axial-vector (lower panel) nucleon three-point function.
The solid line shows the complete results, the dashed line
is the contribution from the $F_1$ and $G_A$ form factors
only, and the dash-dotted line corresponds to a point-like
nucleon. We have used a nucleon coupling constant $\lambda=2.2\,
{\rm fm}^{-3}$ as well as phenomenological values for the
form factors and coupling constants.}
\end{figure}
 In Fig.~\ref{fig_pheno} we show the nucleon pole contribution
to the vector and axial-vector three-point functions. We have
used the phenomenological values of the iso-vector coupling 
constants
\be
\begin{array}{rclcrcl}
 G_E(0) &=& 1,&\hspace{1cm} & G_M(0) &=& 4.7, \\
 G_A(0) &=& 1.25, & &         G_P(0) &=& \frac{4M^2}{m_\pi^2}g_A.
\end{array}
\ee 
We have parametrized $G_{E,V}$ and $G_A$ by dipole
functions with $m_V=0.88$ GeV  and $m_A=1.1$ GeV. The induced 
pseudoscalar form factor is parametrized as a pion propagator 
multiplied by a dipole form factor with dipole mass $m_A$. 
\begin{figure}
\begin{center}
\leavevmode
\includegraphics[width=6cm,angle=0,clip=true]{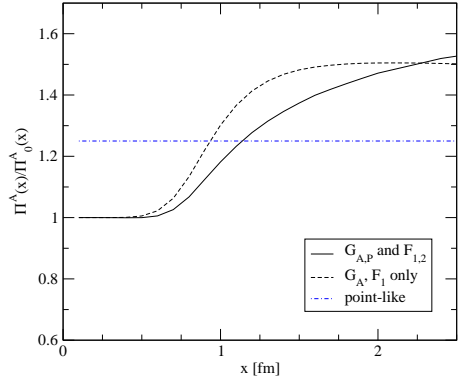}
\end{center}
\caption{\label{fig_pheno_rat}
Ratio of the phenomenological parameterizations of the
axial-vector and vector three-point functions. We have
added a short distance continuum contribution to the
nucleon pole terms. The curves are labeled as in the
previous figure. Note that both the solid and the dashed
line will approach $g_A=1.25$ as $x\to\infty$. Also
note that the solid line is in very good agreement with
the instanton calculation shown in Fig.~\ref{fig_gan}.}
\end{figure}

 We observe that at distances that are accessible in lattice 
or instanton simulations, $x\sim (1-2)$ fm, the typical momentum 
transfer is not small and the correlation function is substantially 
reduced as compared to the result for a point-like nucleon. We also 
observe that the $F_2$ and $G_P$ form factors make substantial 
contributions. Fig.~\ref{fig_pheno_rat} shows that the ratio of 
the axial-vector and vector correlation functions is nevertheless 
close to the value for a point-like nucleon, $g_A/g_V\simeq 1.25$.
We observe that the ratio of point-to-point correlation functions 
approaches this value from above. This is related to the fact 
that the axial radius of the nucleon is smaller than the vector 
radius. As a consequence, the point-to-point correlation function
at finite separation $\tau$ ``sees'' a larger fraction of the 
axial charge as compared to the vector charge. 

\end{Appendix}
\newpage

\begin{Appendix}{}
\section{Instanton contribution to quark three-point functions}
\label{sec_app5}

 In this appendix we provide the results for the traces 
that appear in the single instanton contribution to the 
quark three-point function. Our starting point is the 
expression 
\be 
(P^{con}_{A/VQQ})^{\mu\nu} =
  P_{NZNZ}^{\mu\nu} + c_{A/V}
   \left(P^{\mu\nu}_{ZMm} + P^{\mu\nu}_{mZM}
   \right),
\ee
see equ.~(\ref{pi_3pt_z/nz}). Due to the Dirac structure of the non-zero 
mode part of the propagator, the $NZNZ$ term is the same for both 
the vector and axial-vector correlation functions. It has 4 parts 
stemming from combinations of the two terms of non-zero mode propagator 
equ.~\ref{NZ_prop}
\bea
P_{NZNZ_{11}}^{\mu\nu} &=& \frac{2H(x,z,y)(x-z)^\alpha(z-y)^\beta}
  {\pi^4(x-z)^4(z-y)^4}
S^{\nu\alpha\mu\beta} \nonumber \\
& &\x 
\left[
  1+\frac{\rho^2}{z^2}
  \left(\frac{x\cdot z}{x^2} + \frac{z \cdot y}{y^2}
  \right) +
  \frac{\rho^4 x\cdot y}{x^2 z^2 y^2} 
\right] ,\\
P_{NZNZ_{12}}^{\mu\nu} &=& \frac{H(x,z,y)(x-z)^\alpha y^\beta}
   {2\pi^4(x-z)^4(z-y)^2z^2 y^2}
  [(z^{\alpha_0} + \frac{\rho^2}{x^2} x^{\alpha_0} ) 
  S^{\alpha_0\sigma\sigma_0\beta}
  \pm \frac{\rho^2}{x^2}x^{\alpha_0}\epsilon^{\alpha_0\sigma\sigma_0\beta}
  ]\nonumber\\
& &\times\left[
   \frac{\rho^2(z-y)^{\sigma_0}}{\rho^2+z^2}(S^{\nu\alpha\mu\sigma}
    \pm\epsilon^{\nu\alpha\mu\sigma})
    +\frac{\rho^2(z-y)^{\sigma}}{\rho^2+y^2}(S^{\nu\alpha\mu\sigma_0}
   \mp\epsilon^{\nu\alpha\mu\sigma_0})
\right],\nonumber\\ \\
P_{NZNZ_{21}}^{\mu\nu} &=& \frac{H(x,z,y)(z-y)^\alpha x^\beta}
   {2\pi^4(x-z)^2(z-y)^4z^2 x^2}
   [(z^{\alpha_0} + \frac{\rho^2}{y^2} y^{\alpha_0} ) 
   S^{\beta\sigma\sigma_0\alpha_0}
   \pm \frac{\rho^2}{y^2}y^{\alpha_0}\epsilon^{\beta\sigma\sigma_0\alpha_0} ]
 \nonumber \\
& &\times\left[
 \frac{\rho^2(x-z)^{\sigma_0}}{\rho^2+x^2}
  (S^{\nu\sigma\mu\alpha}\pm\epsilon^{\nu\sigma\mu\alpha})
  +\frac{\rho^2(x-z)^{\sigma}}{\rho^2+z^2}
   (S^{\nu\sigma_0\mu\alpha}\mp\epsilon^{\nu\sigma_0\mu\alpha})
    \right],\nonumber\\
\eea
\bea
P_{NZNZ_{22}}^{\mu\nu} &=& 
    \frac{H(x,z,y) x^{\alpha_0} y^{\beta} }
  {4\pi^4(x-z)^2(z-y)^2 z^2 y^2 x^2}
  \frac{\rho^4}{(\rho^2+z^2)}
 T_{\pm}[\alpha_0,\alpha,\sigma,\alpha_1,\sigma_1,\beta]\nonumber\\
& &\x 
  \left[\frac{(x-z)^\sigma (z-y)^{\sigma_1}}{(\rho^2+x^2)}
(S^{\nu\alpha\mu\alpha_1} \pm \epsilon^{\nu\alpha\mu\alpha_1}) 
\right. \nonumber\\
& & \hspace{1cm}+\left.
\frac{(x-z)^\alpha (z-y)^{\alpha_1}}{(\rho^2+y^2)}
(S^{\nu\sigma\mu\sigma_1} \mp \epsilon^{\nu\sigma\mu\sigma_1}) 
\right],
\eea
where 
\be
H(x,z,y)=\left\{  \left( 1+\frac{\rho^2}{z^2} \right)
 \sqrt{ \left(1+\frac{\rho^2}{x^2}\right) 
        \left(1+\frac{\rho^2}{y^2}\right) }  \right\}^{-1}
\ee
and the Dirac trace $T_{\pm}$ is defined as
\bea
\label{DTrace6mp}
 T_{\mp}[\mu,2,3,4,5,6] &\equiv&
 \left(
  g^{\mu 2}S^{3456} - g^{\mu 3}S^{2456} + g^{\mu 4} S^{2356} - g^{\mu 5
  }S^{2346} + g^{\mu 6} S^{2345}
 \right)
 \nonumber \\
&& \hspace{-2.9cm}\mp   \left(
  g^{\mu 2}\epsilon^{3456} - g^{\mu 3}\epsilon^{2456} +
  g^{23}\epsilon^{\mu 456} + g^{45}\epsilon^{\mu 236} -
  g^{46}\epsilon^{\mu 235} + g^{56}\epsilon^{\mu 234}
 \right) ,
\eea
where $2,3,\ldots$ is short for $\mu_2,\mu_3,\ldots$.
The $S_{ZM}S_m$ term is easily seen to be
\bea
P^{\mu\nu}_{ZMm}&=&{\rm Tr}[-\Psi_0(x)\Psi_0^{+}(z)\gamma^{\mu}
(-\Delta_{\pm}(z,y)\gamma_{\pm})\gamma^{\nu}]\nonumber\\
&=& \frac{\varphi(x)\varphi(z)x^{\alpha_0}z^{\beta_0}}
{8\pi^2(z-y)^2}
\frac{1}
{\sqrt{(1+\frac{\rho^2}{z^2})(1+\frac{\rho^2}{y^2})}}
T_{\mp}[\alpha_0,\sigma_1,\sigma,\beta_0,\mu,\nu]
\nonumber\\
& & \times
\left[
\delta^{\sigma_1,\sigma}+\frac{\rho^2z^\alpha y^\beta}{z^2y^2}
(S^{\sigma_1\sigma\alpha\beta}\pm\epsilon^{\sigma_1\sigma\alpha\beta}
 )\right]
\eea
with $\varphi(x)=\rho/(\pi\sqrt{x}(x^2+\rho^2)^{3/2})$.
The $S_mS_{ZM}$ term is obtained similarly
\bea
P^{\mu\nu}_{mZM}&=&
{\rm Tr}[(-\Delta_{\pm}(x,z)\gamma_{\pm})\gamma^{\mu}
(-\Psi_0(z)\Psi_0^{+}(y))\gamma^{\nu}]\nonumber\\
&=& \frac{\varphi(z)\varphi(y)z^{\alpha_0}y^{\beta_0}}
{8\pi^2(x-z)^2}
\frac{1}
{\sqrt{(1+\frac{\rho^2}{x^2})(1+\frac{\rho^2}{z^2})}}
T_{\mp}[\alpha_0,\sigma_1,\sigma,\beta_0,\nu,\mu]
\times
\nonumber\\
& & \times
\left[
\delta^{\sigma_1,\sigma}+\frac{\rho^2x^\alpha z^\beta}{x^2z^2}
(S^{\sigma_1\sigma\alpha\beta}\pm\epsilon^{\sigma_1\sigma\alpha\beta}
 )\right].
\eea
For all the above formulas, we need to add instanton and anti-instanton
contribution and integrate over the position of the instanton, which was 
suppressed above. As usual, for an instanton at position $z_I$ we have
to shift $x,y,z$ according to $x\to (x-z_I)$, etc.

\subsection{Computation of path exponent}


The correlation functions of operators at different points are made 
gauge invariant by introducing the connector, the Wilson line.  
In this appendix we compute the Wilson line in instanton background.

In order to perform the computation, we need to evaluate the path exponent
\be
P.e.=Pexp(i\int_{x1}^{x2} A^{\nu,a}\tau^a dx^\nu)
\ee
For a general gauge configuration that is not an easy task as the path exponent
is defined as follows.  Parametrize the path by $s\in [0,1]$. The integral
in the exponent is then a limit of a sum of matrices, which might not commute. 
Expand the exponential in power series and reorder the matrices in each term
in decreasing value of parameter s. The sum obtained this way is the well-known Wilson line. 

For the path exponent along a straight line in the field of an instanton the computation is actually much simpler. Let us parametrize the line by $x^\mu = x_1^\mu + s(x_2-x_1)^\mu$, so that the end points $x_1$ and $x_2$ correspond to $s=0$ and $s=1$ respectively.
The integral then reads:
\bea
\int_{x_1}^{x_2} A^{a,\mu}_\pm \tau^a dx^\mu &=&
\int_0^1 \ebe^a_{\mu\nu}\tau^a \frac{\rho^2\, x^\nu(s)}{x^2(s)(x^2(s)+\rho^2)}
\frac{dx^\mu}{ds} ds\nonumber\\
&=&
\int_0^1 \ebe^a_{\mu\nu}\tau^a 
\frac{\rho^2
\left[x_1^\nu(x_2-x_1)^\mu + s(x_2-x_1)^\nu (x_2-x_1)^\mu
\right]}
{(x_1+s(x_2-x_1))^2(x_1+s(x_2-x_1))^2+\rho^2)}
ds , \nonumber
\eea
with $\tau^a$ Pauli matrices.

The second term in the numerator vanishes due to the anti-symmetry of t'Hooft's tensor $\eta^a_{\mu\nu}=-\eta^a_{\nu\mu}$.
Then the integrand is in fact a constant matrix multiplied by a function of $s$
 at any point on the line. As a result, no path-ordering is necessary since the matrices at any point commute. Therefore we can integrate first and then expand the exponential to get:
\bea
\int_{x_1}^{x_2}A^{a,\mu}_{\pm} \tau^a dx^\mu 
&\equiv& \tau^a K^a_{\pm}
=\tau^a \ebe^a_{\mu\nu} x_1^\nu (x_2-x_1)^\mu
 \nonumber\\
&\x& \left\{
\Omega_1(x_1,x_2)
\left[
\tan^{-1}\left( x_2\cdot(x_2-x_1)\Omega_1(x_1,x_2)\right)
 \right.
 \right.
 \nonumber\\
& & \hspace{1.2cm}
\left.
-
\tan^{-1}\left( x_1\cdot(x_2-x_1)\Omega_1(x_1,x_2) \right)
\right]
\nonumber\\
&-&
\Omega_2(x_1,x_2,\rho)
\left[
\tan^{-1}\left(x_2\cdot(x_2-x_1)\Omega_2(x_1,x_2,\rho)\right)
\right.
\nonumber\\
& & \hspace{1.2cm}
\left.
\left.
-
\tan^{-1}\left(x_1\cdot(x_2-x_1)\Omega_2(x_1,x_2,\rho)\right)
\right]
\right\}
\eea
where
\bea
\Omega_1(x_1,x_2)&=&\frac{1}{\sqrt{x_1^2(x_2-x_1)^2 - (x_1\cdot(x_2-x_1))^2}}
 \nonumber\\
\Omega_2(x_1,x_2,\rho)&=&
\frac{1}{\sqrt{(\rho^2 + x_1^2)(x_2-x_1)^2 - (x_1\cdot(x_2-x_1))^2}}
 \nonumber
\eea
The expansion of the exponent of Pauli matrices is then easy:
\bea
Pexp
\left(
i\int_{x_1}^{x_2} A^{a,\mu}_{\pm}\tau^a dx^{\mu} 
\right) 
&=&
\cos\left[
\left|
K_{\pm}(x_1,x_2)
\right|
\right]\nonumber\\
& & \hspace{-2.85cm}
+ i \sin \left[
\left|
K_{\pm}(x_1,x_2)
\right|
\right]
\Omega_1(x_1,x_2)
\ebe^a_{\mu\nu}x_1^\nu(x_2-x_1)^\mu
\tau^a
\eea
where the modulus of $K^a(x_1,x_2)$ is 
\bea
\left|
K(x_1,x_2)
\right|
&=&\sqrt{K^aK^a}\nonumber\\
&=&
\left\{
\tan^{-1}\left(x_2\cdot(x_2-x_1)\Omega_1(x_1,x_2)\right)
\right.
\nonumber\\
& & \hspace{1.5cm}-
\tan^{-1}\left( x_1\cdot(x_2-x_1)\Omega_1(x_1,x_2)\right)
\nonumber\\
& &-
\frac{\Omega_2(x_1,x_2,\rho)}{\Omega_1(x_1,x_2)}
\left(
\tan^{-1}\left( x_2\cdot(x_2-x_1)\Omega_2(x_1,x_2,\rho) \right)
\right.
\nonumber\\
& &\hspace{1.5cm}-
\left.
\left.
\tan^{-1}\left( x_1\cdot(x_2-x_1)\Omega_2(x_1,x_2,\rho) \right)
\right)
\right\}
\eea
In the above formulas we took the instanton to be centered at 0. The generalization to instanton center at z is straight-forwardly obtained by the replacement 
$x_1\to x_1-z$ and $x_2\to x_2-z$.

For example, for the case of an infinite line in 4-direction we get:
\bea
Pexp
\left(
i\int_{-\infty}^{+\infty} A^{a,4}_{\pm}\tau^a dx^{4}
\right)
&=& 
\cos{ 
\left[
\pi (1-\frac{|\vec{z}|}{\sqrt{|\vec{z}|^2 + \rho^2}} )
\right]
}\nonumber\\
&-& 
i \sin{
\left[ \pi (1-\frac{|\vec{z}|}{\sqrt{|\vec{z}|^2 + \rho^2}} )
\right]
}
\frac{\vec{z}}{|\vec{z}|}\cdot \vec{\tau}
\eea
in agreement with formula (206) of \cite{Schafer:1996wv}.
\end{Appendix}
\newpage

\begin{Appendix}{}
\section{Euclidean matrices, conventions, ..}

We use the following conventions:
\bea
g^{\mu\nu}&=&\delta^{\mu\nu}=diag(1,1,1,1)=g_{\mu\nu},\nonumber\\
\epsilon^{1234}&=&\epsilon_{1234}=+1 \nonumber
\eea
The contractions of $\epsilon$ therefore are
\be
\epsilon_{\rho\nu\alpha\beta}
\epsilon^{\rho\mu\sigma\sigma'}
=\left |
\begin{tabular}{c c c}
$\delta^{\mu}_\nu$ & $\delta^{\mu}_\alpha$ & $\delta^{\mu}_\beta$\\
$\delta^{\sigma}_\nu$ & $\delta^{\sigma}_\alpha$ & 
$\delta^{\sigma}_\beta$ \\
$\delta^{\sigma'}_\nu$ & $\delta^{\sigma'}_\alpha$ & 
$\delta^{\sigma'}_\beta$
\end{tabular}
\right |
\ee
and $\epsilon_{\rho\rho'\alpha\beta} \epsilon^{\rho\rho'\mu\nu}
=2(\delta^\mu_\alpha \delta^\nu_\beta -
\delta^\nu_\alpha \delta^\mu_\beta )$,
$\epsilon_{\rho\rho'\alpha\beta} 
\epsilon^{\rho\rho'\alpha\nu}=6\delta^\nu_\beta$
and $\epsilon_{\rho\rho'\alpha\beta} \epsilon^{\rho\rho'\alpha\beta}=24$

Euclidean gamma matrices are defined as
$\{\gamma^\mu,\gamma^\nu\}=2g^{\mu\nu}$. We take
$\gamma^5=\gamma^1\gamma^2\gamma^3\gamma^4$ and 
$\sigma^{\mu\nu}=\frac{1}{2}[\g^\mu,\g^\nu]$.
The following relations apply: $\gamma_\mu^\dag = \gamma_\mu$,
$\gf\sigma^{\mu\nu}=
-\frac{1}{2}\epsilon^{\mu\nu\alpha\beta}\sigma_{\alpha\beta}$
One can easily compute the traces:
\bea
Tr[\gamma^\mu\gamma^\nu] &=& 4g^{\mu\nu}\nonumber \\
Tr[\gamma^\mu\gamma^\sigma\gamma^\nu\gamma^\rho] &=& 4 
(g^{\mu\sigma}g^{\nu\rho} -g^{\mu\nu}g^{\rho\sigma} +
g^{\mu\rho}g^{\sigma\nu} ) \equiv 4 S^{\mu\sigma\nu\rho}\nonumber \\
Tr[\gamma^\mu\gamma^\sigma\gamma^\nu\gamma^\rho\gamma^5] &=& 4 
\epsilon^{\mu\sigma\nu\rho} \nonumber \\
Tr[\gamma^\mu\gamma^\sigma\gamma^\rho\gamma^{\rho'}\gamma^{\sigma'}\gamma^\nu] 
&=& 4 \left[
g^{\mu\sigma}S^{\rho\rho'\sigma'\nu} - 
g^{\mu\rho}S^{\sigma\rho'\sigma'\nu} + 
g^{\mu\rho'}S^{\sigma\rho\sigma'\nu} 
\right. \nonumber\\
& &\hspace{1.5cm}
\left.
- 
g^{\mu\sigma'}S^{\sigma\rho\rho'\nu} + 
g^{\mu\nu}S^{\sigma\rho\rho'\sigma'}  
\right] \nonumber \\
Tr[\gamma^\mu\gamma^\sigma\gamma^\rho\gamma^{\rho'}\gamma^{\sigma'}\gamma^\nu
\gamma^5] 
&=& 4 \left[
\epsilon^{\mu\sigma\rho\nu'}S^{\nu'\rho'\sigma'\nu} + 
S^{\mu\sigma\rho\nu'}\epsilon^{\nu'\rho'\sigma'\nu} 
\right] 
\eea
Some useful decomposition:(for transparency we will short our notation 
from $\g^{\mu_1}\to\g^1$)
\be\label{DTrace}
\g^1\g^2\g^3\g^4\g^5\g^6\gmp =
  \gmp(g^{1\mu} + \sigma^{1\mu})T_{\mp}[\mu,2,3,4,5,6]
\ee
where the Dirac trace $T_{\mp}$ was defined in \ref{DTrace6mp}.
The four vector of $SU(2)_c$ matrices is
$\sigma^\mu_{\pm}=(\vec{\sigma},\mp i)$ while  
the color - rotated matrices are given as  
$\sigma^\mu_{\pm}(R)=(R^{ab}{\sigma}^b,\mp i)$. Here $R^{ab}$ is the 
adjoint representation rotation matrix, defined as $R^{ab}\sigma^b = 
U\sigma^aU^\dag$, with $U\in SU(2)$.
The following relations link the $SU(2)$ generators to t'Hooft $\eta$ 
tensor:
\bea
\sigma^\mu_+\sigma^\nu_- &=& g^{\mu\nu} + i\eta^{a,\mu\nu}R^{ab}\sigma^b 
\nonumber \\
\sigma^\mu_-\sigma^\nu_+ &=& g^{\mu\nu} + 
i\bar{\eta}^{a,\mu\nu}R^{ab}\sigma^b
\eea
The definition and useful properties of $\eta$ can be found e.g. in 
\cite{Schafer:1996wv}. For our computations suffices to list the most 
often used formulas:
\bea
\ebe^a_{\mu\nu}\ebe^a_{\rho\lambda}&=&\delta_{\mu\rho}\delta_{\nu\lambda} 
-\delta_{\mu\lambda}\delta_{\nu\rho}\mp 
\epsilon_{\mu\nu\rho\lambda} \\
\epsilon^{abc}\ebe^a_{\alpha\beta}\ebe^b_{\mu\nu}\ebe^c_{\rho\lambda}&=&
\delta_{\mu\rho}\ebe^a_{\alpha\beta}\ebe^a_{\nu\lambda} 
- \delta_{\mu\lambda}\ebe^a_{\alpha\beta}\ebe^a_{\nu\rho}
+\delta_{\nu\lambda}\ebe^a_{\mu\rho}\ebe^a_{\alpha\beta}
-\delta_{\nu\rho}\ebe^a_{\mu\lambda}\ebe^a_{\alpha\beta}\nonumber
\eea
where $\ebe$ is $\bar{\eta}$ ($\eta$) for upper (lower) sign.

After tracing the 
color indices, all the rotation matrices cancel due to orthogonality
$R^{ab}R^{cb}=\delta^{ac}$ and unimodularity $\det R =1$. The following 
traces are therefore also 
valid for color-rotated matrices.  
\be
Tr[\sigma^\mu_+\sigma^\nu_-]=Tr[\sigma^\mu_-\sigma^\nu_+]=2\delta^{\mu\nu} 
\ee
\be
Tr[\sigma^\mu_\pm\sigma^\nu_\mp\sigma^\rho_\pm\sigma^\sigma_\mp]=
2(S^{\mu\nu\rho\sigma} \mp \epsilon^{\mu\nu\rho\sigma})
\ee
As one can easily see, the traces of $SU(2)$ matrices 
$Tr[\sigma^\mu_\pm\cdots]$ 
can be obtained from the traces of corresponding $\gamma$ matrices 
projected to the particular 2x2 submatrix as follows:
$Tr_c[\sigma^\mu_\pm\cdots\sigma^\nu_\mp] = Tr_D[(\gamma^\mu \cdots 
\gamma^\nu)\gmp]$

We also provide 
some useful decompositions for the multiplication of color matrices:
\be
\sigma^\rho_\mp\sigma^\sigma_\pm\sigma^\mu_\mp\sigma^\nu_\pm =
\delta^{\rho\sigma}\delta^{\mu\nu} - \ebe^{a,\rho\sigma}\ebe^{a,\mu\nu}
+i\sigma^a 
\Bigl[
\ebe^{a,\rho\sigma}\delta^{\mu\nu}+\ebe^{a,\mu\nu}\delta^{\rho\sigma} -
\ebe^{b,\rho\sigma}\ebe^{c,\mu\nu}\epsilon^{bca}
\Bigr]
\ee
\bea
\sigma^\mu_\mp\sigma^\nu_\pm \sigma^\alpha_\mp\sigma^\beta_\pm
\sigma^\rho_\mp\sigma^\sigma_\pm &=&
{\bf{1}}\Bigl\{
 \delta^{\mu\nu}\delta^{\alpha\beta}\delta^{\rho\sigma} 
 -\delta^{\mu\nu}\ebe^{a,\alpha\beta}\ebe^{a,\rho\sigma}
 \Bigr. \nonumber\\
 & &\hspace{0.5cm} -\;\Bigl.
  \ebe^{c,\mu\nu}
   \Bigl[\ebe^{c,\alpha\beta}\delta^{\rho\sigma}
     +\ebe^{c,\rho\sigma}\delta^{\alpha\beta}
     -\ebe^{a,\alpha\beta}\ebe^{b,\rho\sigma}\epsilon^{abc}
   \Bigr]
\Bigr\}\nonumber \\
& &+i\sigma^b
  \Bigl\{
   \ebe^{b,\mu\nu}(\delta^{\alpha\beta}\delta^{\rho\sigma}
         - \ebe^{a,\alpha\beta}\ebe^{a,\rho\sigma})
  \Bigr.
  \nonumber\\
& &\hspace{0.5cm}-\;
   \Bigl.
    \epsilon^{ecb}\ebe^{e,\mu\nu}
      [\ebe^{c,\alpha\beta}\delta^{\rho\sigma} 
        +\ebe^{c,\rho\sigma}\delta^{\alpha\beta}
        -\ebe^{a,\alpha\beta}\ebe^{a',\rho\sigma}\epsilon^{aa'c}]
  \Bigr\}\nonumber\\
\eea

\end{Appendix}


\begin{thebibliography}{99}


\bibitem{Belavin:fg}
A.~A.~Belavin, A.~M.~Polyakov, A.~S.~Shvarts and Y.~S.~Tyupkin,
Phys.\ Lett.\ B {\bf 59}, 85 (1975).

\bibitem{'tHooft:fv}
G.~'t Hooft,
Phys.\ Rev.\ D {\bf 14}, 3432 (1976)
[Erratum-ibid.\ D {\bf 18}, 2199 (1978)].

\bibitem{Yung:1987zp}
A.~V.~Yung,
Nucl.\ Phys.\ B {\bf 297}, 47 (1988).

\bibitem{Balitsky:qn}
I.~I.~Balitsky and A.~V.~Yung,
Phys.\ Lett.\ B {\bf 168}, 113 (1986).

\bibitem{Aoyama:1995ca}
H.~Aoyama, T.~Harano, M.~Sato and S.~Wada,
Nucl.\ Phys.\ B {\bf 466}, 127 (1996)
[arXiv:hep-th/9512064].

\bibitem{Affleck:1980mp}
I.~Affleck,
Nucl.\ Phys.\ B {\bf 191}, 429 (1981).

\bibitem{Nielsen}
N.K.Nielsen, M.Nielsen,
Phys.\ Rev.\ D {\bf 61}, 105020 {2000}

\bibitem{Faccioli:2001ug}
P.~Faccioli and E.~V.~Shuryak,
Phys.\ Rev.\ D {\bf 64}, 114020 (2001)
[arXiv:hep-ph/0106019].

\bibitem{Novikov:dq}
V.~A.~Novikov, L.~B.~Okun, M.~A.~Shifman, A.~I.~Vainshtein, M.~B.~Voloshin 
and V.~I.~Zakharov,
Phys.\ Rept.\  {\bf 41}, 1 (1978).

\bibitem{Shifman:nx}
M.~A.~Shifman,
Z.\ Phys.\ C {\bf 4}, 345 (1980)
[Erratum-ibid.\ C {\bf 6}, 282 (1980)].

\bibitem{Appelquist:zd}
T.~Appelquist and H.~D.~Politzer,
Phys.\ Rev.\ Lett.\  {\bf 34}, 43 (1975);
A.~De Rujula and S.~L.~Glashow,
Phys.\ Rev.\ Lett.\  {\bf 34}, 46 (1975).

\bibitem{Bodwin:1994jh}
G.~T.~Bodwin, E.~Braaten and G.~P.~Lepage,
Phys.\ Rev.\ D {\bf 51}, 1125 (1995)
[Erratum-ibid.\ D {\bf 55}, 5853 (1997)]
[hep-ph/9407339].

\bibitem{Gottfried:1977gp}
K.~Gottfried,
Phys.\ Rev.\ Lett.\  {\bf 40}, 598 (1978).

\bibitem{Voloshin:hc}
M.~B.~Voloshin,
Nucl.\ Phys.\ B {\bf 154}, 365 (1979).

\bibitem{Brodsky:1981kj}
S.~J.~Brodsky and G.~P.~Lepage,
Phys.\ Rev.\ D {\bf 24}, 2848 (1981).

\bibitem{Chernyak:1983ej}
V.~L.~Chernyak and A.~R.~Zhitnitsky,
Phys.\ Rept.\  {\bf 112}, 173 (1984).

\bibitem{Anselmino:yg}
M.~Anselmino, M.~Genovese and D.~E.~Kharzeev,
Phys.\ Rev.\ D {\bf 50}, 595 (1994)
[hep-ph/9310344].

\bibitem{Bjorken:2000ni}
J.~D.~Bjorken,
preprint, hep-ph/0008048.

\bibitem{Balitsky:1993jd}
I.~I.~Balitsky and V.~M.~Braun,
Phys.\ Lett.\ B {\bf 314} (1993) 237
[hep-ph/9305269].

\bibitem{Moch:1996bs}
S.~Moch, A.~Ringwald and F.~Schrempp,
Nucl.\ Phys.\ B {\bf 507} (1997) 134
[hep-ph/9609445].

\bibitem{Kochelev:2001pp}
N.~I.~Kochelev and V.~Vento,
Phys.\ Rev.\ Lett.\  {\bf 87}, 111601 (2001)
[hep-ph/0101337].

\bibitem{Kochelev:1999tc}
N.~I.~Kochelev, V.~Vento and A.~V.~Vinnikov,
Phys.\ Lett.\ B {\bf 472}, 247 (2000)
[hep-ph/9905438].

\bibitem{Diakonov:1995ea}
D.~Diakonov,
Proceedings of the International School of Physics, 
'Enrico Fermi', Course 80: Selected Topics in Nonperturbative QCD, Varenna,
Italy, 1995, 
[hep-ph/9602375].

\bibitem{Schafer:1996wv}
T.~Sch{\"a}fer and E.~V.~Shuryak,
Rev.\ Mod.\ Phys.\  {\bf 70}, 323 (1998)
[hep-ph/9610451].

\bibitem{Callan:1978ye}
C.~G.~Callan, R.~F.~Dashen, D.~J.~Gross, F.~Wilczek and A.~Zee,
Phys.\ Rev.\ D {\bf 18}, 4684 (1978).

\bibitem{Callan:1977gz}
C.~G.~Callan, R.~F.~Dashen and D.~J.~Gross,
Phys.\ Rev.\ D {\bf 17}, 2717 (1978).

\bibitem{'tHooft:up}
G.~'t Hooft,
Phys.\ Rev.\ Lett.\ {\bf 37}, 8 (1976).

\bibitem{Shifman:nz}
M.~A.~Shifman, A.~I.~Vainshtein and V.~I.~Zakharov,
Nucl.\ Phys.\ B {\bf 165}, 45 (1980).

\bibitem{Shifman:uw}    
M.~A.~Shifman, A.~I.~Vainshtein and V.~I.~Zakharov,
Nucl.\ Phys.\ B {\bf 163}, 46 (1980).

\bibitem{PDG}
K.~Hagiwara {\it et al.}  [Particle Data Group Collaboration],
Phys.\ Rev.\ D {\bf 66}, 010001 (2002).

\bibitem{Narison:2002hk}
S.~Narison,
preprint, hep-ph/0202200.

\bibitem{Feldmann:1999uf}
T.~Feldmann,
Int.\ J.\ Mod.\ Phys.\ A {\bf 15}, 159 (2000)
[hep-ph/9907491].

\bibitem{Novikov:va}
V.~A.~Novikov, M.~A.~Shifman, A.~I.~Vainshtein and V.~I.~Zakharov,
Nucl.\ Phys.\ B {\bf 165}, 67 (1980).

\bibitem{Narison:1996fm}
S.~Narison,
Nucl.\ Phys.\ B {\bf 509}, 312 (1998)
[hep-ph/9612457].

\bibitem{Schafer:1994fd}
T.~Sch{\"a}fer and E.~V.~Shuryak,
Phys.\ Rev.\ Lett.\  {\bf 75}, 1707 (1995)
[hep-ph/9410372].

\bibitem{Michael:1995br}
C.~Michael and P.~S.~Spencer,
Phys.\ Rev.\ D {\bf 52}, 4691 (1995)
[hep-lat/9503018].

\bibitem{Smith:1998wt}
D.~A.~Smith and M.~J.~Teper  [UKQCD collaboration],
Phys.\ Rev.\ D {\bf 58}, 014505 (1998)
[hep-lat/9801008].

\bibitem{deForcrand:1997sq}
P.~de Forcrand, M.~Garcia Perez and I.~O.~Stamatescu,
Nucl.\ Phys.\ B {\bf 499}, 409 (1997)
[hep-lat/9701012].

\bibitem{DeGrand:1997gu}
T.~DeGrand, A.~Hasenfratz and T.~G.~Kovacs,
Nucl.\ Phys.\ B {\bf 505}, 417 (1997)
[hep-lat/9705009].

\bibitem{Sexton:1995kd}
J.~Sexton, A.~Vaccarino and D.~Weingarten,
Phys.\ Rev.\ Lett.\  {\bf 75}, 4563 (1995)
[hep-lat/9510022].

\bibitem{Lee:1999kv}
W.~J.~Lee and D.~Weingarten,
Phys.\ Rev.\ D {\bf 61}, 014015 (2000)
[hep-lat/9910008].

\bibitem{Minkowski:1998mf}
P.~Minkowski and W.~Ochs,
Eur.\ Phys.\ J.\ C {\bf 9}, 283 (1999)
[hep-ph/9811518].

\bibitem{Close:1996yc}
F.~E.~Close, G.~R.~Farrar and Z.~p.~Li,
Phys.\ Rev.\ D {\bf 55}, 5749 (1997)
[hep-ph/9610280].

\bibitem{Franz:2000ee}
M.~Franz, M.~V.~Polyakov and K.~Goeke,
Phys.\ Rev.\ D {\bf 62}, 074024 (2000)
[hep-ph/0002240].

\bibitem{Novikov:uy}
V.~A.~Novikov, M.~A.~Shifman, A.~I.~Vainshtein and V.~I.~Zakharov,
Nucl.\ Phys.\ B {\bf 165}, 55 (1980).

\bibitem{Partridge:1980vk}
R.~Partridge {\it et al.},
Phys.\ Rev.\ Lett.\  {\bf 45}, 1150 (1980).

\bibitem{Baltrusaitis:1985mr}
R.~M.~Baltrusaitis {\it et al.}  [Mark-III Collaboration],
Phys.\ Rev.\ D {\bf 33}, 629 (1986).

\bibitem{Shuryak:1992ke}
E.~V.~Shuryak and J.~J.~M.~Verbaarschot, 
Nucl.\ Phys.\ B {\bf 410}, 55 (1993)
[hep-ph/9302239].

\bibitem{Anselmino:1993bd}
M.~Anselmino and S.~Forte,
Phys.\ Lett.\ B {\bf 323}, 71 (1994)
[hep-ph/9311365].

\bibitem{Shifman:1988zk}
M.~A.~Shifman,
Phys.\ Rept.\  {\bf 209}, 341 (1991).

\bibitem{Jaminon:ac}
M.~Jaminon, M.~Mathot and B.~van den Bossche, 
Nucl.\ Phys.\ A {\bf 662}, 157 (2000)

\bibitem{Jin:2002up}
H.~y.~Jin, X.~m.~Zhang,
Phys.\ Rev.\ D {\bf 66}, 057505 (2002)
[hep-ph/0208120].

\bibitem{Ringwald:1989ee}
A.~Ringwald,
Nucl.\ Phys.\ B {\bf 330} (1990) 1.

\bibitem{Espinosa:qn}
O.~Espinosa,
Nucl.\ Phys.\ B {\bf 343} (1990) 310.

\bibitem{Nowak:2000de}
M.~A.~Nowak, E.~V.~Shuryak and I.~Zahed,
Phys.\ Rev.\ D {\bf 64}, 034008 (2001)
[hep-ph/0012232].




\bibitem{Ashman:1987hv}
J.~Ashman {\it et al.}  [European Muon Collaboration],
Phys.\ Lett.\ B {\bf 206}, 364 (1988).

\bibitem{Filippone:2001ux}
B.~W.~Filippone and X.~D.~Ji,
Adv.\ Nucl.\ Phys.\  {\bf 26}, 1 (2001)
[hep-ph/0101224].

\bibitem{Dorokhov:ym}
A.~E.~Dorokhov, N.~I.~Kochelev and Y.~A.~Zubov,
Int.\ J.\ Mod.\ Phys.\ A {\bf 8}, 603 (1993).

\bibitem{Bass:1993bs}
S.~D.~Bass and A.~W.~Thomas,
Prog.\ Part.\ Nucl.\ Phys.\  {\bf 33}, 449 (1994)
[hep-ph/9310306].

\bibitem{Bass:2003vp}
S.~D.~Bass,
Acta Phys.\ Polon.\ B {\bf 34}, 5893 (2003)
[hep-ph/0311174].

\bibitem{Altarelli:1988nr}
G.~Altarelli and G.~G.~Ross,
Phys.\ Lett.\ B {\bf 212}, 391 (1988).

\bibitem{Jaffe:1989jz}
R.~L.~Jaffe and A.~Manohar,
Nucl.\ Phys.\ B {\bf 337}, 509 (1990).

\bibitem{Shifman:zn}
M.~A.~Shifman, A.~I.~Vainshtein and V.~I.~Zakharov,
Phys.\ Lett.\ B {\bf 78}, 443 (1978).

\bibitem{Anselm:1992wz}
A.~Anselm,
Phys.\ Lett.\ B {\bf 291}, 455 (1992).

\bibitem{Kuhn:1990df}
J.~H.~Kuhn and V.~I.~Zakharov,
Phys.\ Lett.\ B {\bf 252}, 615 (1990).

\bibitem{Narison:hv}
S.~Narison, G.~M.~Shore and G.~Veneziano,
Nucl.\ Phys.\ B {\bf 433}, 209 (1995)
[hep-ph/9404277].


\bibitem{Shuryak:1994rr}
E.~V.~Shuryak and J.~J.~Verbaarschot,
Phys.\ Rev.\ D {\bf 52}, 295 (1995)
[hep-lat/9409020].

\bibitem{Schafer:2000hn}
T.~Sch{\"a}fer and E.~V.~Shuryak,
preprint, hep-lat/0005025.

\bibitem{Schafer:1993ra}
T.~Sch{\"a}fer, E.~V.~Shuryak and J.~J.~Verbaarschot,
Nucl.\ Phys.\ B {\bf 412}, 143 (1994)
[hep-ph/9306220].

\bibitem{Diakonov:2002fq}
D.~Diakonov,
Prog.\ Part.\ Nucl.\ Phys.\  {\bf 51}, 173 (2003)
[hep-ph/0212026].

\bibitem{Brown:1977eb}
L.~S.~Brown, R.~D.~Carlitz, D.~B.~Creamer and C.~K.~Lee,
Phys.\ Rev.\ D {\bf 17}, 1583 (1978).

\bibitem{Andrei:xg}
N.~Andrei and D.~J.~Gross,
Phys.\ Rev.\ D {\bf 18}, 468 (1978).

\bibitem{Nason:1993ak}
P.~Nason and M.~Porrati,
Nucl.\ Phys.\ B {\bf 421}, 518 (1994)
[hep-ph/9302211].

\bibitem{Geshkenbein:vb}
B.~V.~Geshkenbein and B.~L.~Ioffe,
Nucl.\ Phys.\ B {\bf 166}, 340 (1980).

\bibitem{Dorokhov:2003kf}
A.~E.~Dorokhov and W.~Broniowski,
Eur.\ Phys.\ J.\ C {\bf 32}, 79 (2003)
[hep-ph/0305037].

\bibitem{Dubovikov:bf}
M.~S.~Dubovikov and A.~V.~Smilga,
Nucl.\ Phys.\ B {\bf 185}, 109 (1981).

\bibitem{Schafer:1994nv}
T.~Sch{\"a}fer, E.~V.~Shuryak and J.~J.~M.~Verbaarschot,
Phys.\ Rev.\ D {\bf 51}, 1267 (1995)
[hep-ph/9406210].

\bibitem{Schafer:1995pz}
T.~Sch{\"a}fer and E.~V.~Shuryak,
Phys.\ Rev.\ D {\bf 53}, 6522 (1996)
[hep-ph/9509337].

\bibitem{Schafer:1995uz}
T.~Sch{\"a}fer and E.~V.~Shuryak,
Phys.\ Rev.\ D {\bf 54}, 1099 (1996)
[hep-ph/9512384].

\bibitem{Isgur:2000ts}
N.~Isgur and H.~B.~Thacker,
Phys.\ Rev.\ D {\bf 64}, 094507 (2001)
[hep-lat/0005006].

\bibitem{Weinberg:gf}
S.~Weinberg,
Phys.\ Rev.\ Lett.\  {\bf 67}, 3473 (1991).

\bibitem{Vogl:1991qt}
U.~Vogl and W.~Weise,
Prog.\ Part.\ Nucl.\ Phys.\  {\bf 27}, 195 (1991).

\bibitem{Steininger:ed}
K.~Steininger and W.~Weise,
Phys.\ Rev.\ D {\bf 48}, 1433 (1993).

\bibitem{Dorokhov:2001pz}
A.~E.~Dorokhov,
preprint, hep-ph/0112332.

\bibitem{Kochelev:1997ux}
N.~I.~Kochelev,
Phys.\ Rev.\ D {\bf 57}, 5539 (1998)
[hep-ph/9711226].

\bibitem{Ioffe:kw}
B.~L.~Ioffe,
Nucl.\ Phys.\ B {\bf 188}, 317 (1981)
[Erratum-ibid.\ B {\bf 191}, 591 (1981)].

\bibitem{Rapp:1997zu}
R.~Rapp, T.~Sch{\"a}fer, E.~V.~Shuryak and M.~Velkovsky,
Phys.\ Rev.\ Lett.\  {\bf 81}, 53 (1998)
[hep-ph/9711396].

\bibitem{Leinweber:1994nm}
D.~B.~Leinweber,
Phys.\ Rev.\ D {\bf 51}, 6383 (1995)
[nucl-th/9406001].

\bibitem{Anselmino:1992vg}
M.~Anselmino, E.~Predazzi, S.~Ekelin, S.~Fredriksson and D.~B.~Lichtenberg,
Rev.\ Mod.\ Phys.\  {\bf 65}, 1199 (1993).

\bibitem{Dolgov:1998js}
D.~Dolgov, R.~Brower, J.~W.~Negele and A.~Pochinsky,
Nucl.\ Phys.\ Proc.\ Suppl.\  {\bf 73}, 300 (1999)
[hep-lat/9809132].

\bibitem{Dong:1995rx}
S.~J.~Dong, J.~F.~Lagae and K.~F.~Liu,
Phys.\ Rev.\ Lett.\  {\bf 75}, 2096 (1995)
[hep-ph/9502334].

\bibitem{Forte:1990xb}
S.~Forte and E.~V.~Shuryak,
Nucl.\ Phys.\ B {\bf 357}, 153 (1991).

\bibitem{Hutter:1995cs}
M.~Hutter,
preprint, hep-ph/9509402.

\bibitem{Kacir:1996qn}
M.~Kacir, M.~Prakash and I.~Zahed,
Acta Phys.\ Polon.\ B {\bf 30}, 287 (1999)
[hep-ph/9602314].

\bibitem{Diakonov:1995qy}
D.~Diakonov, M.~V.~Polyakov and C.~Weiss,
Nucl.\ Phys.\ B {\bf 461}, 539 (1996)
[hep-ph/9510232].

\bibitem{Diakonov:1987ty}
D.~Diakonov, V.~Y.~Petrov and P.~V.~Pobylitsa,
Nucl.\ Phys.\ B {\bf 306}, 809 (1988).

\bibitem{Brodsky:1988ip}
S.~J.~Brodsky, J.~R.~Ellis and M.~Karliner,
Phys.\ Lett.\ B {\bf 206}, 309 (1988).

\bibitem{Blotz:1993am}
A.~Blotz, M.~V.~Polyakov and K.~Goeke,
Phys.\ Lett.\ B {\bf 302}, 151 (1993).

\bibitem{Faccioli:2003yy}
P.~Faccioli,
preprint,
hep-ph/0312019.

\bibitem{Maiani:ca}
L.~Maiani and M.~Testa,
Phys.\ Lett.\ B {\bf 245}, 585 (1990).



\bibitem{Creutz1}
Michael Creutz: Quarks, Gluons and Lattices, Cambridge University Press, 
Cambridge, 1983

\bibitem{Shifman2}M.A.Shifman, A.I.Vainshtein, V.I.Zakharov, Nucl.Phys. 
{\bf B 163}: 46-56 (1980)

\bibitem{Zahed}S.Chernyshev, M.A.Nowak, I. Zahed, Phys.Rev. {\bf D 
53}: 5176 - 5184 (1996)

\bibitem{Prosen}T.Prosen, T.H.Seligman, H.A.Weidenm\"{u}ller, 
Europhys.Lett. {\bf 55}(1): 12-18 (2001)

\bibitem{Eriksson}K.E.Eriksson, N.Svartholm, R.S.Skagerstam, 
J.Math.Phys. {\bf 22}: 2276 (1981)

\bibitem{Fateev}V.A.Fateev, E.Onofri, Lett.Math.Phys {\bf 5}: 367 
(1981)

\bibitem{Brower}R.Brower, M.Nauenberg, Nucl.Phys. {\bf B 180}: 221 
(1981)

\bibitem{Creutz2}
M. Creutz,  J.Math.Phys. {\bf 19}(10): 2043-2046 (1978), Rev.Mod.Phys. 
{\bf 50}(3): 561-571 (1978)

\bibitem{Jaap}Jaap Hoek, Phys.Lett. {\bf B 102}: 129 (1981)

\bibitem{Bars}I.Bars, Physica Scripta {\bf 23}: 983-986 (1981) and 
J.Mth.Phys. {\bf 21}(11): 2676-2681 (1980)

\bibitem{Balantekin}A.B.Balantekin, Phys.Rev. {\bf D 62}: 085017 
(2000) 
and A.B.Balantekin, P.Cassak, J.Math.Phys. {\bf 43}:604-620 (2002)

\bibitem{CORE}
V.I.Borodulin, R.N.Rogalyov, S.R.Slabospitsky, hep-ph/9507456

\bibitem{General}
A.J.Macfarlane, A.Sudbery, P.H.Weisz, Commun. math. Phys. {\bf 
11}: 77-90(1968)

\bibitem{Invar}P.Dittner, Commun. math. Phys. {\bf 22}: 238-252 (1971)

\end{thebibliography}
\end{document}